\documentclass[preprintnumbers, floatfix, letterpaper, twocolumn,aps,prd,epsfig,nofootinbib,natbib,longbibliography]{revtex4-1}

\usepackage{graphicx}
\usepackage{epstopdf}
\usepackage{latexsym}
\usepackage{amssymb}
\usepackage{amsmath}
\usepackage{color}
\usepackage{mathrsfs}
\usepackage{xparse}
\usepackage{bbding}
\usepackage{pifont}
\usepackage{comment}
\usepackage{ulem}
\usepackage{float}
\usepackage[inline]{enumitem}
\usepackage{soul}
\usepackage[caption=false]{subfig}

\let\Right\right
\let\Left\left
\makeatletter
\def\right#1{\Right#1\@ifnextchar){\!\right}{}}
\def\left#1{\Left#1\@ifnextchar({\!\left}{}}
\makeatother

\usepackage[
            pdfstartview=FitH,
            bookmarksnumbered=true,
            bookmarksopen=true,
            colorlinks,
            linkcolor=blue,
            anchorcolor=green,
            citecolor=blue
            ]{hyperref}
\begin{document}

  \renewcommand\arraystretch{2}
 \newcommand{\bq}{\begin{equation}}
 \newcommand{\eq}{\end{equation}}
 \newcommand{\bqn}{\begin{eqnarray}}
 \newcommand{\eqn}{\end{eqnarray}}
 \newcommand{\nb}{\nonumber}
 \newcommand{\lb}{\label}
 
\newcommand{\La}{\Lambda}
\newcommand{\va}{\scriptscriptstyle}
\newcommand{\be}{\nopagebreak[3]\begin{equation}}
\newcommand{\ee}{\end{equation}}

\newcommand{\ba}{\nopagebreak[3]\begin{eqnarray}}
\newcommand{\ea}{\end{eqnarray}}

\newcommand{\la}{\label}
\newcommand{\n}{\nonumber}

\newcommand{\R}{\mathbb{R}}

 \newcommand{\cb}{\color{blue}}
    \newcommand{\cc}{\color{cyan}}
        \newcommand{\cm}{\color{magenta}}
\newcommand{\rc}{\rho^{\scriptscriptstyle{\mathrm{I}}}_c}
\newcommand{\rd}{\rho^{\scriptscriptstyle{\mathrm{II}}}_c} 
\NewDocumentCommand{\evalat}{sO{\big}mm}{%
  \IfBooleanTF{#1}
   {\mleft. #3 \mright|_{#4}}
   {#3#2|_{#4}}%
}
\newcommand{\PRL}{Phys. Rev. Lett.}
\newcommand{\PL}{Phys. Lett.}
\newcommand{\PR}{Phys. Rev.}
\newcommand{\CQG}{Class. Quantum Grav.}

\title{Periodic orbits and their gravitational wave radiations in $\gamma$-metric}
 
\author{Chao Zhang${}^{a, b}$}
\email{ phyzc@cup.edu.cn}

\author{Tao Zhu${}^{c, d}$}
\email{corresponding author:  zhut05@zjut.edu.cn}

\affiliation{${}^{a}$Basic Research Center for Energy Interdisciplinary, College of Science, China University of Petroleum-Beijing, Beijing 102249, China\\
${}^{b}$Beijing Key Laboratory of Optical Detection Technology for Oil and Gas, China University of Petroleum-Beijing, Beijing 102249, China\\
${}^{c}$ Institute for Theoretical Physics and Cosmology, Zhejiang University of Technology, Hangzhou, 310032, China\\
${}^{d}$ United Center for Gravitational Wave Physics (UCGWP), Zhejiang University of Technology, Hangzhou, 310032, China}

\date{\today}

\begin{abstract}

The $\gamma$-metric, also known as Zipoy-Voorhees spacetime, is a static, axially symmetric vacuum solution to Einstein’s field equations characterized by two parameters: mass and the deformation parameter $\gamma$. It reduces to the Schwarzschild metric when $\gamma = 1$. In this paper, we explore potential signatures of the $\gamma$-metric on periodic orbits and their gravitational-wave radiation. Periodic orbits are classified by a rotational number specified by three topological numbers $(z, w, v)$, each triple corresponding to characteristic zoom-whirl behavior. We show that deviations from $\gamma=1$ alter the radii and angular momentum of bound orbits and thereby shift the $(z, w, v)$ taxonomy. We also compute representative gravitational waveforms for certain periodic orbits and demonstrate that $\gamma \neq 1$ can induce phase shifts and amplitude modulations correlated with changes in the zoom–whirl structure. In particular, larger zoom numbers lead to increasingly complex substructures in the waveforms, and finite deviations from $\gamma=1$ can significantly modify these features. Our results indicate that precise measurements of waveform morphology from extreme-mass-ratio inspirals may constrain deviations from spherical symmetry encoded in $\gamma$.

\end{abstract}


\maketitle
\section{Introduction}
\renewcommand{\theequation}{1.\arabic{equation}} \setcounter{equation}{0}
\label{secI}

The $\gamma$-metric considered in this work was originally introduced by Zipoy and Voorhees \cite{Zipoy:1966btu, Voorhees:1970ywo} and is commonly referred to as the Zipoy-Voorhees spacetime (see, e.g., \cite{Benavides-Gallego:2018htf}). The Zipoy-Voorhees solution belongs to the Weyl class, and represents a two-parameter family of static, axisymmetric, and asymptotically flat vacuum solutions to the Einstein equations, characterized by a mass-like parameter $m$ and an ``oblateness" parameter $\gamma$ that quantifies departures from spherical symmetry \cite{Lukes-Gerakopoulos:2012qpc, Kodama:2003ch, Saito:2024hzc, Chakrabarty:2022fbd}. In this sense, it can be viewed as a generalization of the Schwarzschild geometry that admits prolate or oblate deformations \cite{Toshmatov:2019bda}. Over the past decade, this spacetime has attracted considerable attention as a tractable model for deviations from spherical symmetry of the Schwarzschild metric. 

Recently, its properties and applications have been extended to areas such as charged compact bodies \cite{Yunusov:2025chw}, Shapiro time delay \cite{Utepova:2025jkx}, tidal force \cite{Heidmann:2025yzd}, accretion disk \cite{Turimov:2023lbn}, naked singularities \cite{Gurtug:2023pfp}, etc., and a higher-dimensional version of it \cite{Hajibarat:2021ngf}. A related family of solutions, the $q$-metric, has also been extensively investigated \cite{Faraji:2025aeg, Momynov:2024bhn, Lora-Clavijo:2023ukh, Boshkayev:2015jaa, Idrissov:2025ugs}. Other notable studies include the analysis of nonintegrability and chaotic features \cite{Lukes-Gerakopoulos:2012qpc, Destounis:2023khj, Destounis:2023cim}, the sensitivity of test particle trajectories to perturbations in $\gamma$ \cite{Herrera:1998rj}, the dynamics of magnetized particles \cite{Abdujabbarov:2020hdp}, 
a limitation of that which leads to the Levi-Civita spacetime \cite{Herrera:1998eq}, and the gravitational lensing on neutrino oscillations \cite{Chakrabarty:2021bpr, Toshmatov:2019qih}, etc.

When $\gamma = 1$ the solution reduces to the Schwarzschild one, while $\gamma > 1$ ($\gamma < 1$) describes an oblate (prolate) source \cite{Chakrabarty:2021bpr}. Crucially, for $\gamma \neq 1$ the hypersurface $r = 2m$ is not an event horizon but a genuine curvature singularity \cite{Benavides-Gallego:2018htf}, so the $\gamma$-metric provides a simple setting in which to study naked singularities and horizonless compact objects \cite{Gurtug:2023pfp, Shaikh:2021cvl, Bambi:2015kza}. The presence of a nonvanishing quadrupole moment for $\gamma \neq 1$ further enriches its phenomenology and observational signatures \cite{Benavides-Gallego:2018htf}. These properties motivate using the $\gamma$-metric as a testbed to explore how departures from spherical symmetry of the Schwarzschild metric affect strong-field orbital dynamics and gravitational-wave (GW) emission.

The direct detection of GWs emitted from binary black holes (BHs) etc. by Advanced LIGO has opened an era of experimental strong-field gravity \cite{LIGOScientific:2016aoc}. Ground-based networks and planned next-generation observatories (Cosmic Explorer, Einstein Telescope, etc.), together with future space-based missions, will expand accessible frequency bands and increase sensitivity, stimulating theoretical work on extreme mass-ratio inspiral (EMRI) waveforms, quasi-normal mode (QNM) spectra, and environmental influences on GW emission \cite{Zhang:2022rfr, Tu:2023xab, Shen:2023erj, Zhang:2021bdr, Zhang:2022roh, Zi:2023pvl, Zi:2024lmt}. In particular, EMRIs are especially attractive targets because their dynamics are typically ``clean" (weakly affected by internal structure or strong external perturbations), making them precise probes of the background spacetime.

On the other hand, observational evidence for supermassive BHs (SMBHs) at galactic centers — for example, the Event Horizon Telescope images of M87* and Sgr A* \cite{EventHorizonTelescope:2019dse, EventHorizonTelescope:2022wkp} — combined with the long inspiral timescales of EMRIs, make these systems excellent laboratories for precise tests of gravity. In an EMRI, a stellar-mass compact object follows an approximately geodesic orbit in the background spacetime of the central object, emitting GWs at harmonics of the orbital frequencies. Because the energy and angular momentum radiated away by the smaller object are almost negligible per orbit, the inspiral is adiabatic on long timescales compared with the orbital period; consequently, space-based detectors such as LISA, TianQin, and Taiji are expected to track these signals over many cycles and thereby map the central geometry with high fidelity \cite{LISA:2017pwj, Gair:2004iv, Schutz:1999xj, Danzmann:1997hm, Liu:2020eko, Shi:2019hqa, Ruan:2018tsw, Gong:2021gvw, Zi:2021pdp, Zi:2023geb, Bian:2025ifp}. Furthermore, SMBHs can significantly modify the distribution of dark matter in their vicinity, leaving imprints on the GW signals.

For bound equatorial motion of an EMRI system, in general, it can be well approximated by periodic orbits, a typical orbit that can be decomposed into periodic radial oscillations combined with secular azimuthal advance. Periodic (or near-periodic) geodesics often exhibit zoom-whirl behavior and have been analyzed thoroughly in Schwarzschild and Kerr spacetimes by using a characterization scheme described by three topological numbers \cite{Levin:2008ci, Levin:2009sk, Bambhaniya:2020zno, Rana:2019bsn, Levin:2008mq}. With this scheme, extensive research has been carried out on periodic orbits within various spacetimes with compact objects, including naked singularities \cite{Babar:2017gsg}, Kerr-Sen BHs \cite{Liu:2018vea}, and hairy BHs in Horndeski's theory \cite{Lin:2023rmo}, etc. For the studies of periodic orbits in other BHs, see refs.~\cite{Yao:2023ziq, Lin:2022llz, Chan:2025ocy, Wang:2022tfo, Lin:2023eyd, Haroon:2025rzx, Habibina:2022ztd, Zhang:2022psr, Lin:2022wda, Gao:2021arw, Lin:2021noq, Deng:2020yfm, Tu:2023xab, Zhou:2020zys, Gao:2020wjz, Deng:2020hxw, Azreg-Ainou:2020bfl, Wei:2019zdf, Pugliese:2013xfa,Zhang:2022zox, Healy:2009zm, Wang:2025wob, Alloqulov:2025bxh, Wei:2025qlh} and references therein. The periodic orbits also give rise to characteristic waveform features, which has been explored in a large number of spacetimes \cite{Tu:2023xab, Yang:2024lmj, Shabbir:2025kqh, Junior:2024tmi,  Jiang:2024cpe, Yang:2024cnd, Li:2024tld, QiQi:2024dwc, Haroon:2025rzx, Alloqulov:2025ucf, Wang:2025hla, Lu:2025cxx, Zare:2025aek, Gong:2025mne, Li:2025sfe, Choudhury:2025qsh, Chen:2025aqh, Deng:2025wzz, Li:2025eln, Zahra:2025tdo, Sharipov:2025yfw, Ahmed:2025azu} and references therein. 

Given the richer multipolar structure and horizonless character of the $\gamma$-metric, it is natural to ask how its periodic orbit taxonomy and associated GW signatures differ from the Schwarzschild case (viz., a BH \cite{Babar:2017gsg}). Such differences could help distinguish compact-object models observationally and quantify how sensitive EMRI waveforms are to departures from spherical symmetry.

In this paper, we study equatorial periodic orbits in the $\gamma$-metric and compute representative \textcolor{black}{gravitational-wave signals produced by a point-like mass around a massive compact object on such orbits.} The rest of the paper is organized as follows. In Sec.\ref{secII} will briefly review some of the important features of the $\gamma$-metric, and a comparison to that of the Schwarzschild one is presented. After that, we shall calculate the effective potential, the marginally bound orbit as well as the innermost stable circular orbit for the center super massive body (which acts as the background spacetime characterized by the $\gamma$-metric) and the small object (which can be treated as a test particle) in Sec.\ref{secIII}. The periodic orbits for different choices of parameters will be discussed in Sec.\ref{secIV} and some of the corresponding gravitational waveforms will be exhibited in Sec.\ref{secV}. Finally, we shall address some concluding remarks and outlooks in Sec.\ref{secVI}.
 
In this paper, we adopt the unit system so that $c=G_N=1$, where $c$ denotes the speed of light and $G_N$ is the Newtonian gravitational constant. In this way, we are left with one degree of freedom for choosing the unit for \{length, time, mass\}. We adopt the standard convention: Greek indices ($\mu,\nu,\dots$) run from $0$ to $3$, and Latin indices ($i,j,k,\dots$) run from $1$ to $3$. Throughout the paper, we use the signature $(-\;+\;+\;+)$.

\section{The spacetime described by the $\gamma$-metric}
\renewcommand{\theequation}{2.\arabic{equation}} \setcounter{equation}{0}
\label{secII}

In this section, we provide a brief introduction to the $\gamma$ spacetime, which is characterized by two parameters, a mass-like parameter $m$ and an ``oblateness" parameter $\gamma$ \cite{Abdikamalov:2019ztb}. It is treated as the background supermassive compact body (in the role of a SMBH \footnote{\textcolor{black}{It is worth mentioning here that, as pointed out in last section, the $\gamma \neq 1$ case actually correspond to a description to a compact body with naked singularity, in lieu of a BH (see, e.g., \cite{Babar:2017gsg} for a similar scenario, in which the GWs emitted from EMRI systems were also discussed). It becomes a BH only for the $\gamma =1 $ case, namely, the Schwarzschild one. As to be shown in the following, in this paper we mainly focus on the $\gamma \neq 1$ case. We are free to compare our results with the Schwarzschild one and study the EMRI system with the $\gamma$-metric since in paper we are solely bothered by the exterior space of the compact object. In contrast, the interior nature of the compact object can be ignored for the current study. Because of this, the compact body described by the $\gamma$-metric is often considered as a BH mimicker, as mentioned in \cite{Benavides-Gallego:2018htf, Abdikamalov:2019ztb}. These facts enable us treating the central compact body described by the $\gamma$-metric as playing the role of a SMBH.} }). With the help of the Erez-Rosen coordinate system \cite{Erez1959}, such a solution is given in the form of a line element as
\bqn
\lb{ds2}
d s^2 &=& -F dt^2 \nb\\
&& + F^{-1} [G dr^2 + H d \theta^2 + r(r-2 m) \sin^2\theta d\phi^2], \nb\\
\eqn
where
\bqn
\lb{FGH}
F(r) &=& \left(1-\frac{2 m}{r}\right)^\gamma,\nb\\
G(r, \theta) &=& \left[\frac{r(r-2 m)}{r^2 -2 m r + m^2 \sin^2\theta}\right]^{\gamma^2-1}, \nb\\
H(r, \theta) &=& \frac{[r(r-2m)]^{\gamma^2}}{(r^2 - 2 m r+m^2 \sin^2 \theta)^{\gamma^2-1}},
\eqn
and $m$ is related to the total mass of the source $M$ by $M=\gamma m$, where $\gamma$ is a dimensionless parameter. 

Here are two crucial features of the metric \eqref{ds2}. First of all, as can be seen from the expression of function $F(r)$, the $r<2m$ region can be a little subtle as we have $1-2m/r<0$ in this region. Fortunately, as will be seen shortly, such a problem will not bother the main discussions in this paper. Secondly, the singularity led by the condition $g^{rr}=0$  (which mimics the event horizon of a BH \cite{Carroll:2004st}) does not always exist. Actually, the $g^{rr}$  is found to be
 \bqn
\lb{grr}
g^{rr} &=& r^{1-\gamma^2-\gamma} [(r-m)^2]^{\gamma^2-1} (r-2m)^{\frac{5}{4}-\left(\gamma-\frac{1}{2}\right)^2}. \nb\\
\eqn
Therefore, one of the allowed regions for this singular point to exist is $\gamma \in (1/2-\sqrt{5}/2, 1/2+\sqrt{5}/2)$. \footnote{\textcolor{black}{This $\gamma$ parameter is required to be within the region of $(0.5, 1)$ under certain nonintegrability restrictions, as pointed out in \cite{Gurtug:2021noy}. However, here we care more about the cases when a healthy solution of $g^{rr}=0$ exists, so a $\gamma$ outside the region of $(0.5, 1)$ is discussed in some parts of this paper. 
In addition to the theoretical values, such a broad discussion provide us with a more comprehensive analysis about the deviations of our results to a Schwarzschild one. 
On the other hand,  to minimize potential contradictions within our analysis or calculations, the regime of $\gamma \in (0.5, 1)$ is also under consideration. By mimicking the treatment of \cite{Abdikamalov:2019ztb}, some of the key examples will be given for this regime of $\gamma$.} 
} 
In that case, the singularity locates at the point $r=r_h\equiv2m$ \footnote{\textcolor{black}{An investigation to the Kretschmann scalar has shown that this $r=r_h$ point does not represent an event horizon \cite{Benavides-Gallego:2018htf}. Instead, it behaves as an infinitely redshifted surface \cite{Abdikamalov:2019ztb}. } }. 


\begin{figure}[h]
\includegraphics[width=0.8
\linewidth]{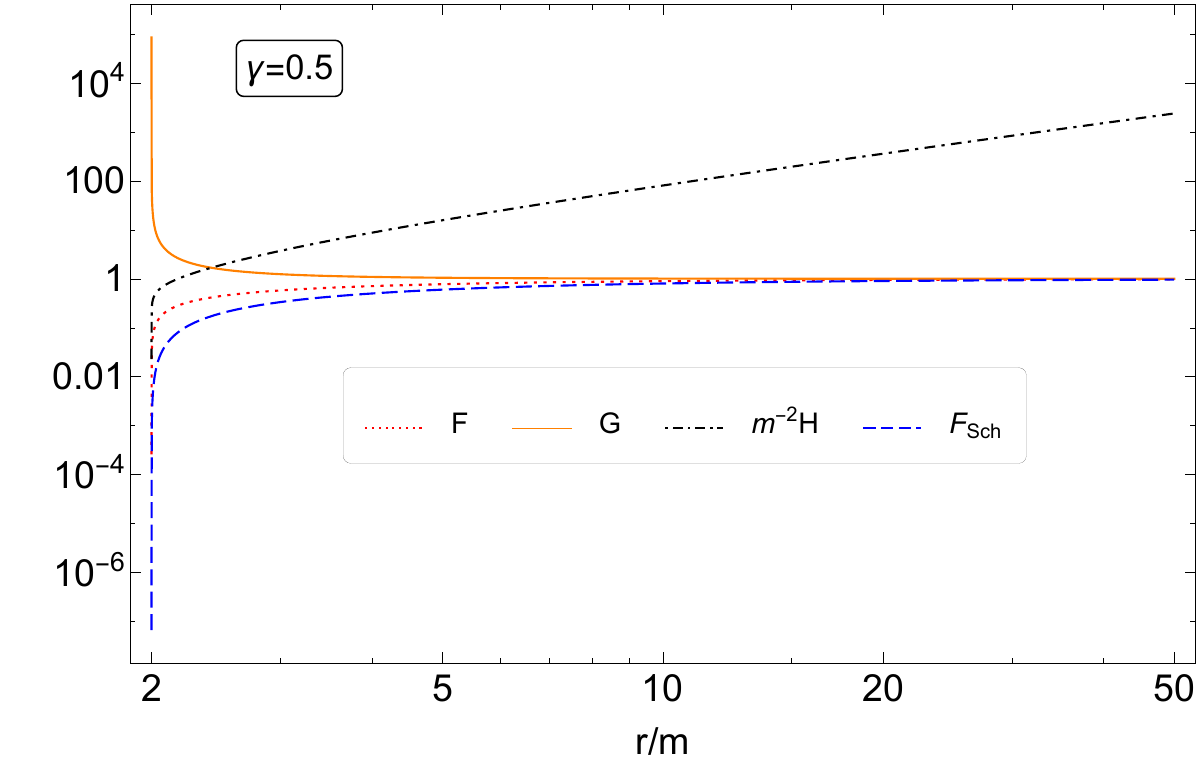} 
\includegraphics[width=0.8
\linewidth]{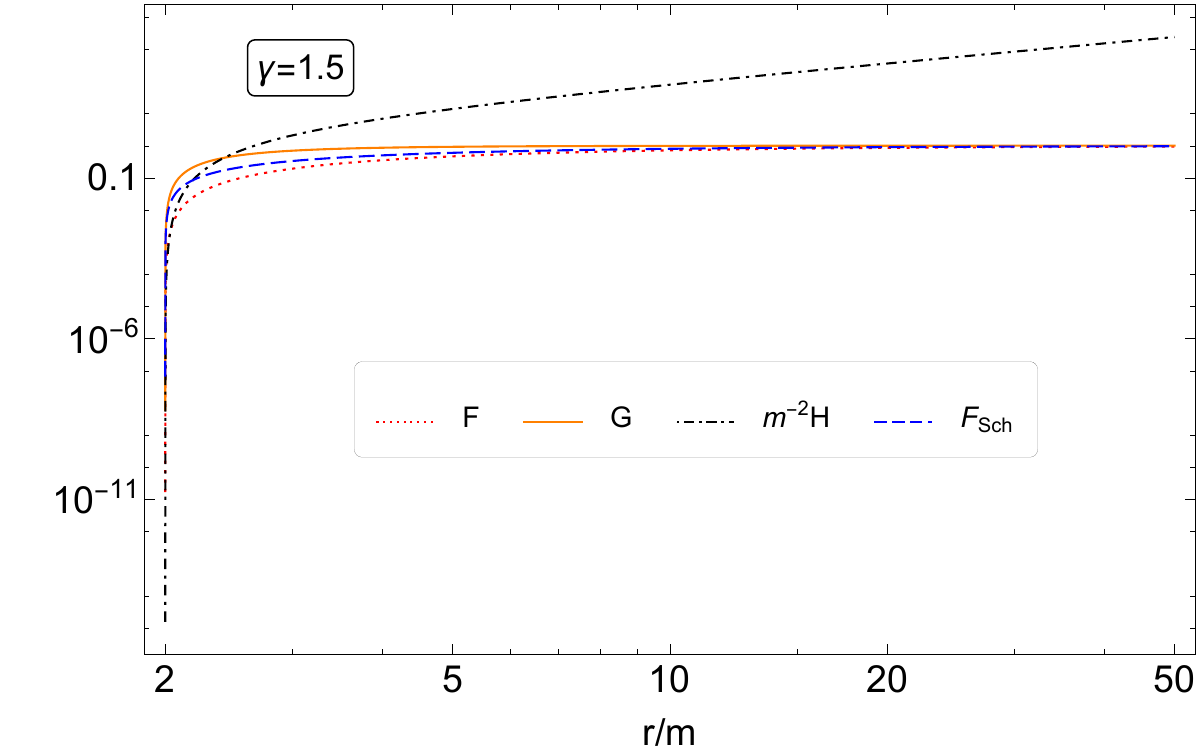} 
\caption{The behaviors of dimensionless quantities $F$, $G$ and $m^{-2}H$ as functions of $r/m$. For the upper panel we choose $\gamma=0.5$ while for the lower panel we choose $\gamma=1.5$. The function $F$ for the Schwarzschild case (denoted as $F_{\text{Sch}}$) is also exhibited in here as a comparison. 
} 
\label{plotFGH}
\end{figure} 

Now let us first explore and plot out the behavior of dimensionless quantities $F$, $G$, and $H/m^2$ within the regime of $r/m \in (2, +\infty)$ at the equatorial plane (i.e., $\theta=\pi/2$). This part of the plot is shown in Fig.\ref{plotFGH}. Within the allowed regime $\gamma \in (1/2-\sqrt{5}/2, 1/2+\sqrt{5}/2)$, we choose two typical values of $\gamma$ for this figure. For its upper panel, we set $\gamma=0.5$ while for the lower panel we set $\gamma=1.5$. By using \eqref{FGH}, it is not difficult to confirm that the functions for different components of the $\gamma$-metric are continuous and evolving in a healthy way outside the singularity $r_h$. Such a feature guarantees that one will not be bothered by the potentially bizarre behaviors (as mentioned earlier) within the singularity.

\section{Determine the bounding radii for the small object}
\renewcommand{\theequation}{3.\arabic{equation}} \setcounter{equation}{0}
\label{secIII}

In this section, we explore the geodesic motion, effective potential, and bound orbits of a small test particle orbiting the $\gamma$ spacetime. The corresponding Lagrangian that governs the test particle's motion is given by
 \bqn
\lb{Lagrangian}
{\mathcal{L}} &=& \frac{1}{2} g_{\mu \nu} \frac{d x^\mu}{d \lambda} \frac{d x^\nu}{d \lambda},
\eqn
where $\lambda$ is the affine parameter of the world line of the particle under consideration. For massless particles we have ${\mathcal{L}}=0$ while for massive ones we have ${\mathcal{L}}<0$. By working on top of this Lagrangian, we obtain the corresponding generalized momentum $p_\mu$, which is given by
\bqn
\lb{pmu}
p_\mu = \frac{\partial {\mathcal{L}}}{\partial {\dot x}^\mu} = g_{\mu \nu} {\dot x}^\nu.
\eqn
Here, a dot overhead denotes the derivative with respect to $\lambda$. It leads to the equations of motion \cite{Wang:2025hla, Tu:2023xab, Anjomshoa:2025tdl}
\bqn
\lb{EOM}
p_t = g_{tt} {\dot t} = -E, \nb\\
p_\phi = g_{\phi  \phi} {\dot \phi} = L, \nb\\
p_r = g_{rr} {\dot r}, \nb\\
p_\theta = g_{\theta \theta} {\dot \theta},
\eqn
where $E$ and $L$ represent the energy and angular momentum, respectively. 

For timelike geodesics, we have 
\bqn
\lb{geodesic}
g_{\mu \nu} {\dot x}^\mu {\dot x}^\nu &=& -1.
\eqn
Given the above equation, we are on the stage of determining the effective potential for our problem. In the following, we shall consider the case in the equatorial plane where $\theta(\lambda) = \pi/2$ (so that $d \theta/d \lambda=0$). 

\subsection{The effective potential}

We can then compute the geodesics. Substituting Eqs.\eqref{ds2}, \eqref{FGH} as well as \eqref{EOM} into the \eqref{geodesic}, we obtain the following expression (see, e.g., \cite{Yang:2024cnd, Azizallahi:2023rrv})
 \bqn
\lb{geodesic2}
{\dot r}^2 &=& \frac{1}{G} \left(E^2 - V_{\text{eff}}\right),
\eqn
for which we have constructed the corresponding effective potential $V_{\text{eff}}$ with its form given by
 \bqn
\lb{Veff}
V_{\text{eff}} &=& F\left[1+ \frac{L^2 F}{r(r-2m)}\right],
\eqn
where $L$ is the angular momentum (as mentioned earlier) with the dimension of $M$.

\begin{figure}[h]
\includegraphics[width=0.9
\linewidth]{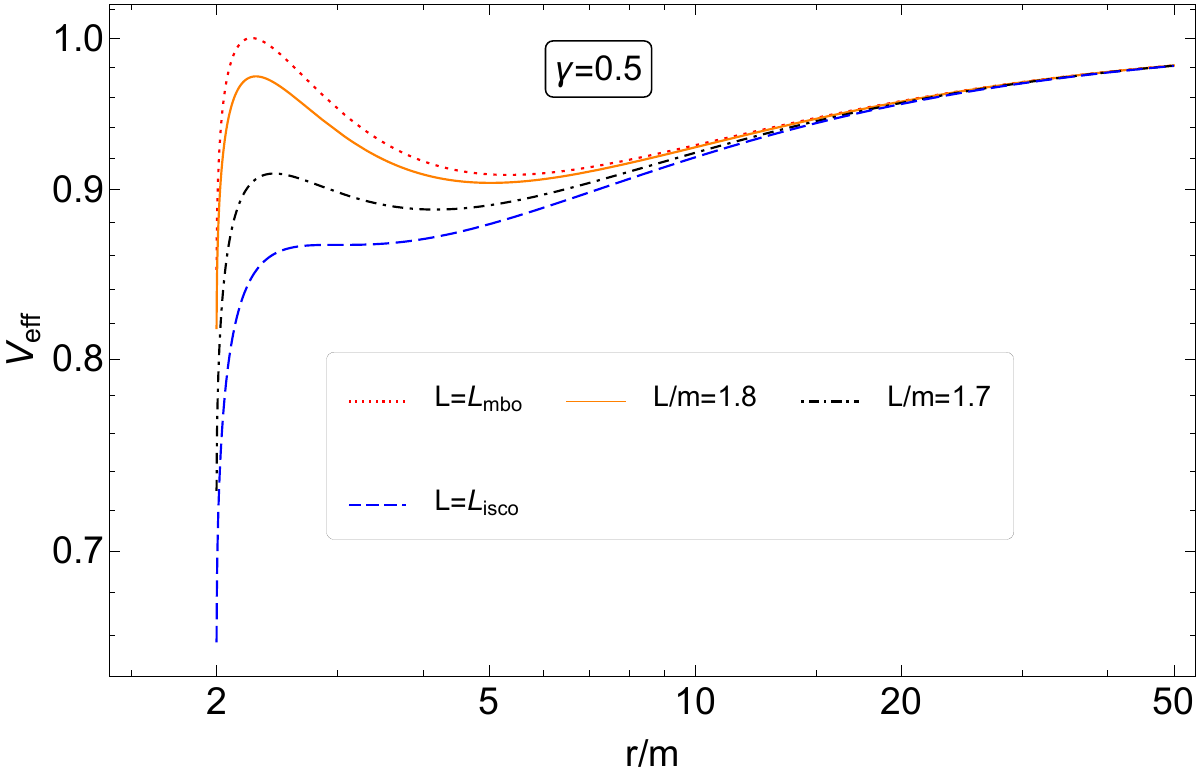} 
\includegraphics[width=0.9
\linewidth]{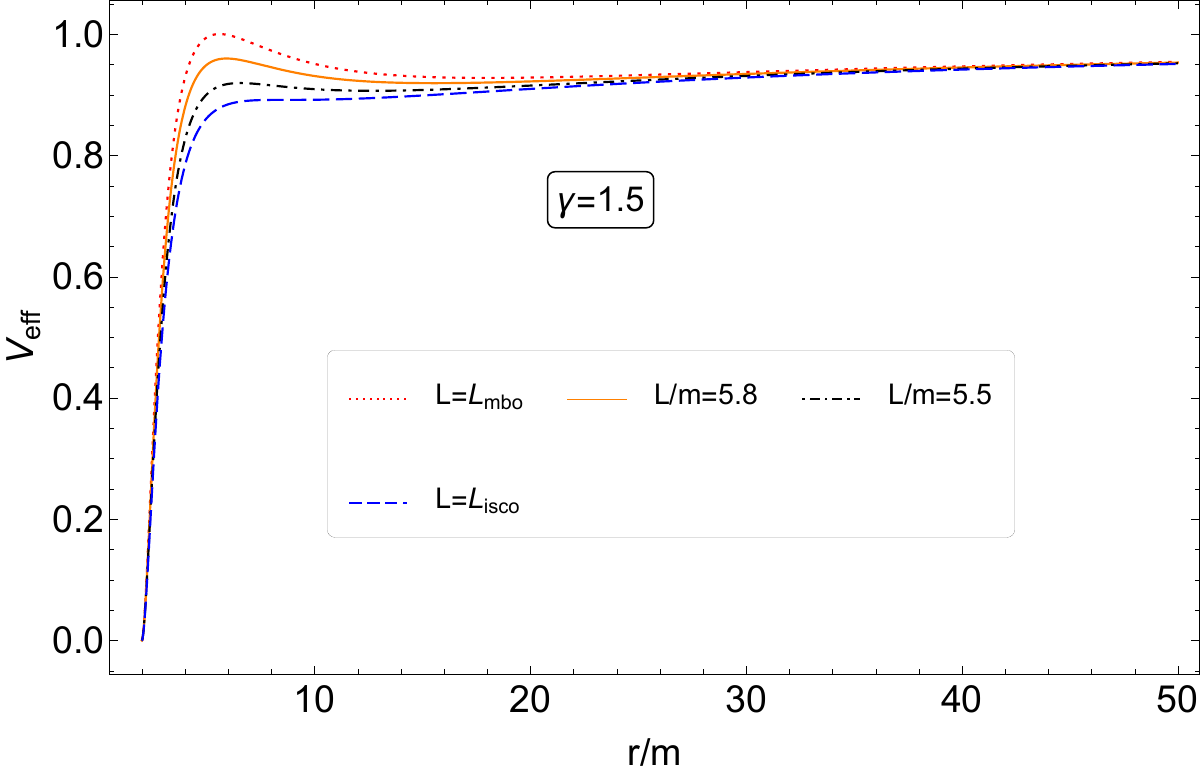} 
\caption{The behaviors of $V_{\text{eff}}$ as a function of the dimensionless variable $r/m$ by choosing different values for $L$. For the upper panel, we set $\gamma=0.5$ while for the lower panel we set $\gamma=1.5$.  
} 
\label{plotveff}
\end{figure} 


For the two representative values of \(\gamma=0.5\) and \(\gamma=1.5\), we plot the effective potential \(V_{\mathrm{eff}}\) as a function of the dimensionless radial coordinate \(r/m\) for several choices of the angular momentum \(L\) in Fig.~\ref{plotveff}. Each curve in Fig.~\ref{plotveff} exhibits identifiable stationary points and inflection points. Qualitatively, the behavior of these curves is consistent with the results reported in the literature (see, e.g., \cite{Tu:2023xab}).

\subsection{The marginally bound orbit}

The marginally bound orbit (MBO) represents the smallest possible circular bound orbit, possessing the minimum radius with energy $E=1$. It is determined by the conditions
 \bqn
\lb{Lrmbo}
\left. V_{\text{eff}} \right|_{r=r_{\text{mbo}}, L=L_{\text{mbo}}} &=& 1, \nb\\
\left. \frac{d V_{\text{eff}} }{dr} \right|_{r=r_{\text{mbo}}, L=L_{\text{mbo}}} &=& 0, 
\eqn
where $L_{\text{mbo}}$ and $r_{\text{mbo}}$ denotes the angular momentum and radius that corresponds to the MBO, respectively. 

\begin{figure}[h]
\includegraphics[width=0.9
\linewidth]{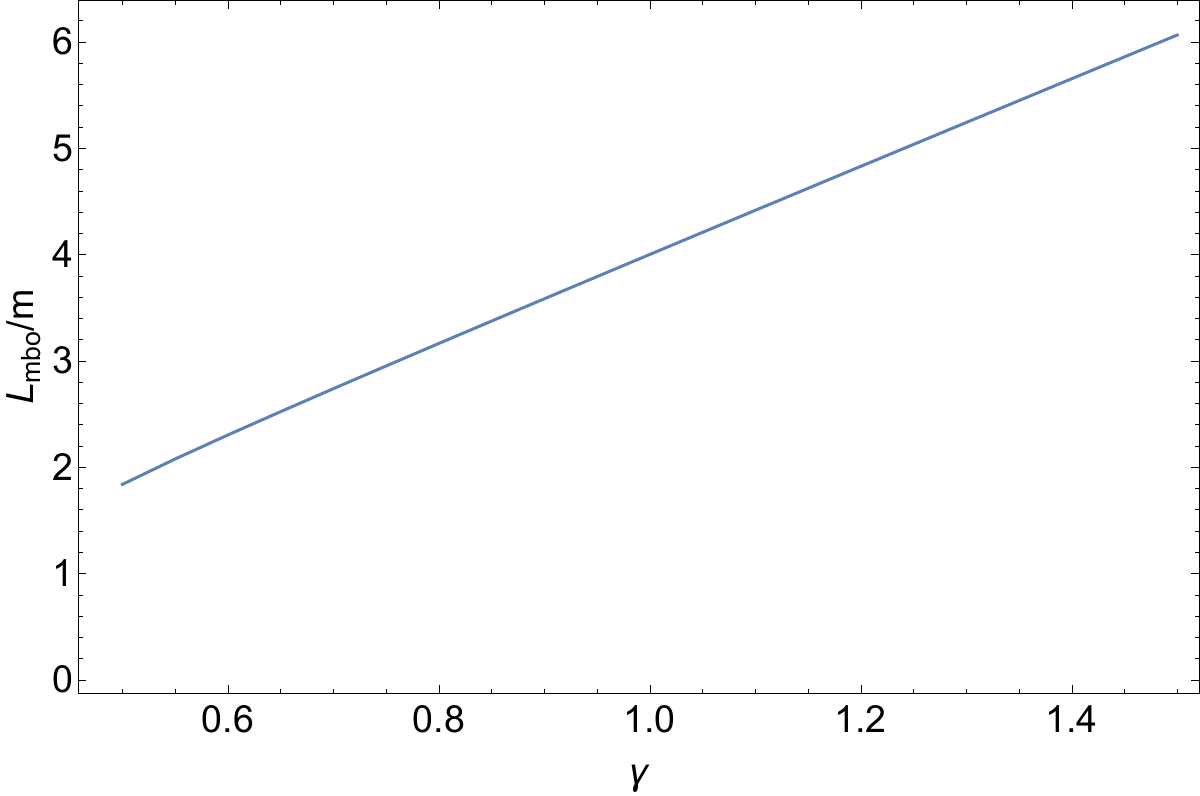} 
\includegraphics[width=0.9
\linewidth]{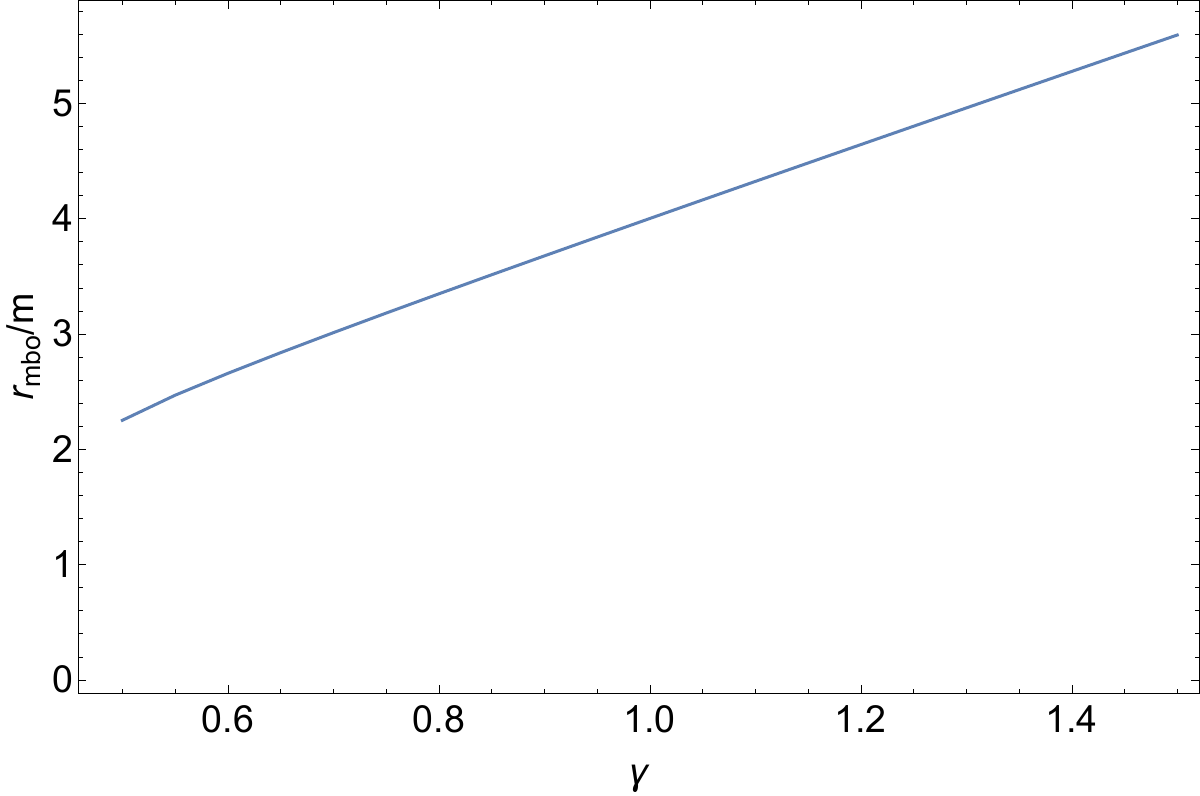} 
\caption{The behaviors of dimensionless quantities $L_{\text{mbo}}/m$ and $r_{\text{mbo}}/m$  for a changing $\gamma$. 
} 
\label{plotmbo}
\end{figure} 


For a given physical value of the parameter \(\gamma\), the corresponding quantities \(L_{\mathrm{mbo}}\) and \(r_{\mathrm{mbo}}\) are determined by solving Eq. (\ref{Lrmbo}).  Owing to the complexity of this equation, closed‑form analytic solutions are not accessible. Instead, we solve Eq. (\ref{Lrmbo}) numerically and present the resulting dimensionless quantities \(L_{\mathrm{mbo}}/m\) and \(r_{\mathrm{mbo}}/m\) as functions of \(\gamma\) in Fig.~\ref{plotmbo}.  Within the parameter range considered here, both \(L_{\mathrm{mbo}}\) and \(r_{\mathrm{mbo}}\) increase monotonically with \(\gamma\).

\subsection{The innermost stable circular orbit}

The  innermost stable circular orbit (ISCO) is determined by the conditions
 \bqn
\lb{ELrisco}
\left. V_{\text{eff}} \right|_{r=r_{\text{isco}}, L=L_{\text{isco}}} &=& E_{\text{isco}}^2, \nb\\
\left. \frac{d V_{\text{eff}} }{dr} \right|_{r=r_{\text{isco}}, L=L_{\text{isco}}} &=& 0,  \nb\\
\left. \frac{d^2 V_{\text{eff}} }{dr^2} \right|_{r=r_{\text{isco}}, L=L_{\text{isco}}} &=& 0, 
\eqn
where $E_{\text{isco}}$, $L_{\text{isco}}$ and $r_{\text{isco}}$ denote the energy, angular momentum, and radius that correspond to the ISCO, respectively.

\begin{figure}[h]
\includegraphics[width=0.9
\linewidth]{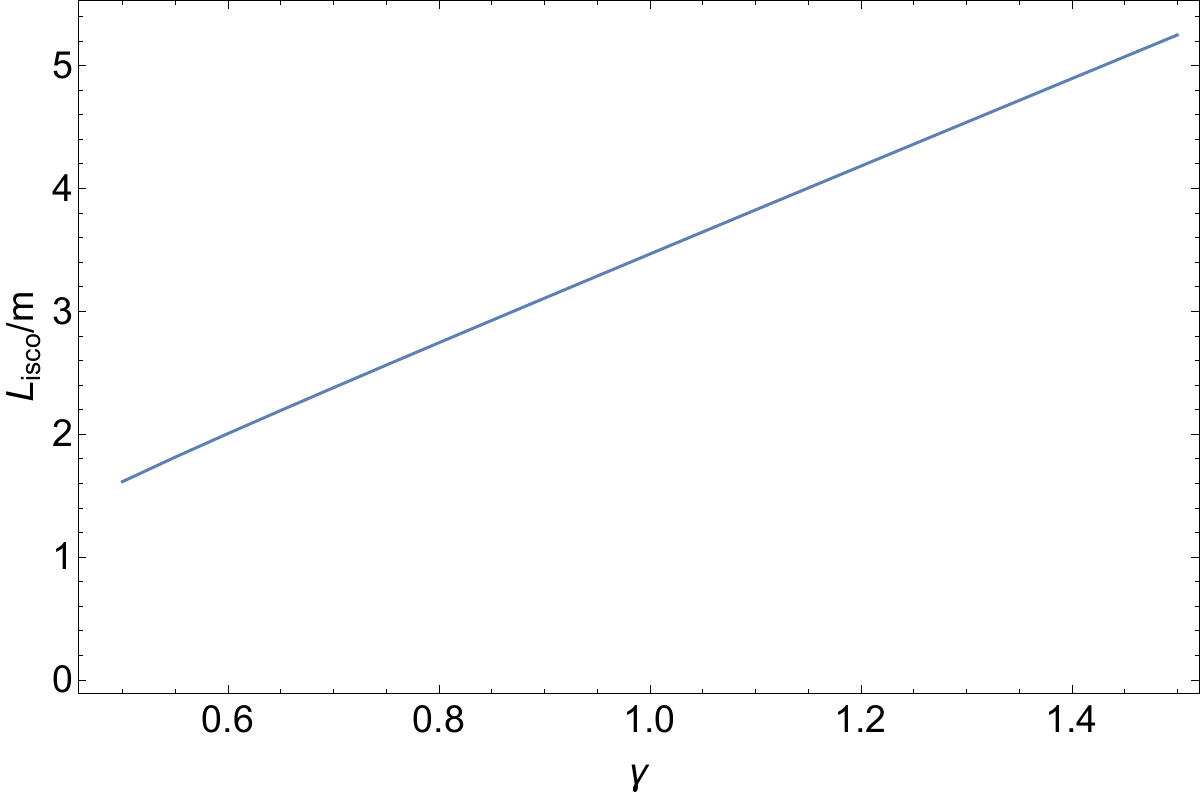} 
\includegraphics[width=0.9
\linewidth]{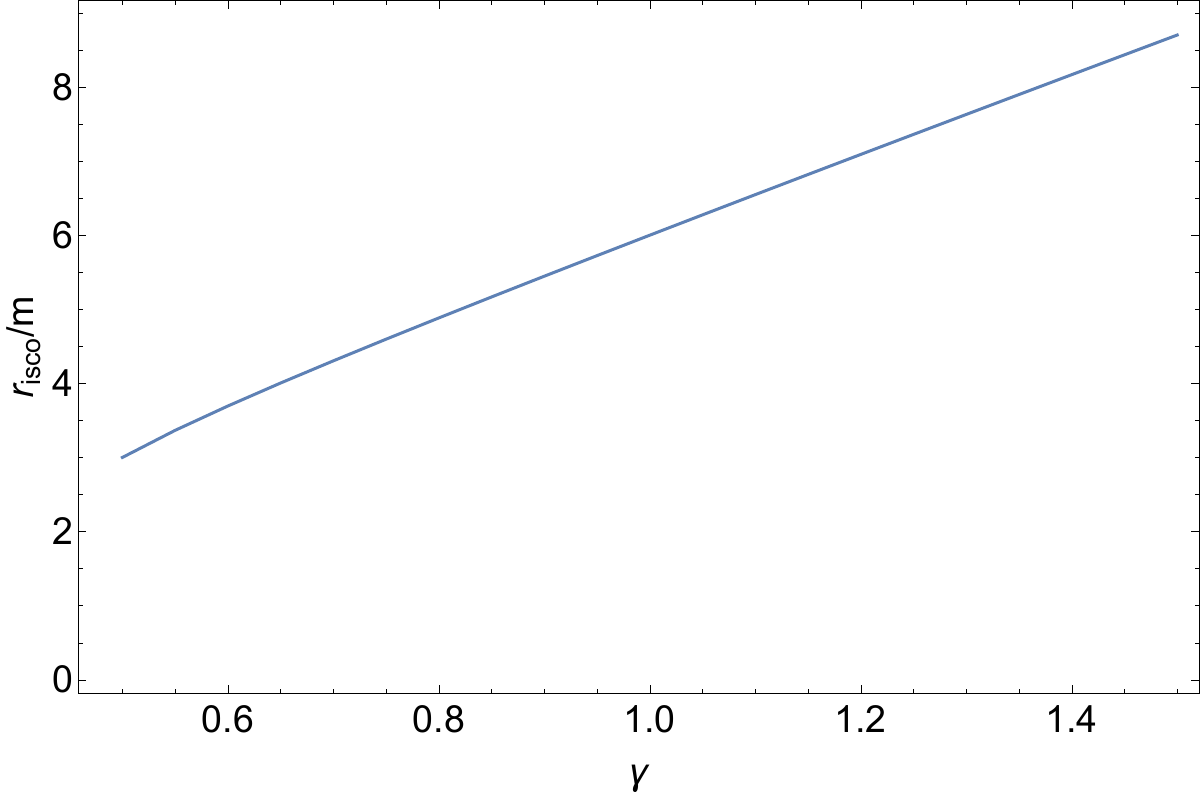} 
\includegraphics[width=0.9
\linewidth]{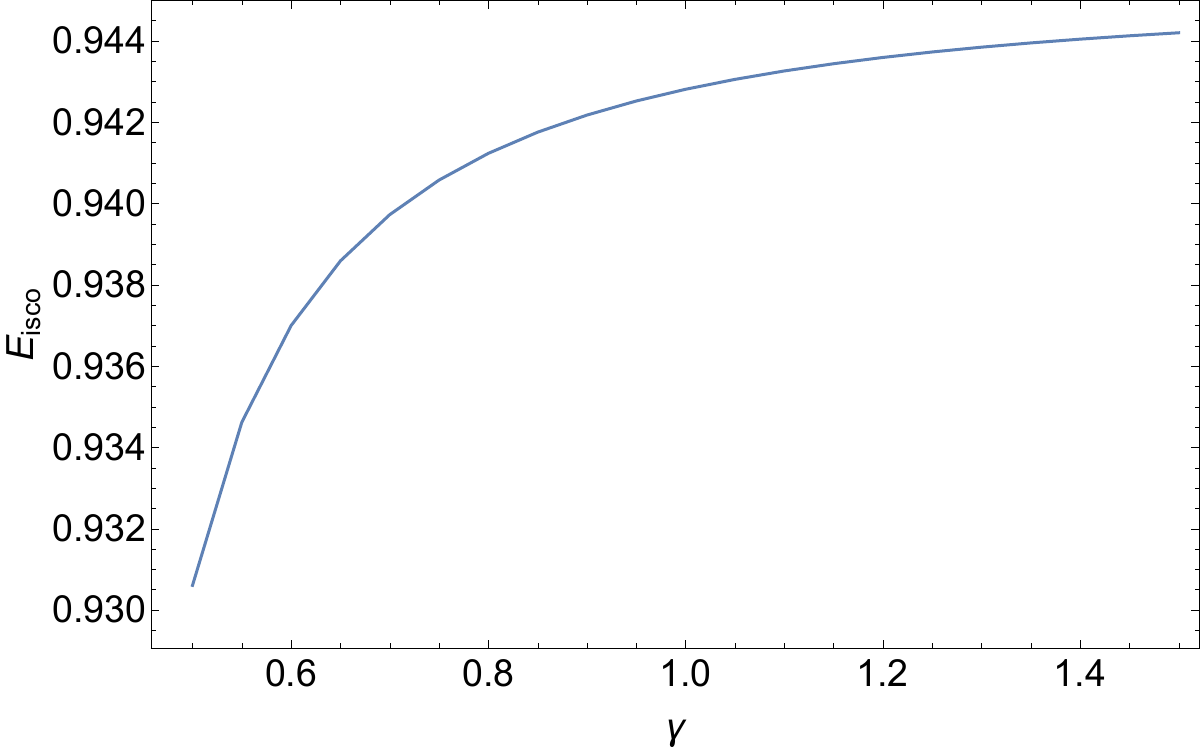} 
\caption{The behaviors of dimensionless quantities $E_{\text{isco}}$, $L_{\text{isco}}/m$ and $r_{\text{isco}}/m$  for a changing $\gamma$. 
} 
\label{plotisco}
\end{figure}


For a given value of the parameter \(\gamma\), the corresponding quantities \(L_{\mathrm{isco}}\) and \(r_{\mathrm{isco}}\) are obtained by solving the last two equations of Eq. (\ref{ELrisco}). Substituting these solutions into the left-hand side of the first equation of Eq.~(\ref{ELrisco}) then yields the energy \(E_{\mathrm{isco}}\). Because these relations are algebraically involved, we do not present closed‑form expressions; instead, we solve the system numerically and plot the dimensionless combinations \(L_{\mathrm{isco}}/m\), \(E_{\mathrm{isco}}\), and \(r_{\mathrm{isco}}/m\) as functions of \(\gamma\) in Fig. \ref{plotisco}. Within the parameter range considered here, all three quantities—\(E_{\mathrm{isco}}\), \(L_{\mathrm{isco}}\), and \(r_{\mathrm{isco}}\)—increase monotonically with \(\gamma\).

\section{Periodic orbits in $\gamma$-metric}
\renewcommand{\theequation}{4.\arabic{equation}} \setcounter{equation}{0}
\label{secIV}


Having explored the ISCOs and MBOs, we now focus on periodic orbits of a test particle near a supermassive compact body, which is described by the $\gamma$-metric. These bound orbits are characterized by their repeating trajectories. Periodicity arises when the frequency ratio between the radial ($\omega_r$) and azimuthal ($\omega_\phi$) oscillations is rational. While a generic orbit may exhibit an irrational frequency ratio $\frac{\omega_\phi}{\omega_r}$, a rational number can always approximate this ratio with arbitrary precision due to its density on the number axis \cite{Rudin:1976pm}. It has also been proven that periodic geodesic orbits exist in the background of a Schwarzschild or Kerr BH, as mentioned in Sec.\ref{secI}.

This ability to approximate generic trajectories by nearby periodic ones renders the study of periodic orbits particularly valuable for understanding the broader orbital dynamics and the associated gravitational‑wave (GW) emission. Analysis of periodic orbits, therefore, yields important insight into the structure of the generally complex GW signals produced by these systems and provides a robust framework for their interpretation.

Considering the equatorial motion, the dynamics of the test particle reduces to radial \(r\)-motion and azimuthal \(\phi\)-motion. The apsidal angle \(\Delta\phi\) is obtained by integrating the change in the azimuthal coordinate over a complete radial oscillation, i.e., from the periastron \(r_1\) to the apastron \(r_2\) and back:
 \bqn
\Delta\phi=\oint d\phi \;=\; 2\int_{\phi_1}^{\phi_2} d\phi
\;=\; 2\int_{r_1}^{r_2}\frac{\dot\phi}{\dot r}\,dr.
 \eqn
Here \(r_1\) and \(r_2\) are the two radial turning points of the motion (\(r_2>r_1\)) \cite{Tu:2023xab}. For a given triplet \(\{E, L,\gamma\}\), the turning points are determined from the effective potential [cf. Eq. \eqref{Veff}] via the condition \(\dot r(\lambda)=0\), i.e. \(E^2=V_{\mathrm{eff}}(r=r_{1,2})\) [cf. Eq. \eqref{geodesic2}].

Substituting Eqs. \eqref{EOM} and \eqref{geodesic2} into the above expression yields
\begin{equation}\label{apsidal}
\Delta\phi \;=\; 2\int_{r_1}^{r_2} \frac{L\,F(r)}{r(r-2m)} 
\sqrt{\frac{G(r)}{E^2-V_{\rm eff}(r)}}\,dr.
\end{equation}
The prefactor of 2 reflects the symmetry of the inbound and outbound segments of the radial motion. The apsidal angle is therefore sensitive not only to the particle’s conserved energy \(E\) and angular momentum \(L\) but also to the spacetime structure encoded in the metric functions (here \(F(r)\) and \(G(r)\)). Consequently, compact objects characterized by different values of the parameter \(\gamma\) will, in general, produce different apsidal angles.

Adopting the taxonomy developed in \cite{Levin:2008mq}, we define the frequency ratio $q$ as the relationship between the radial ($\omega_r$) and azimuthal ($\omega_\phi$) oscillation frequencies. This ratio is uniquely expressed in terms of three non-zero integers $(z, w, v)$, zoom $z$ ($z \ne 0$), whirl $w$, and vertex $v$ ($v<z$) as
\begin{equation}\label{qradial}
q = \frac{\omega_\phi}{\omega_r} - 1 = w + \frac{v}{z}.
\end{equation}
The frequency ratio $q$ is also linked to the apsidal angle $\Delta\phi$ traversed during one radial period by the relation:
\begin{equation}
\label{qradial2}
\frac{\omega_\phi}{\omega_r} = \frac{\Delta\phi}{2\pi},
\end{equation}
where $\Delta\phi$ is given by Eq. (\ref{apsidal}). Therefore, we have $q= {\Delta\phi}/{2\pi}-1$. 
For irrational counterparts of $q$, the particle's trajectory traces a precessing orbit, characterized by a precession angle $(\Delta\phi - 2\pi)$. Rational values of $q$, however, correspond to periodic orbits, where the particle returns to its initial position after a finite time. As shown in \cite{Levin:2008mq}, generic orbits can be interpreted as perturbations of these periodic orbits. Thus, the study of periodic orbits provides critical insights into the behavior of generic orbits and the gravitational radiation emitted near compact objects.

Given the angular momentum for bound orbits lies between the values of ISCO (viz., $L_{\text{isco}}$) and MBO (viz., $L_{\text{mbo}}$), we can express the angular momentum $L$ as
\begin{equation}
\lb{Lexpression}
L = L_{\rm isco} + \epsilon (L_{\rm mbo} - L_{\rm isco}),
\end{equation}
where $\epsilon$ is a parameter ranging from 0 to 1. Specifically, $\epsilon = 0$ corresponds to $L=L_{\rm isco}$, and $\epsilon = 1$ corresponds to $L=L_{\rm mbo}$. The constraint $0 \le \epsilon \le 1$ ensures that we remain within the regime of bound orbits.

\begin{figure*}
\centering
\includegraphics[width=8.5cm]{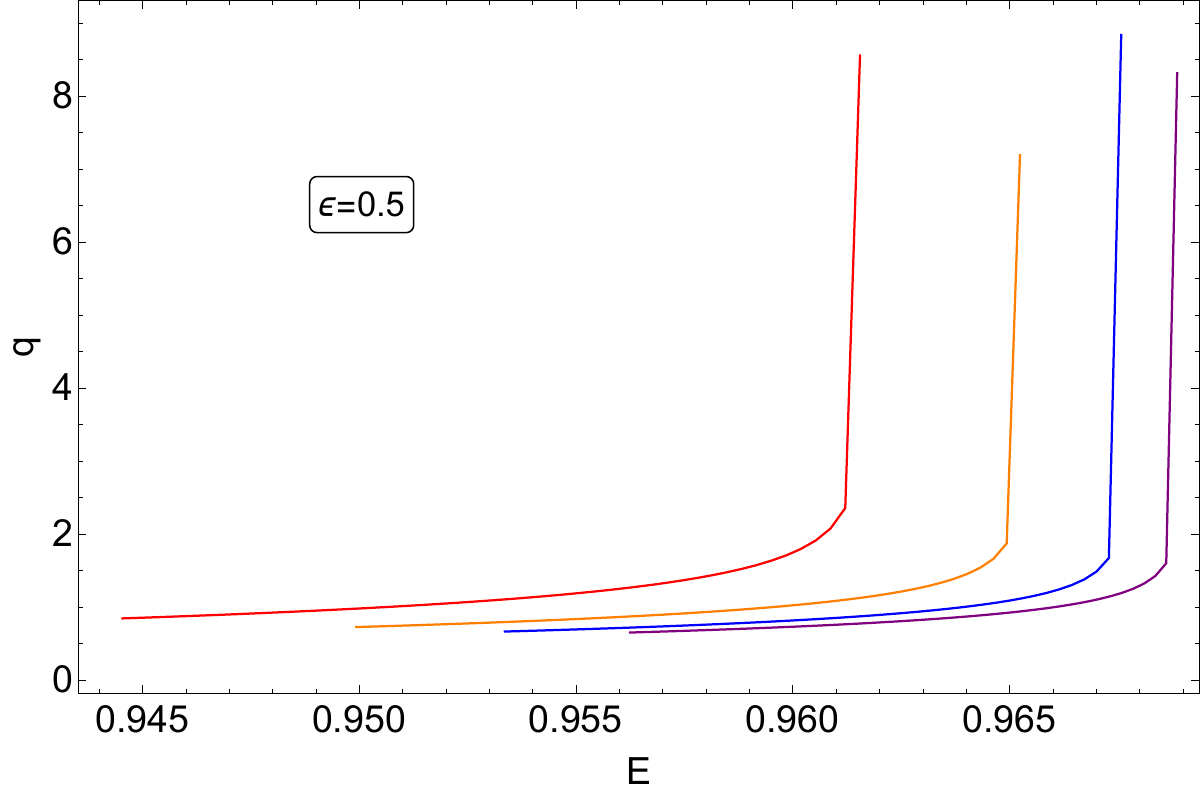}
\includegraphics[width=8.5cm]{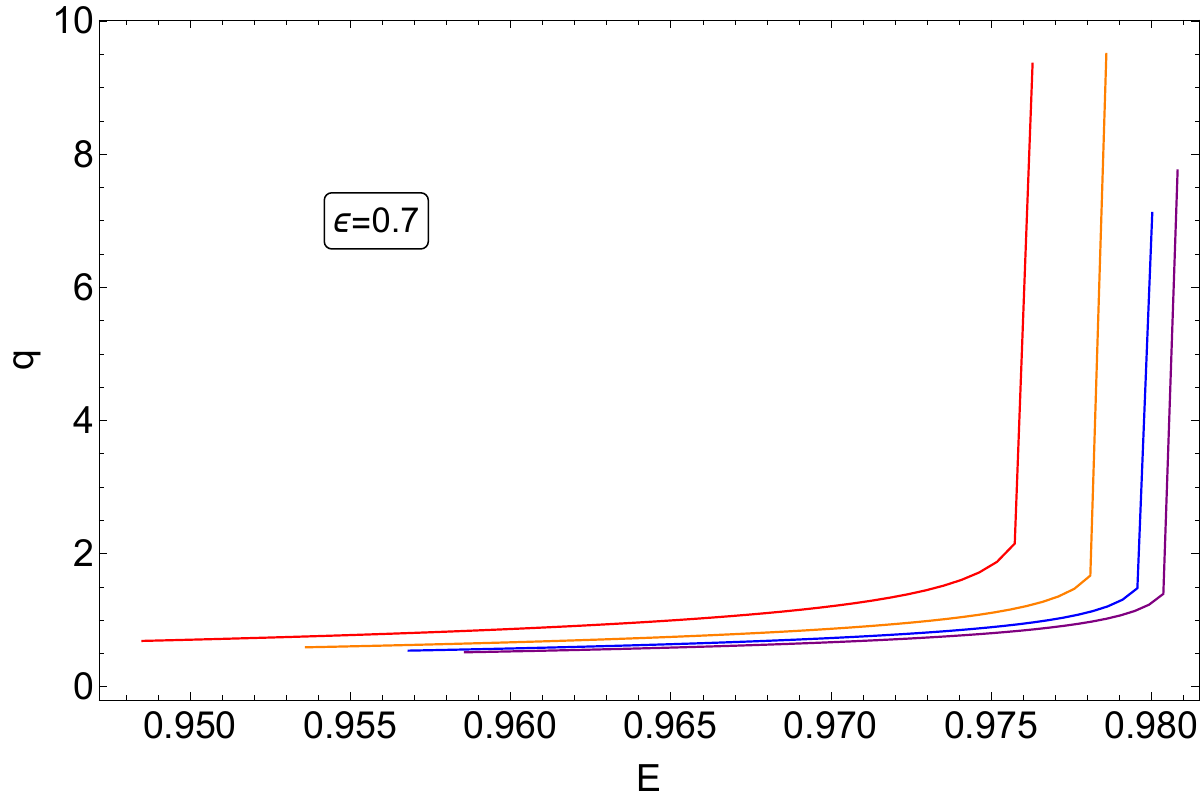} 
\caption{The relation between the rational number $q$ and the energy $E$ for different choices of $\epsilon$ [cf. \eqref{Lexpression}] and $\gamma$. Notice that, for those curves in each panel we have $\gamma=0.5, 0.6, 0.8, 1.2$ respectively from left to right (red, orange, blue and purple curve, respectively).}
\label{qE1}
\end{figure*}

\begin{figure*}
\centering
\includegraphics[width=8.5cm]{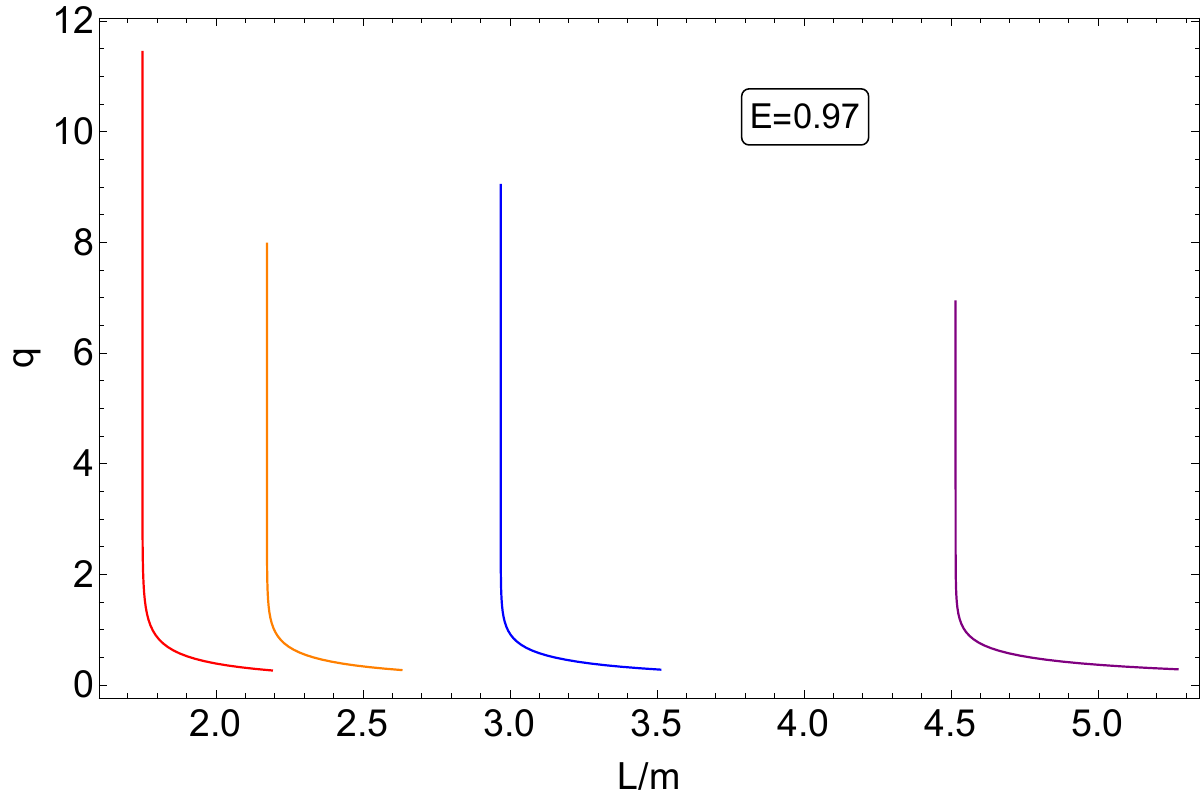} 
\includegraphics[width=8.5cm]{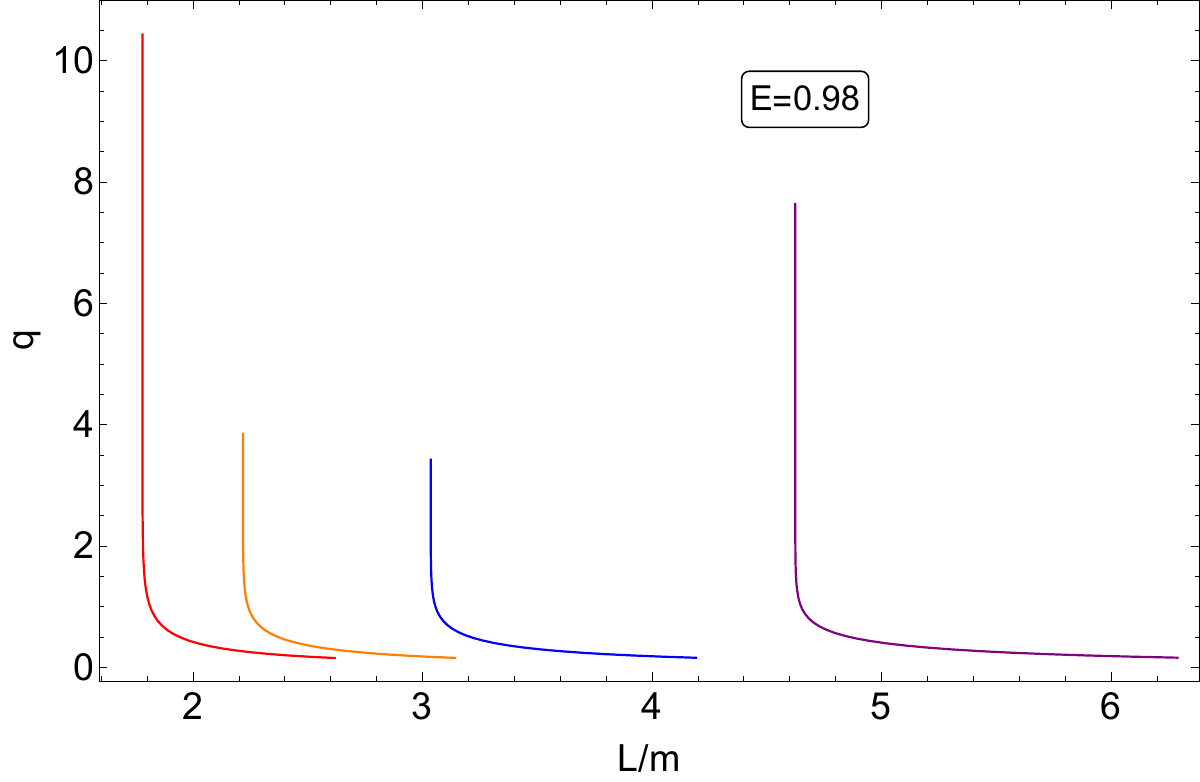} 
\caption{The relation between the rational number $q$ and the dimensionless quantity $L/m$ for different choices of $E$ and $\gamma$. Notice that, for those curves in each panel we have $\gamma=0.5, 0.6, 0.8, 1.2$ respectively from left to right (red, orange, blue and purple curve, respectively).}
\label{qL1}
\end{figure*}

The value of $q$ can be determined through Eqs.\eqref{apsidal}, \eqref{qradial} and \eqref{qradial2}.
To better know the behaviors of $q$, it is plotted out as a function of $E$ in Fig.\ref{qE1}, for which we have selected various values for $\gamma$ and $\epsilon$ (so that $L$). According to Fig.~\ref{qE1}, $q$ is monotonically increasing with $E$. Such a phenomenon is consistent with that observed in the literature \cite{Tu:2023xab, Haroon:2025rzx, Wang:2025hla}. As a result, the $q$ will become extremely large when $E$ increases. Another important fact is that the real solutions for the turning points $r_{1, 2}$ do not always exist for an arbitrary $E$ and selected $L$ when solving  $E^2=V_{\text{eff}}(r=r_{1, 2})$.  Because of this, in Fig.~\ref{qE1} we show only the regimes of $E$ that allow real $r_{1, 2}$ to exist (so that the corresponding integration can be safely executed).

Similarly, it is possible to determine the numerical relation between $q$ and $L$ for a fixed $E$. The behaviors of $q$ as a function of the dimensionless variable $L/m$ is given in Fig.~\ref{qL1}, for which we have selected various values for $\gamma$ and $E$. According to Fig.\ref{qL1}, $q$ is monotonically decreasing with $L$. Such a phenomenon is also consistent with that observed in the literature \cite{Tu:2023xab, Haroon:2025rzx, Wang:2025hla}. As usual, the real solutions for the turning points $r_{1, 2}$ do not always exist for an arbitrary $L$ and selected $E$ when solving  $E^2=V_{\text{eff}}(r=r_{1, 2})$. In Fig.~\ref{qL1} we show only the regimes of $L$ that allow real $r_{1, 2}$ to exist (so that the corresponding integration can be safely executed).

\begin{figure*}
\centering
\includegraphics[width=5.3cm]{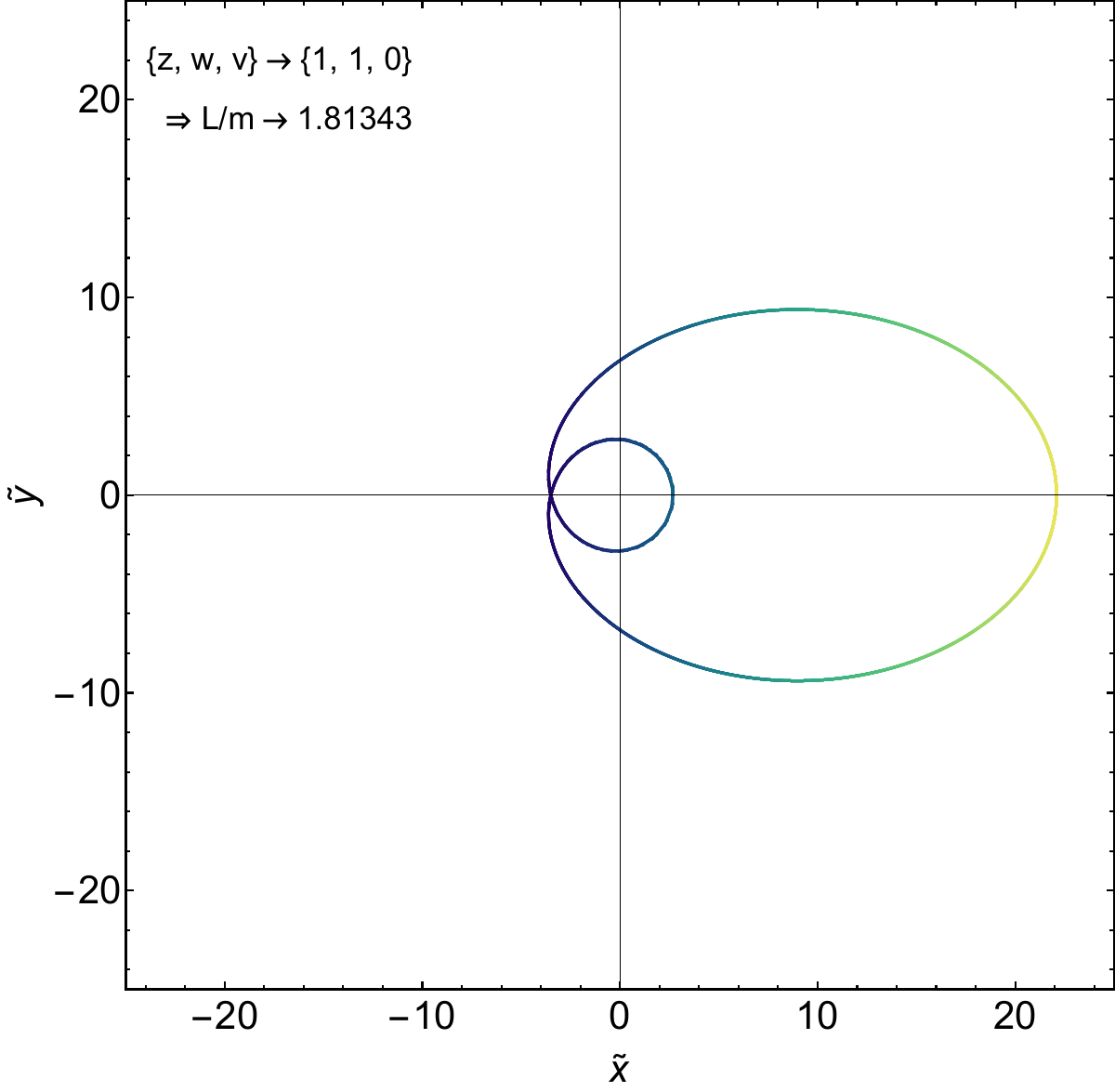} 
\includegraphics[width=5.3cm]{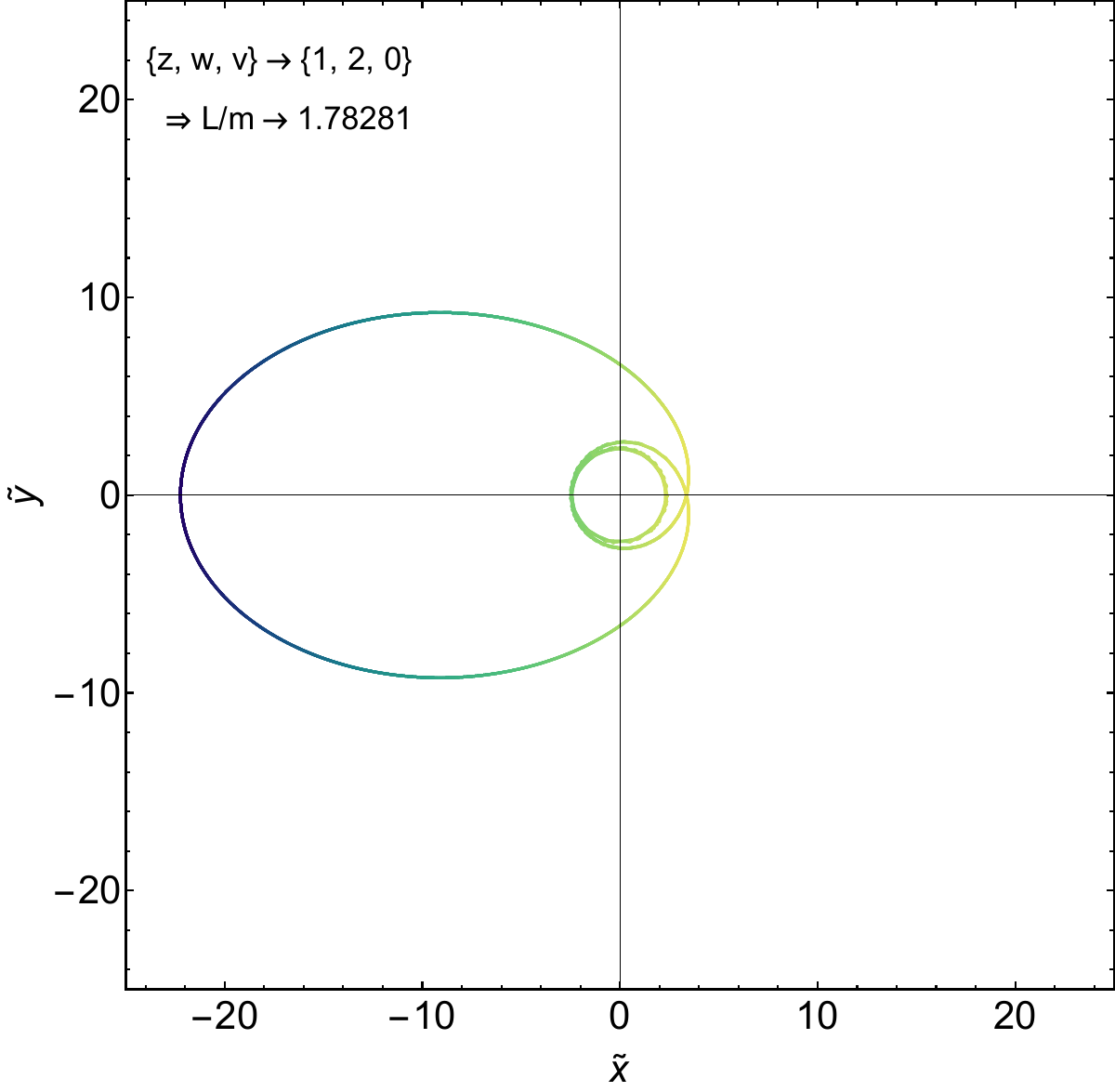}
\includegraphics[width=5.3cm]{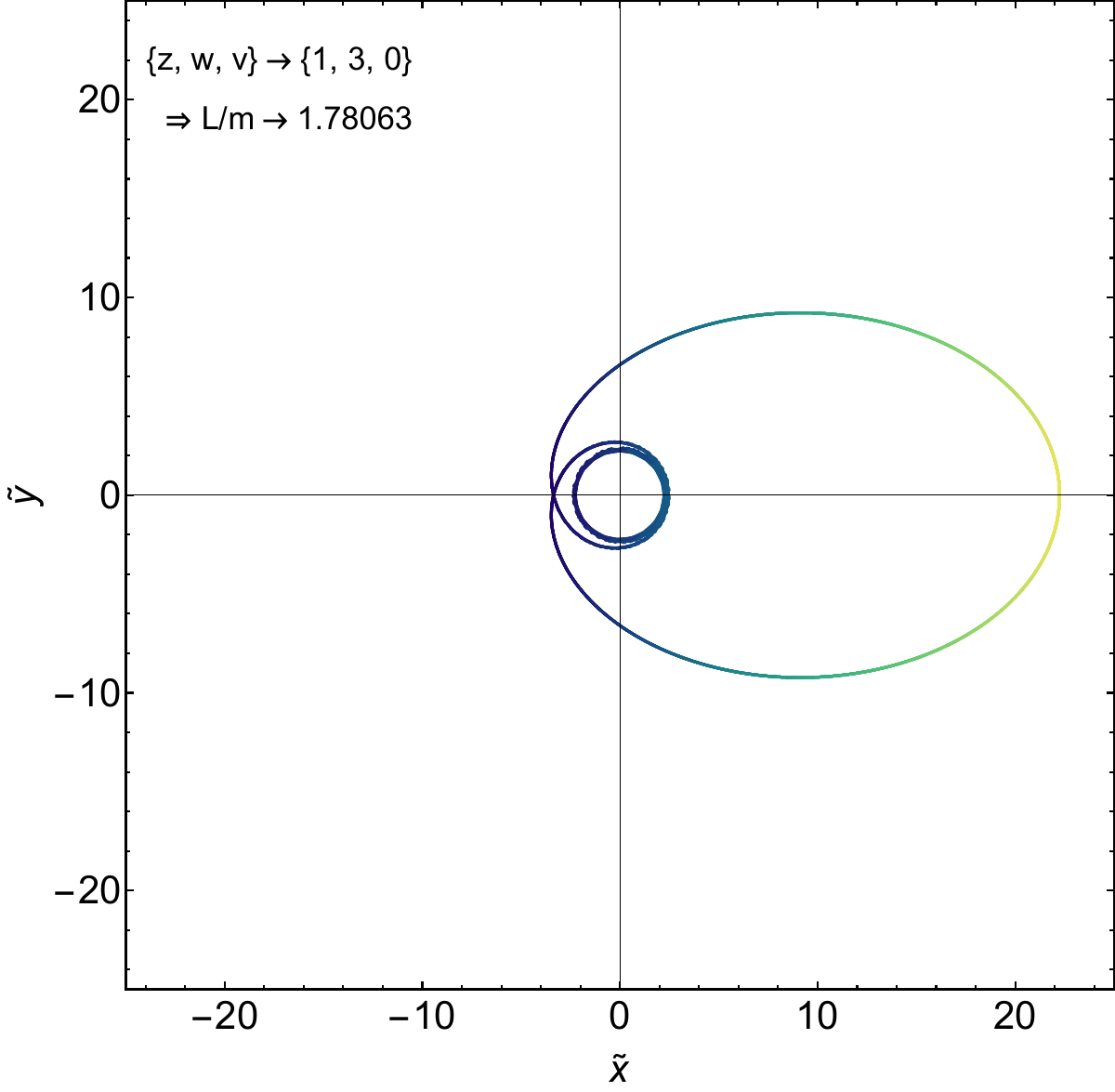} 
\includegraphics[width=5.3cm]{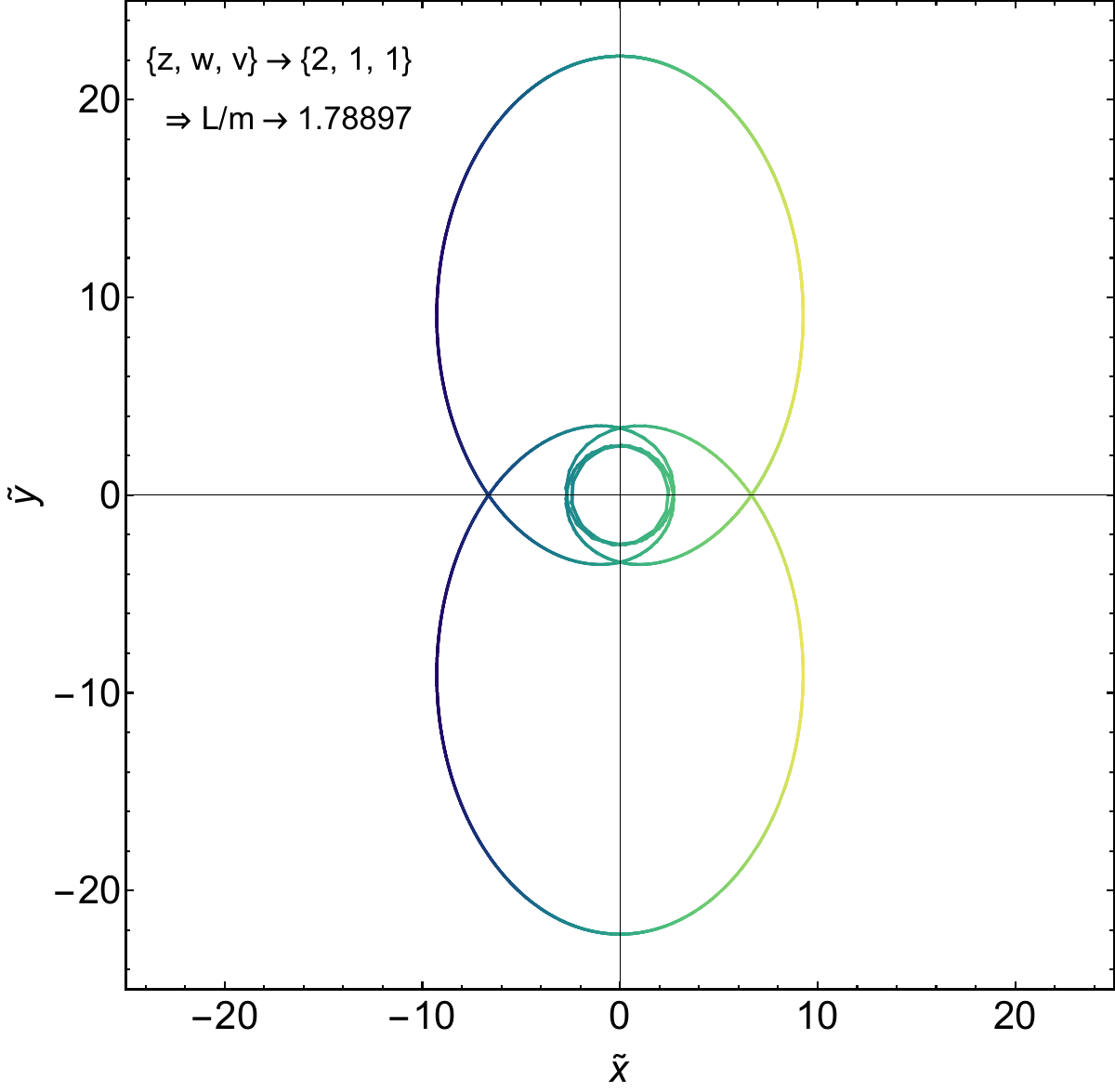} 
\includegraphics[width=5.3cm]{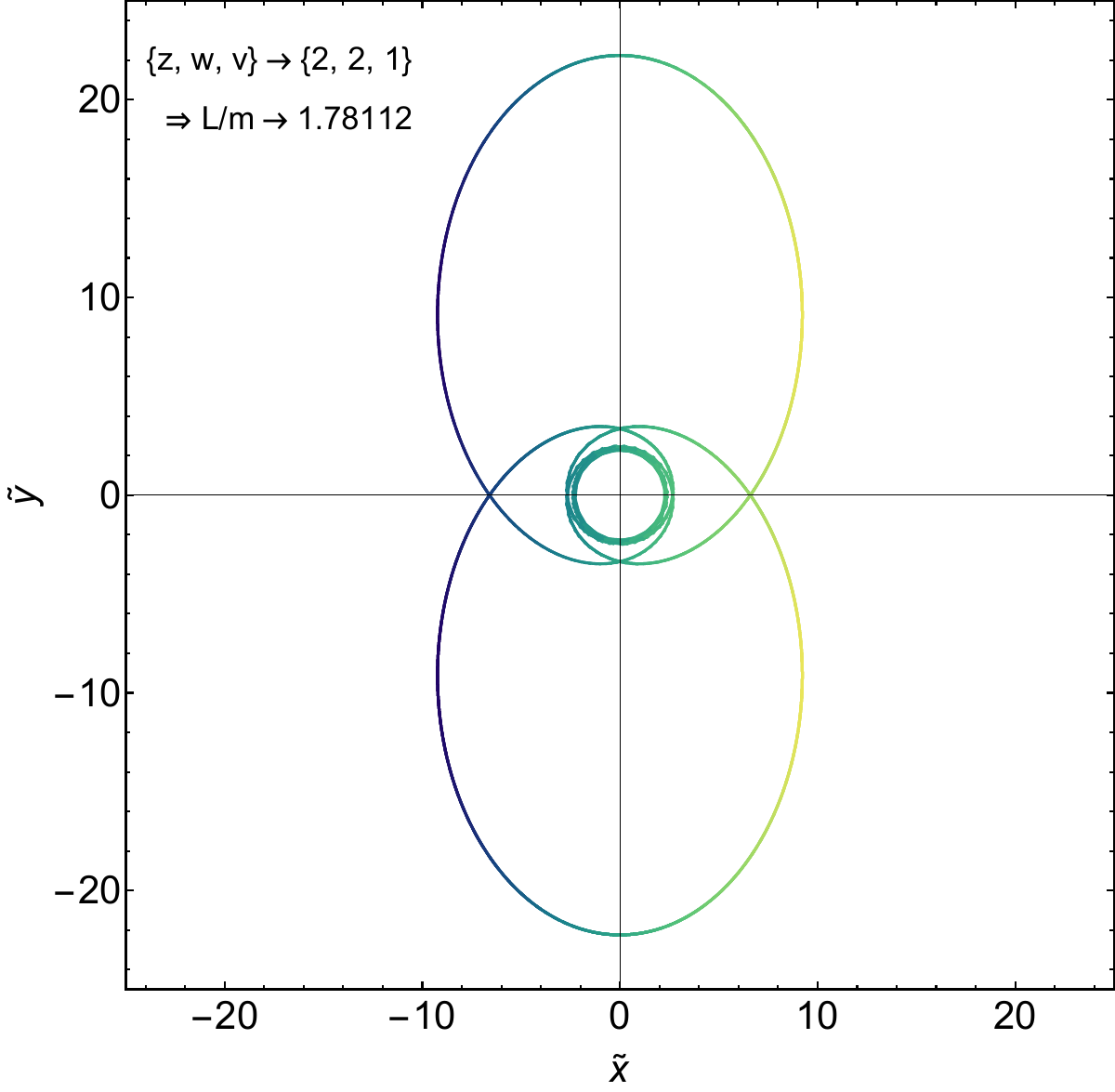} 
\includegraphics[width=5.3cm]{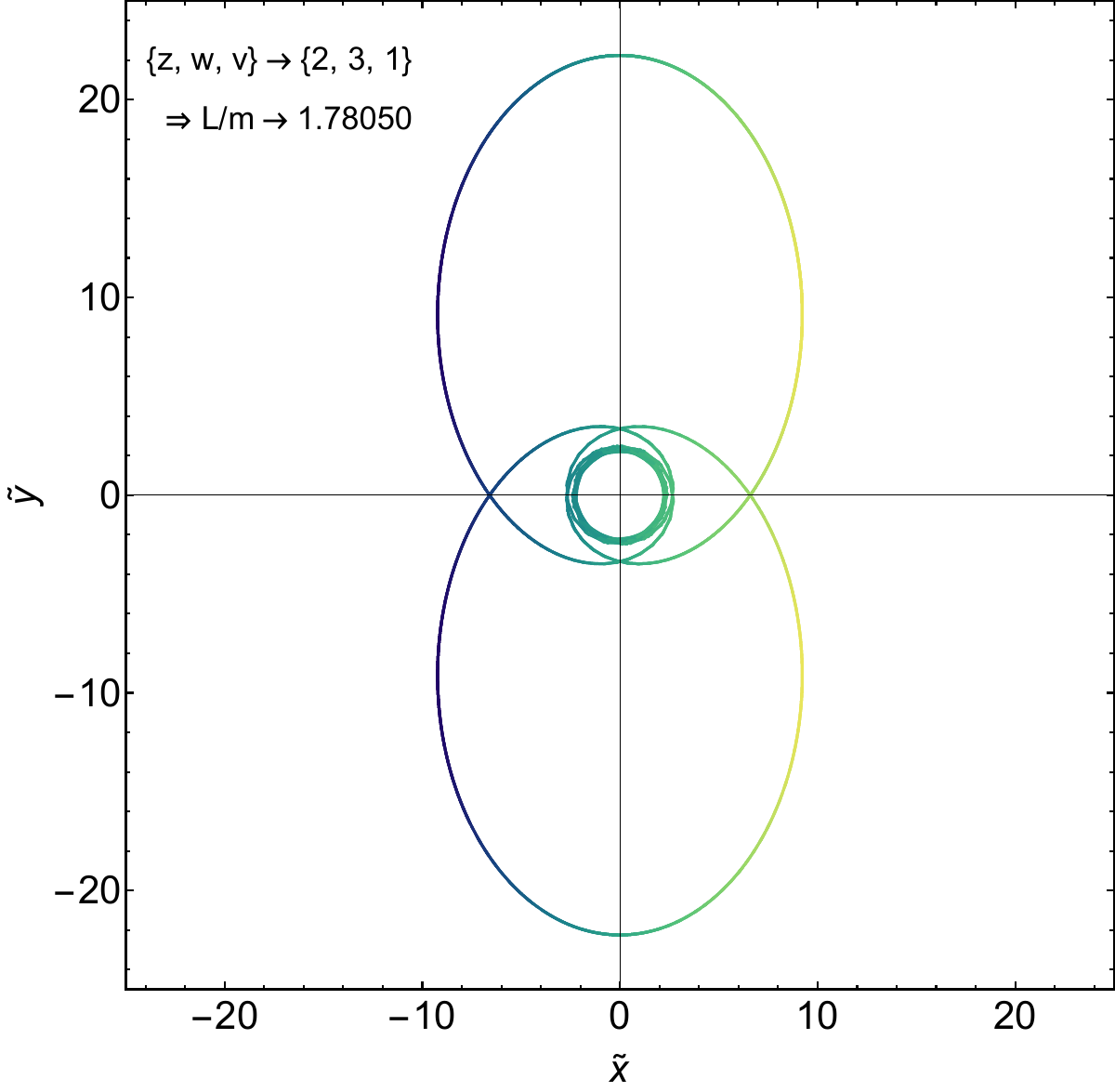} 
\includegraphics[width=5.3cm]{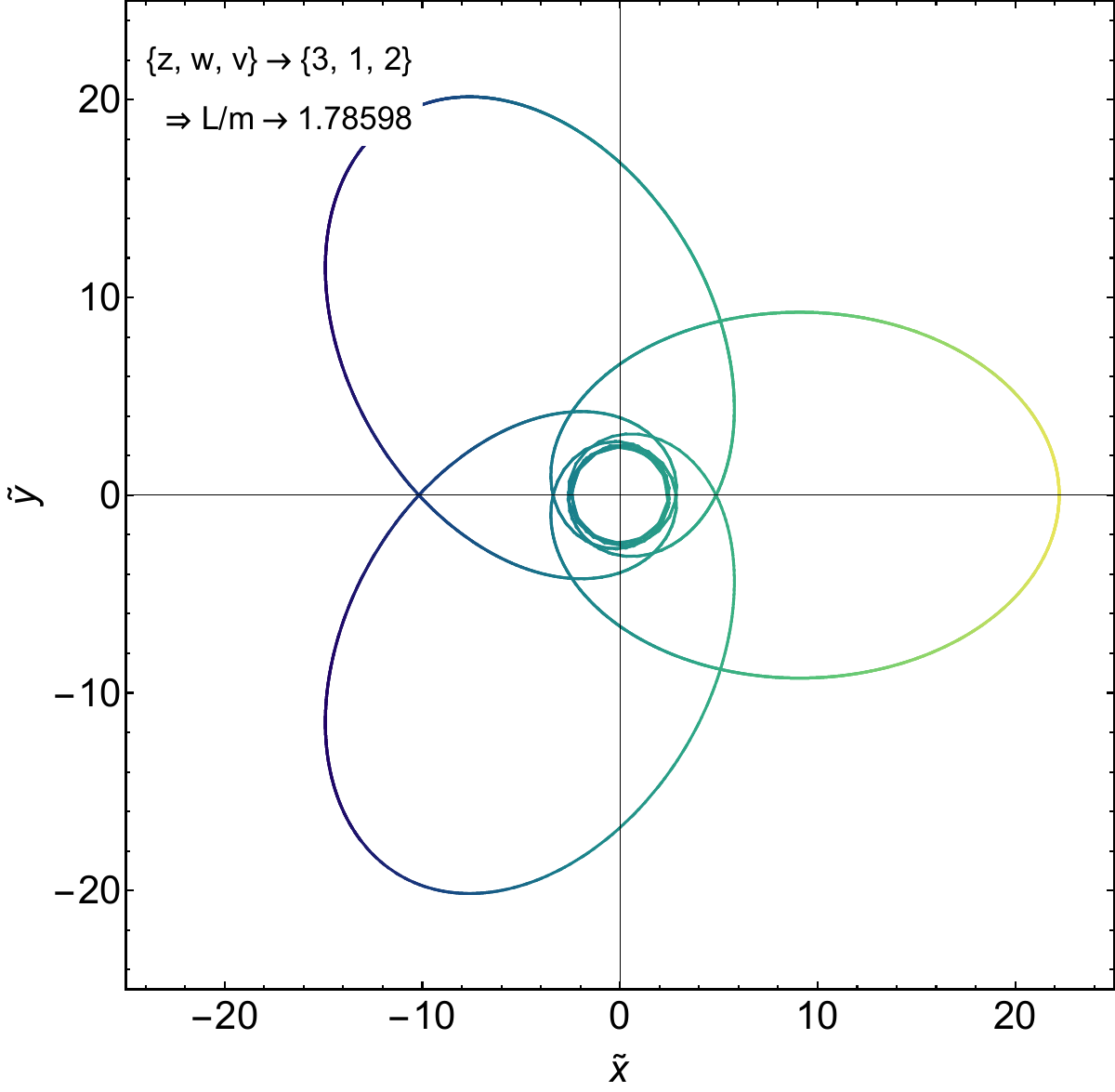} 
\includegraphics[width=5.3cm]{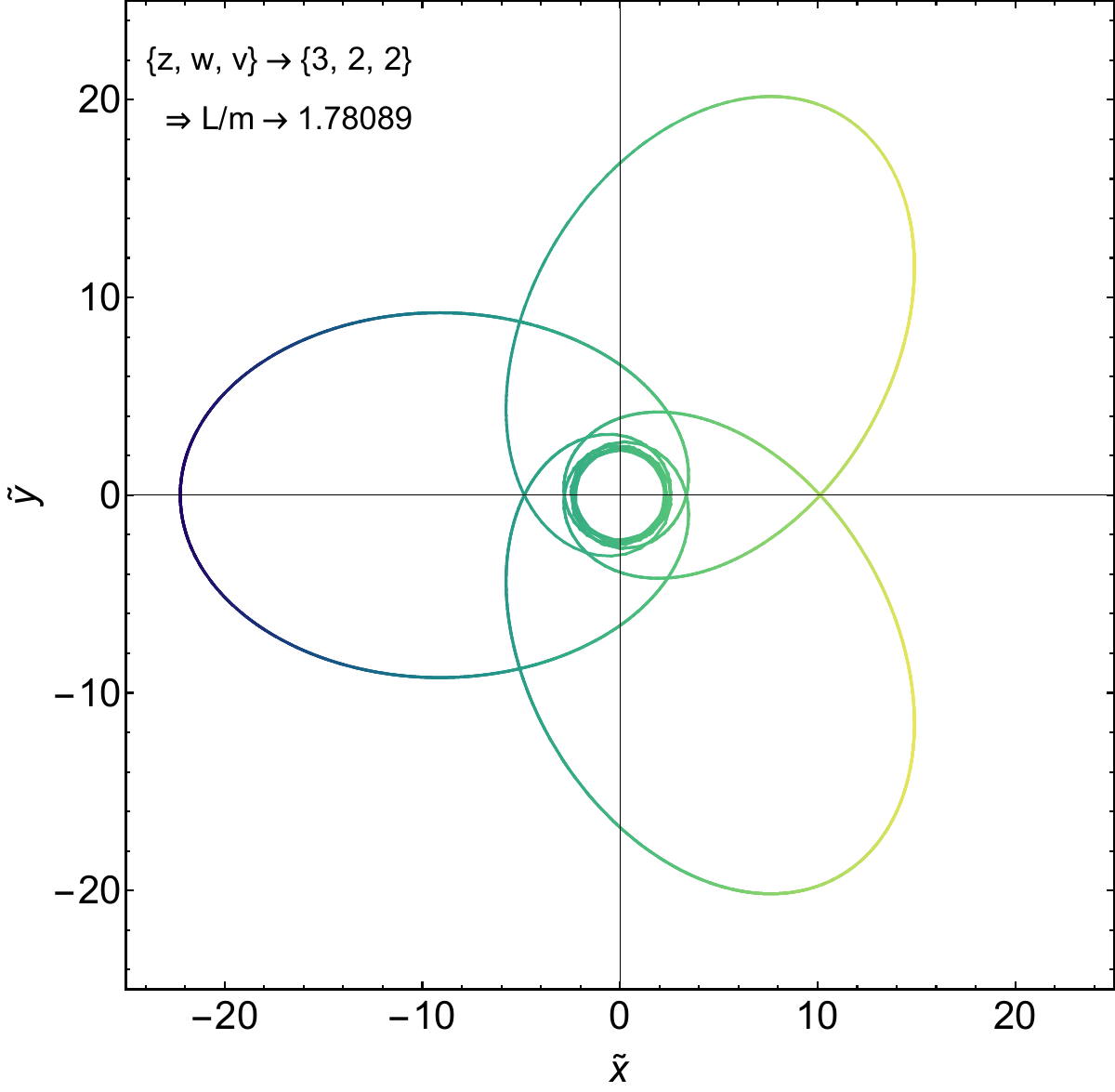}
\includegraphics[width=5.3cm]{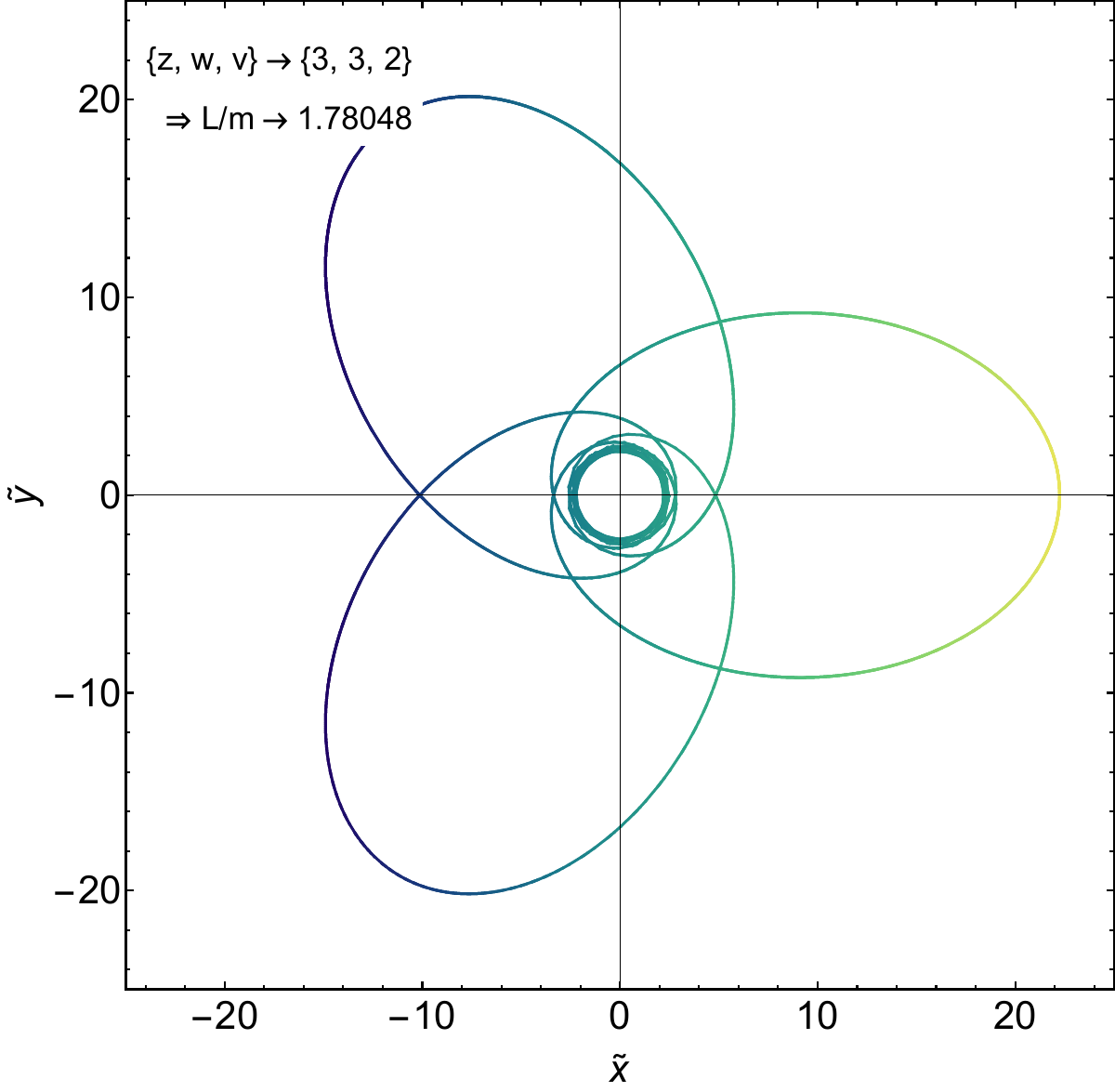} 
\includegraphics[width=5.3cm]{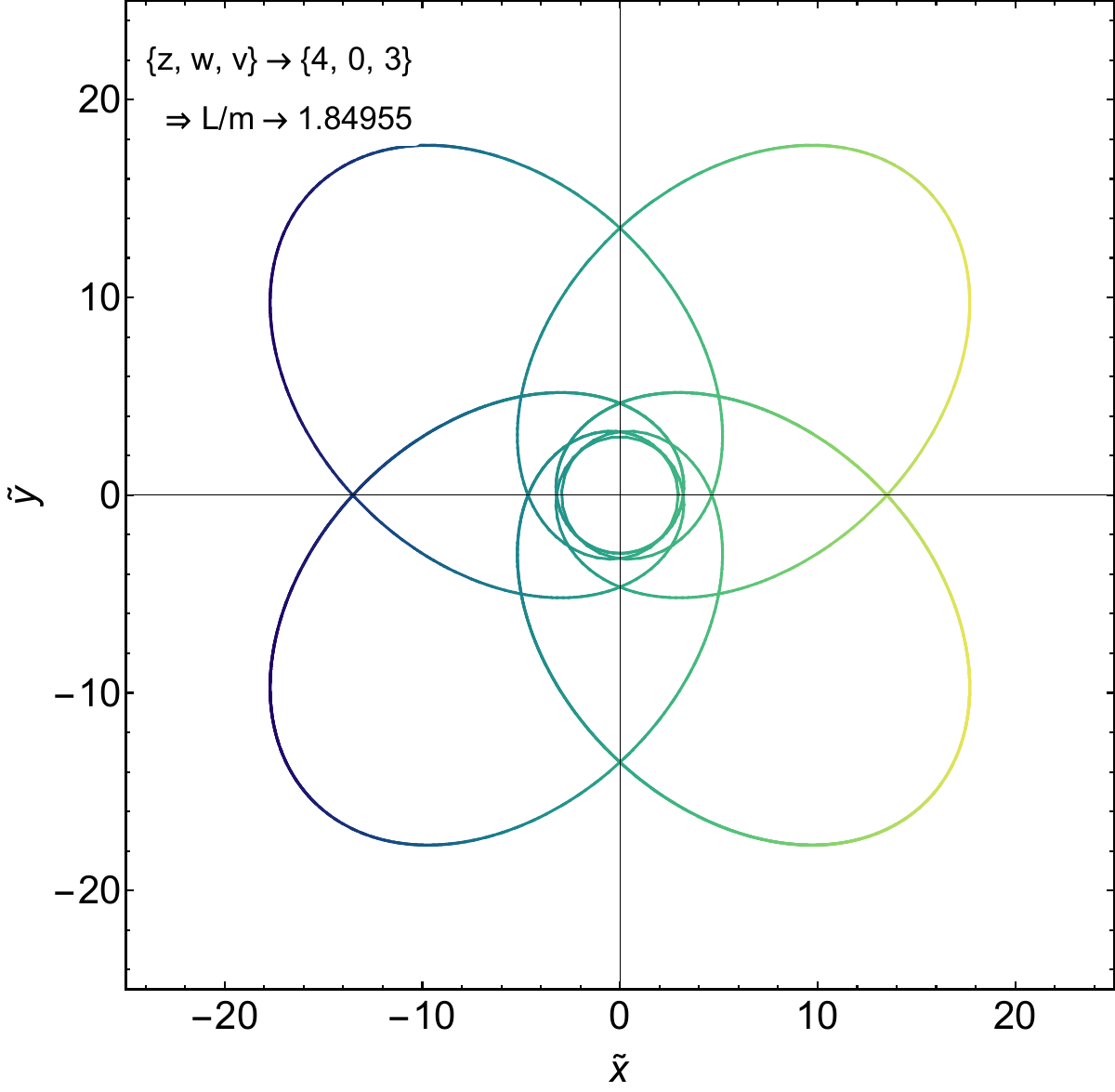} 
\includegraphics[width=5.3cm]{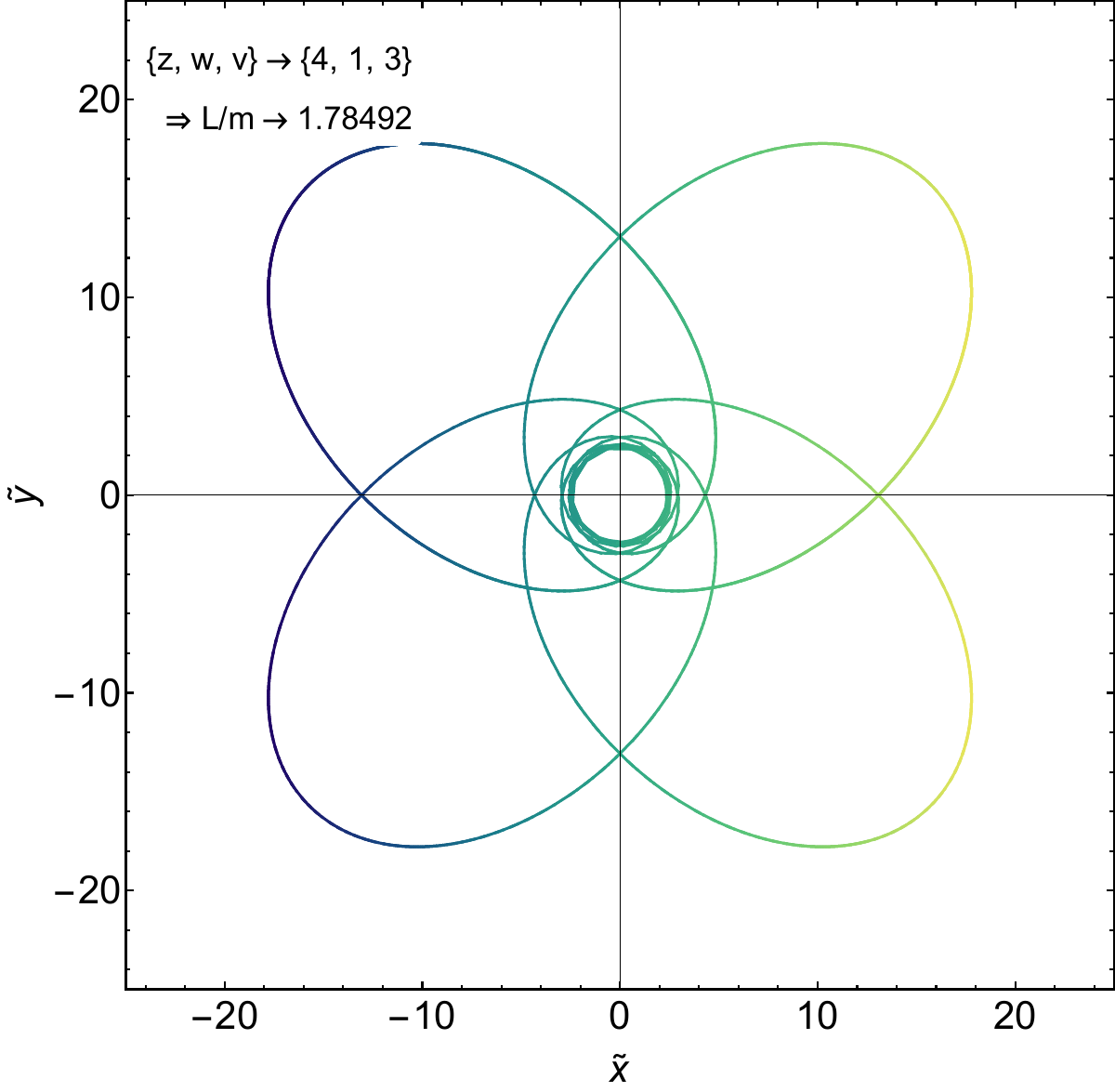} 
\includegraphics[width=5.3cm]{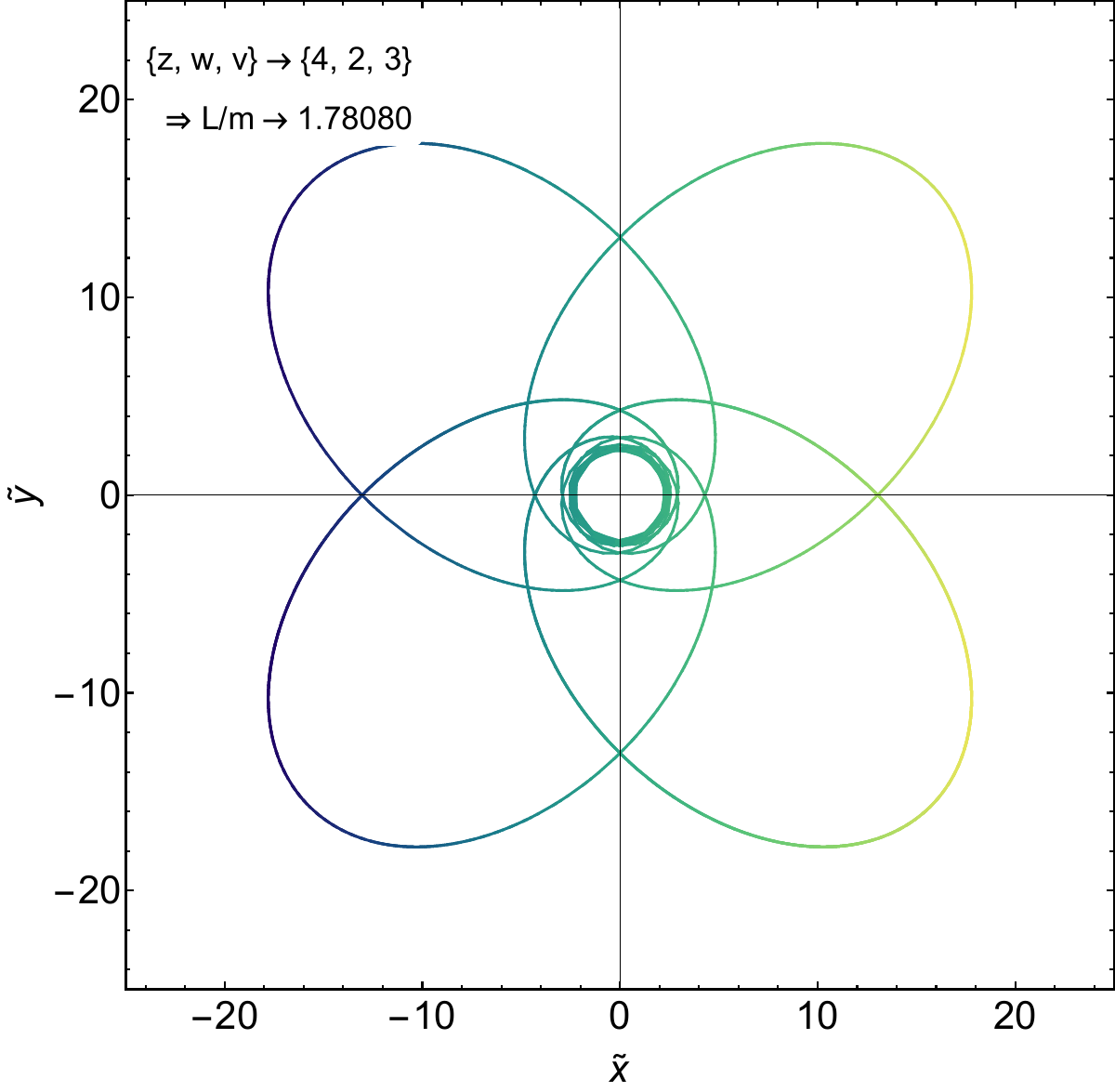} 
\captionsetup{justification=raggedright,singlelinecheck=false}
\caption{Periodic orbits for different values of $(z, w, v)$ (so that different $q$ as well as $L/m$) around a compact object characterized by the $\gamma$-metric \eqref{ds2}. For the above trajectories we are adopting the 2D coordinate system $({\tilde x}, {\tilde y})$, for which we have the dimensionless variables ${\tilde x} \equiv r/m \cos{\phi}$ and ${\tilde y} \equiv r/m \sin{\phi}$. Here we have set $\gamma=0.5$ and $E=0.98$.}
\label{periodic1}
\end{figure*}

\begin{figure*}
\centering
\includegraphics[width=5.3cm]{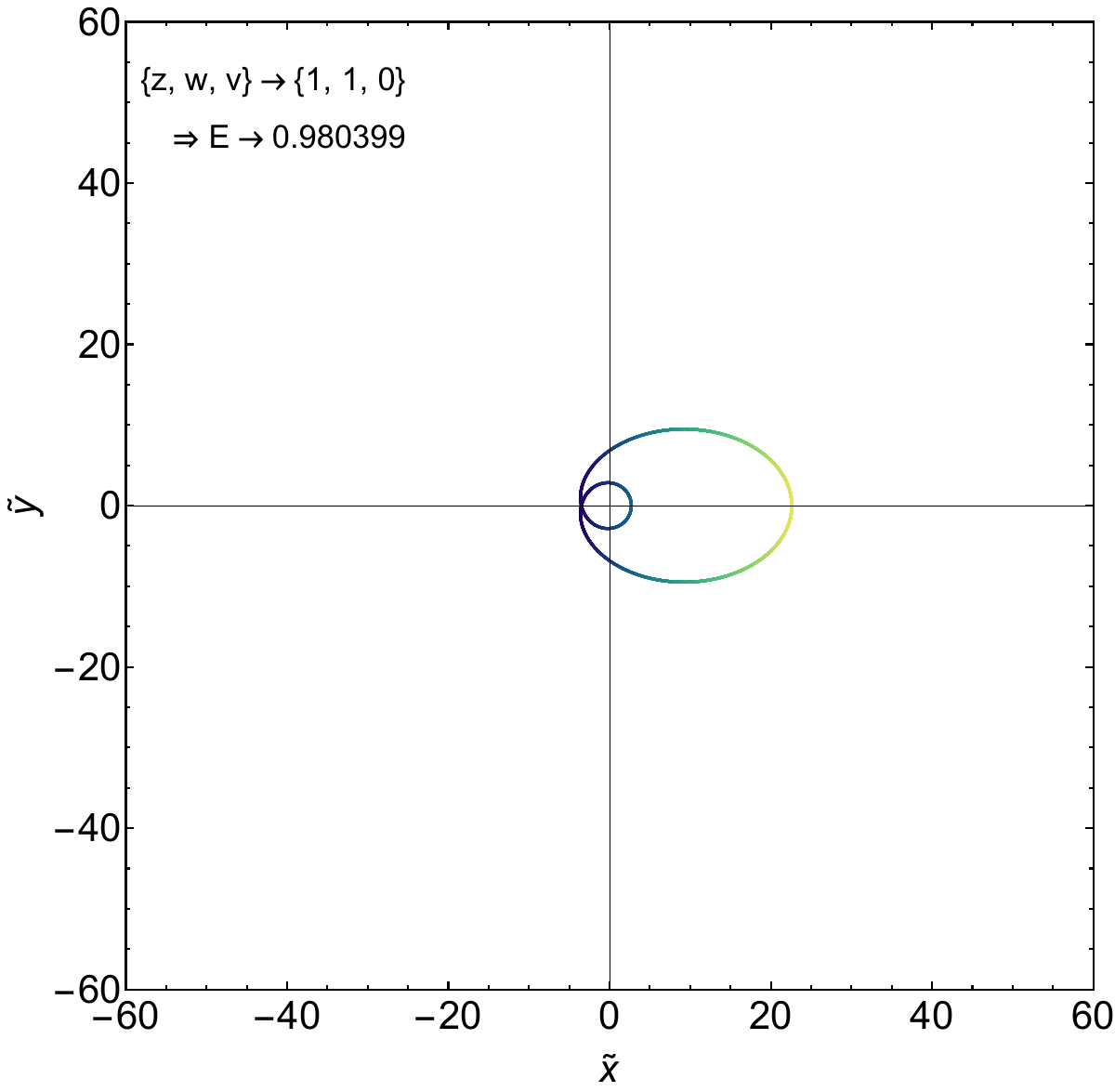} 
\includegraphics[width=5.3cm]{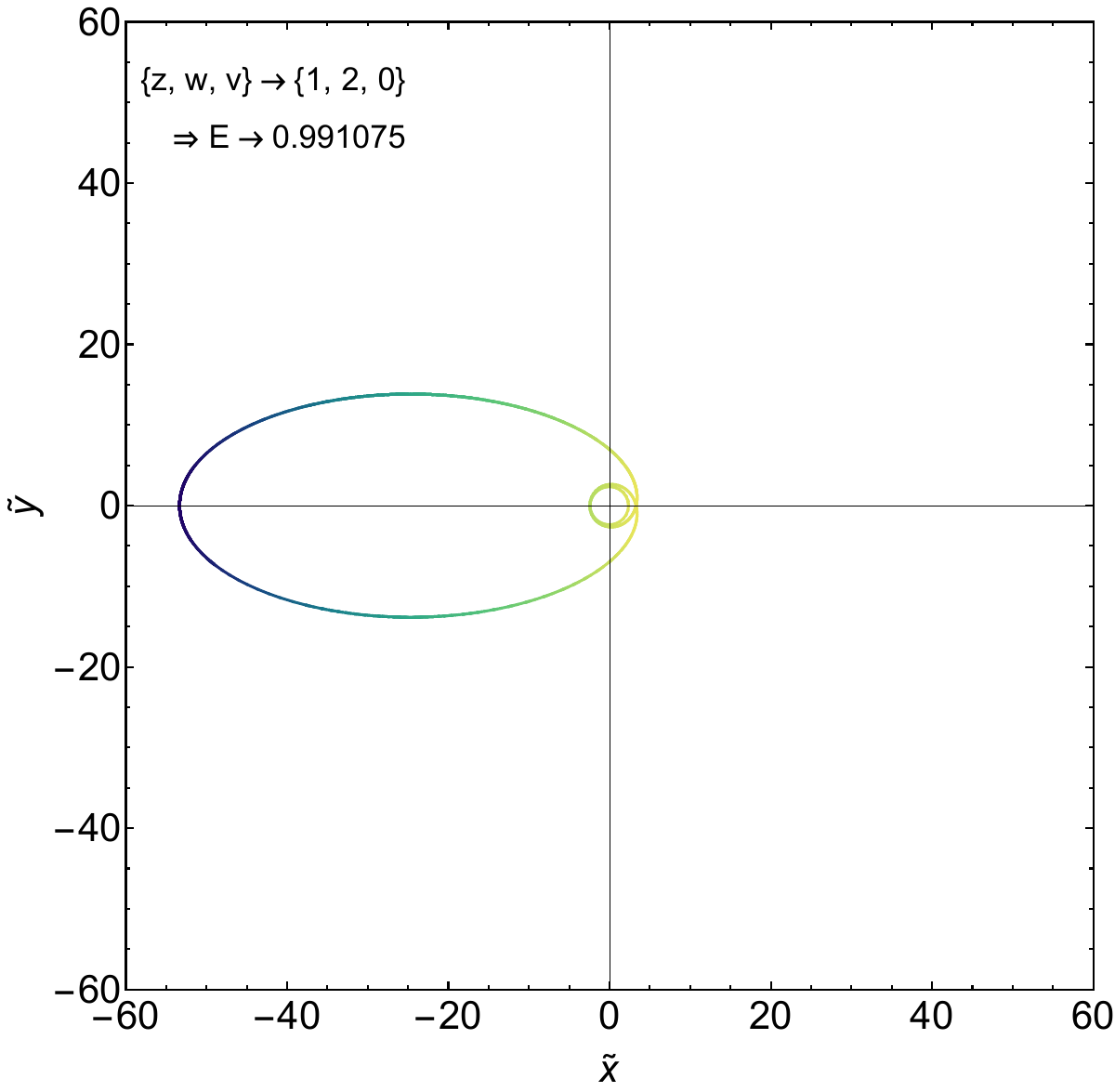}
\includegraphics[width=5.3cm]{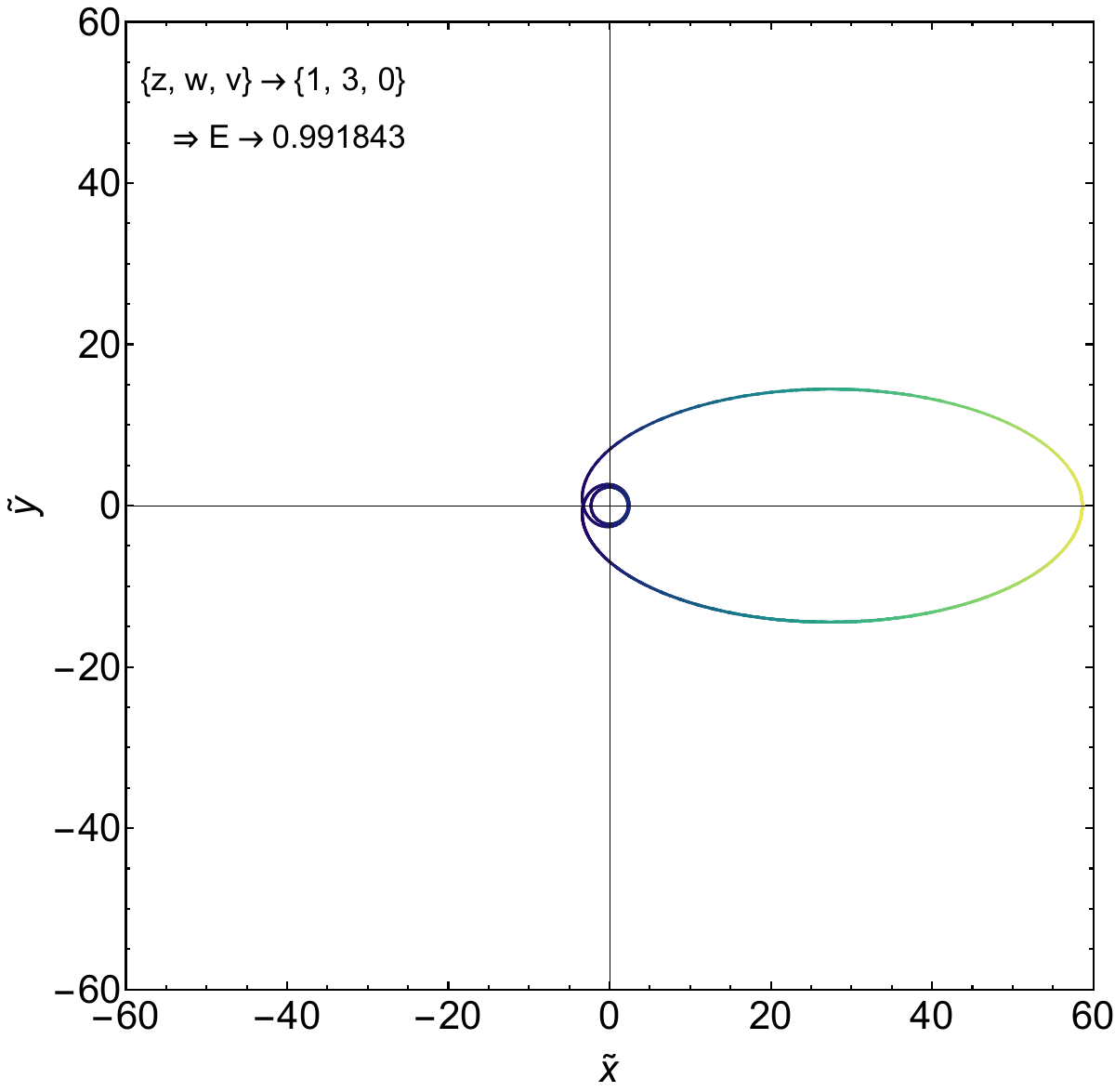} 
\includegraphics[width=5.3cm]{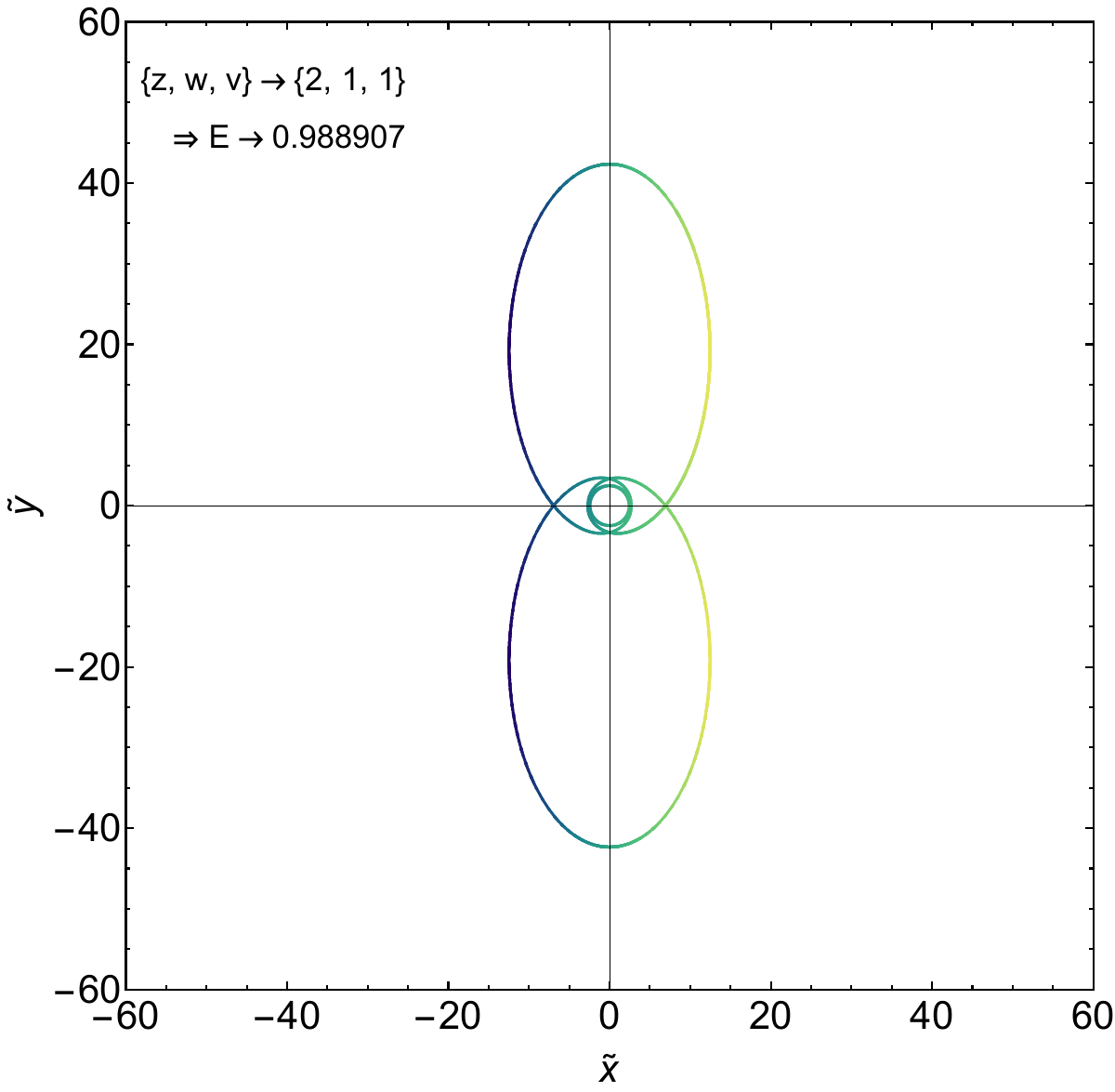} 
\includegraphics[width=5.3cm]{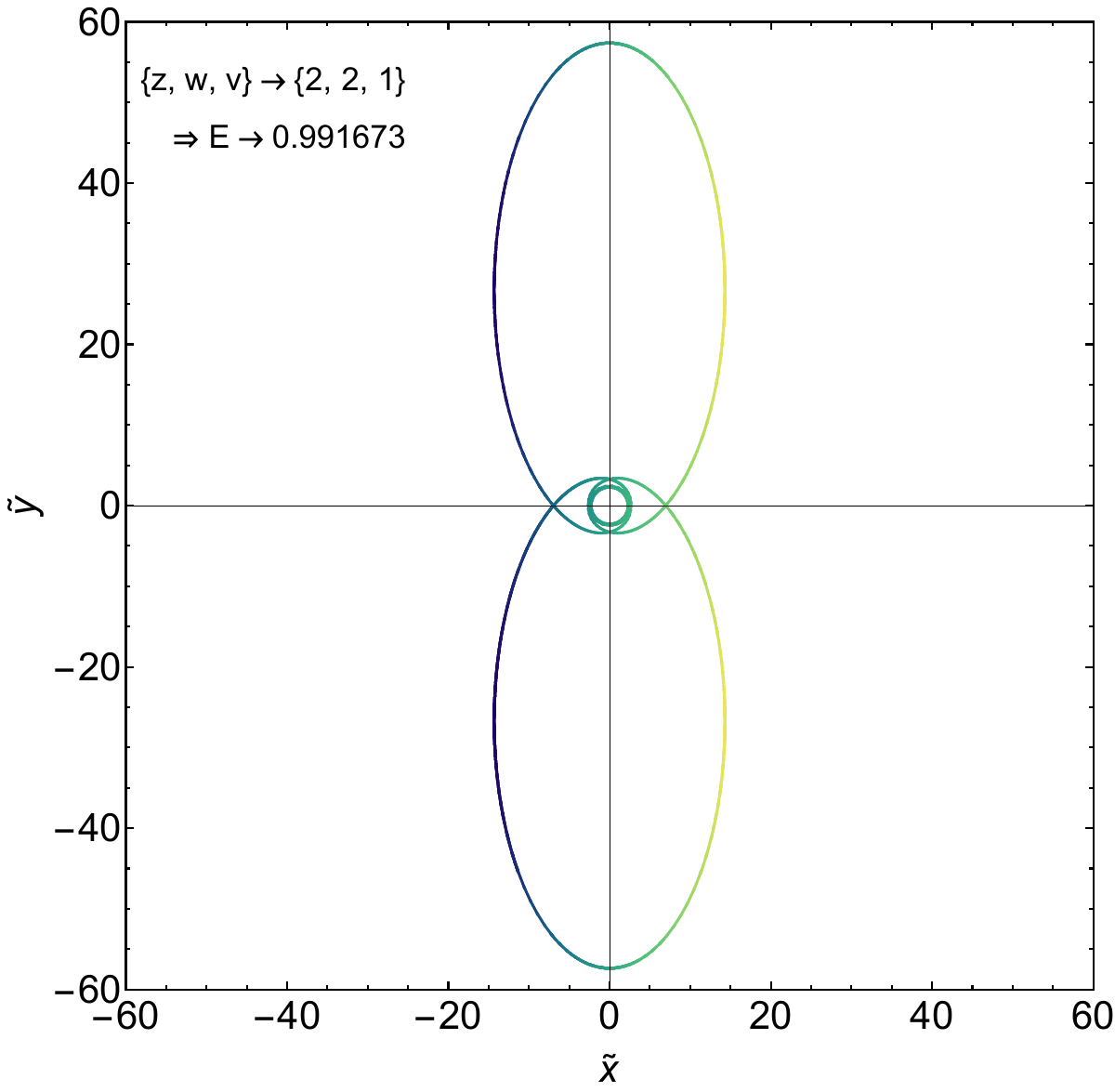} 
\includegraphics[width=5.3cm]{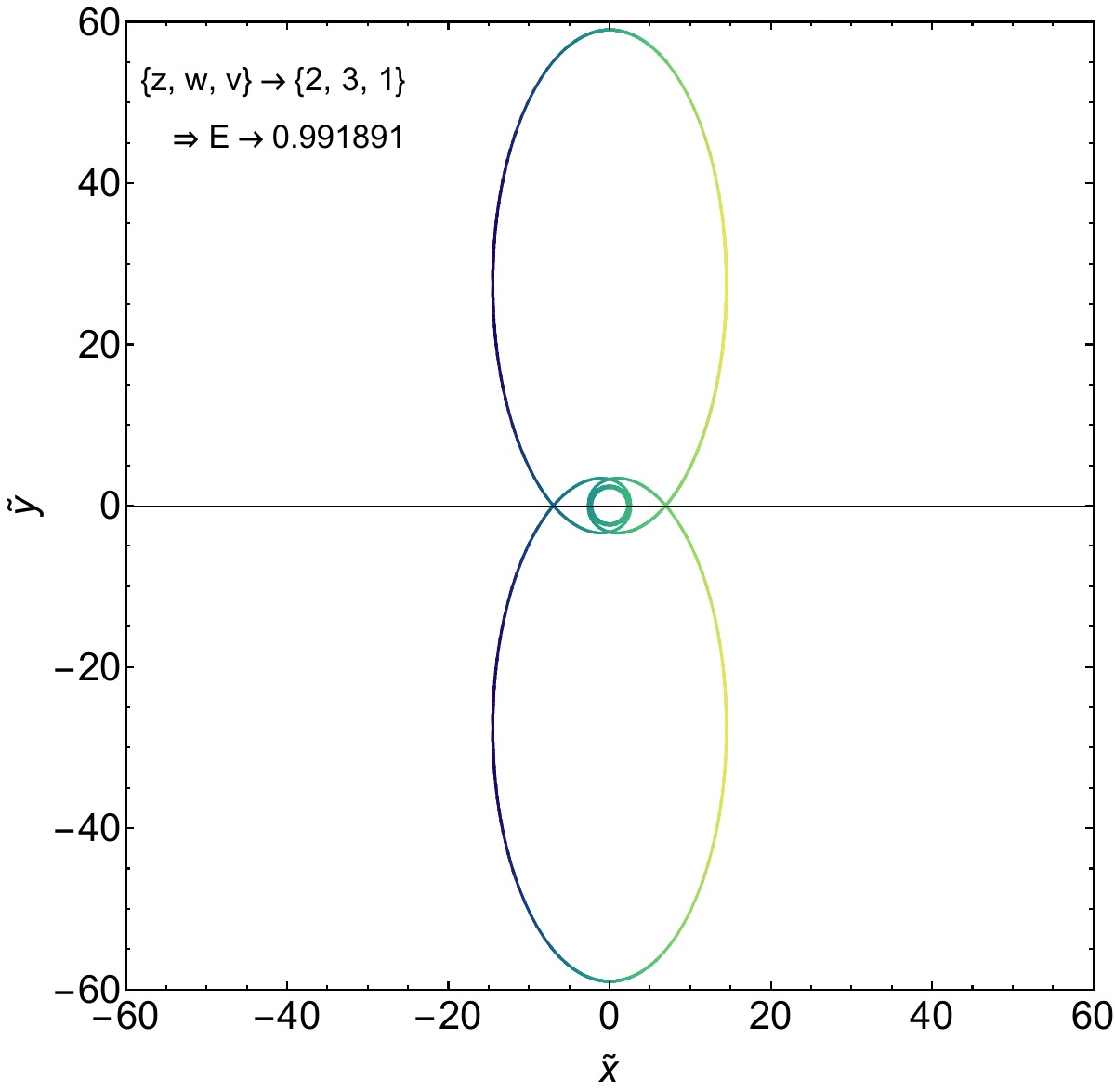} 
\includegraphics[width=5.3cm]{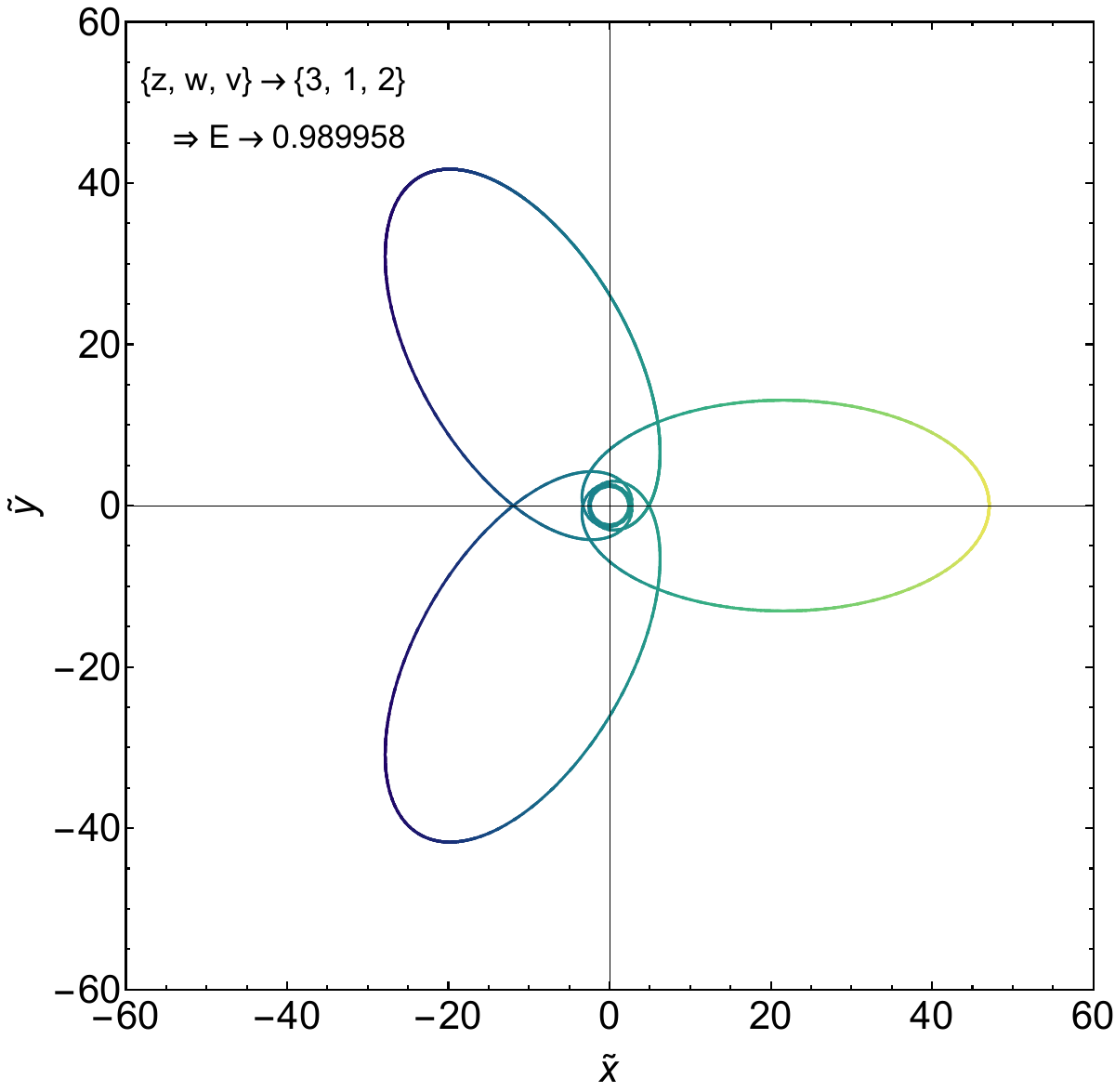} 
\includegraphics[width=5.3cm]{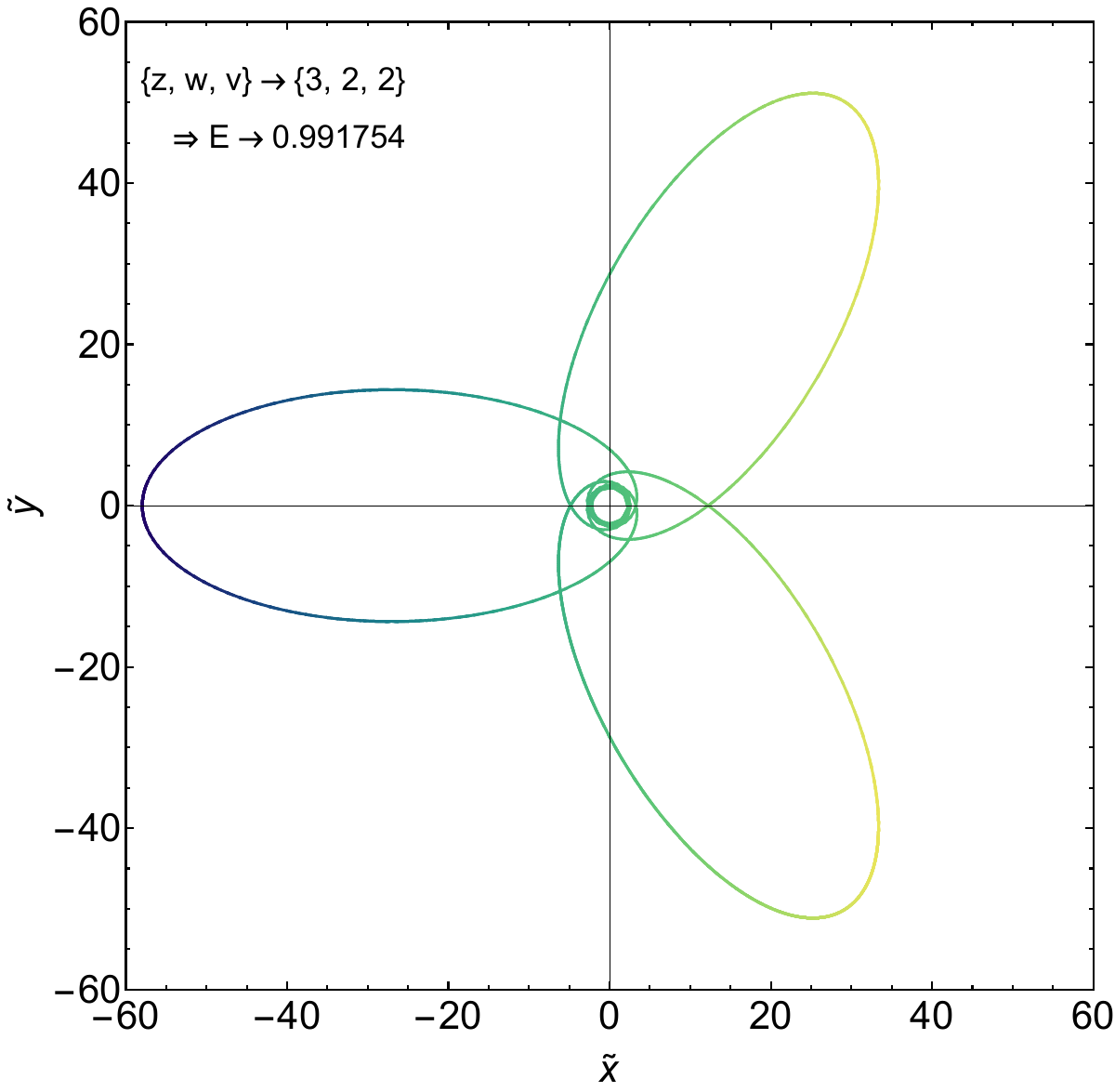}
\includegraphics[width=5.3cm]{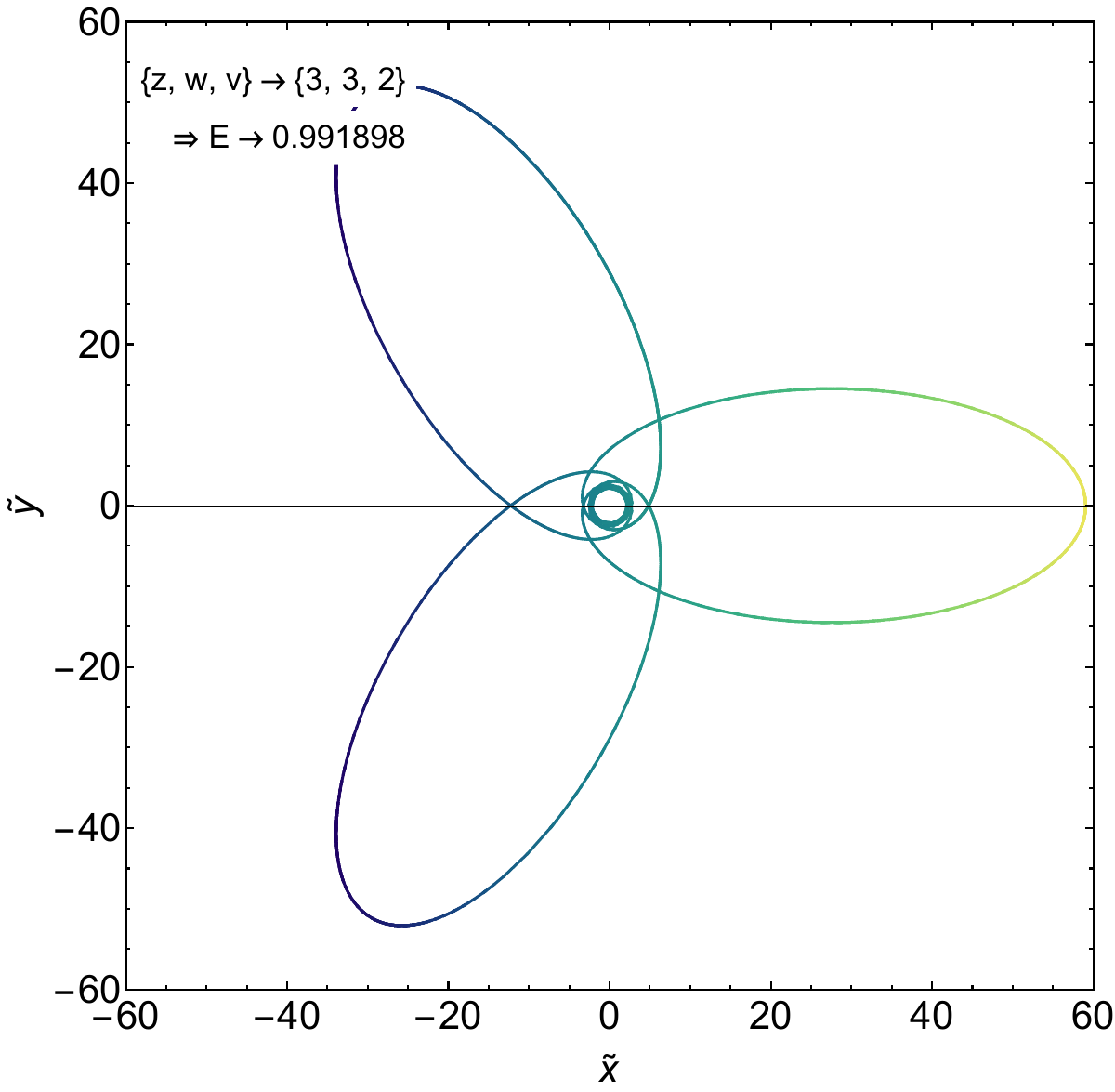} 
\includegraphics[width=5.3cm]{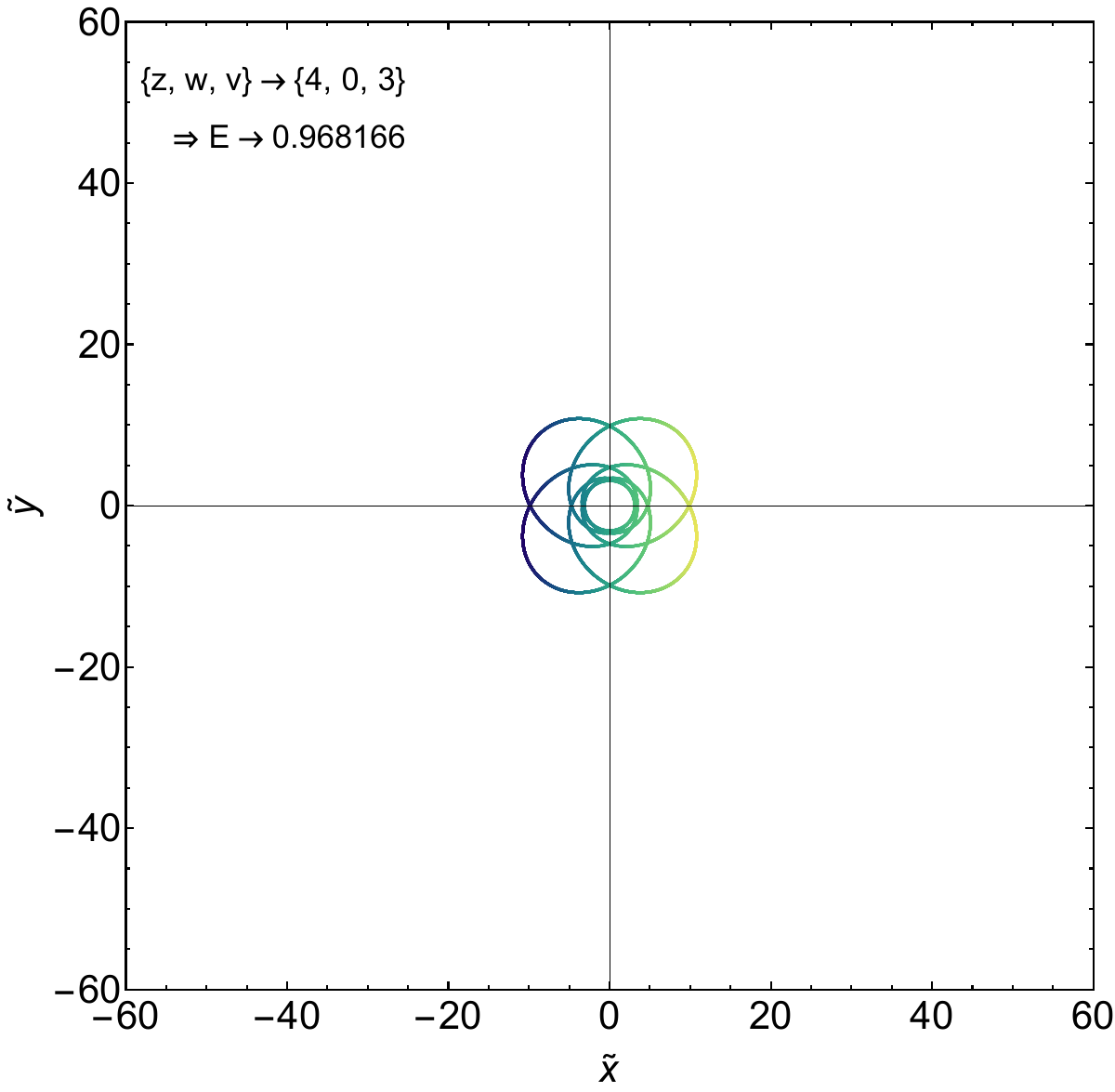} 
\includegraphics[width=5.3cm]{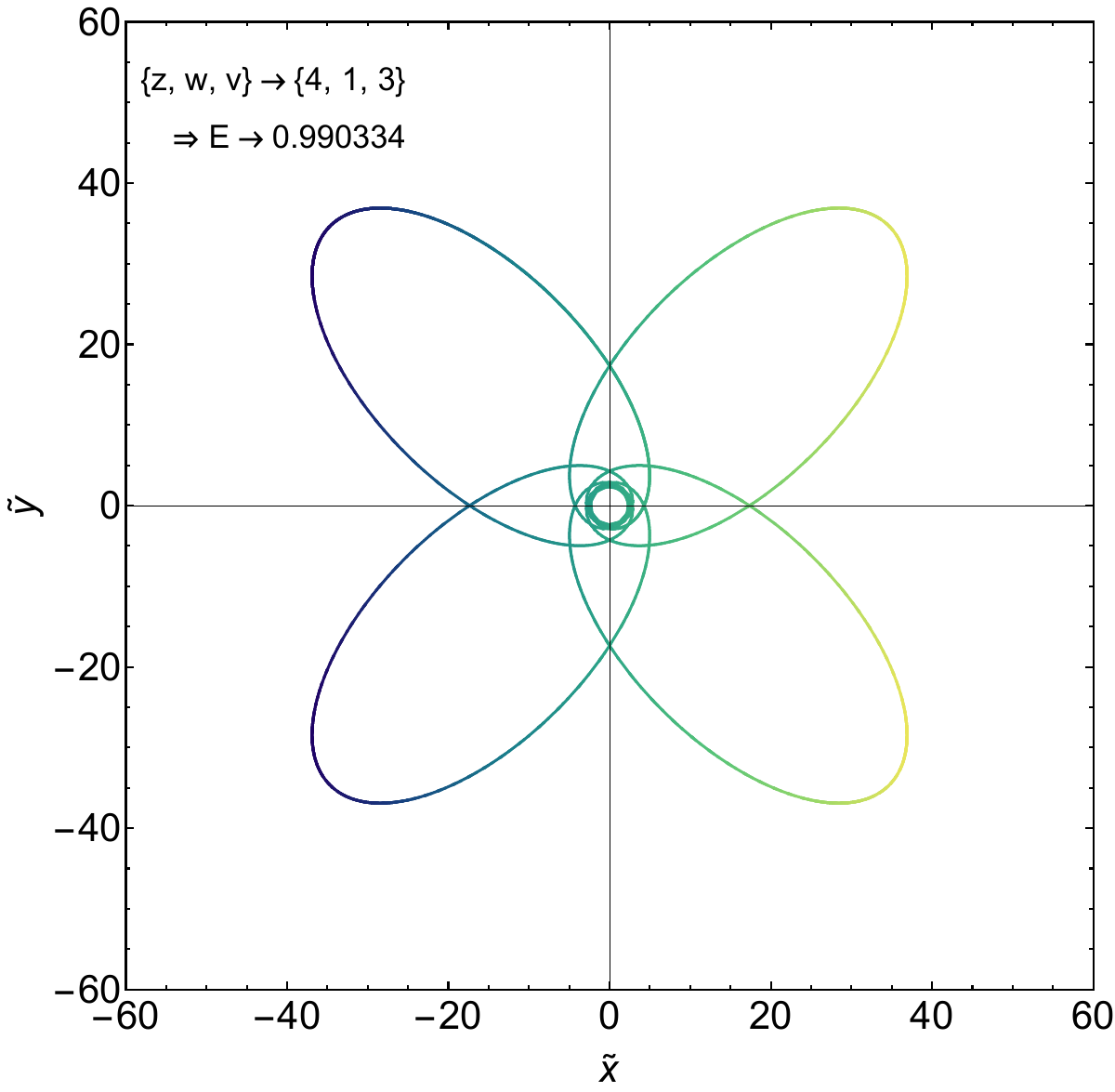} 
\includegraphics[width=5.3cm]{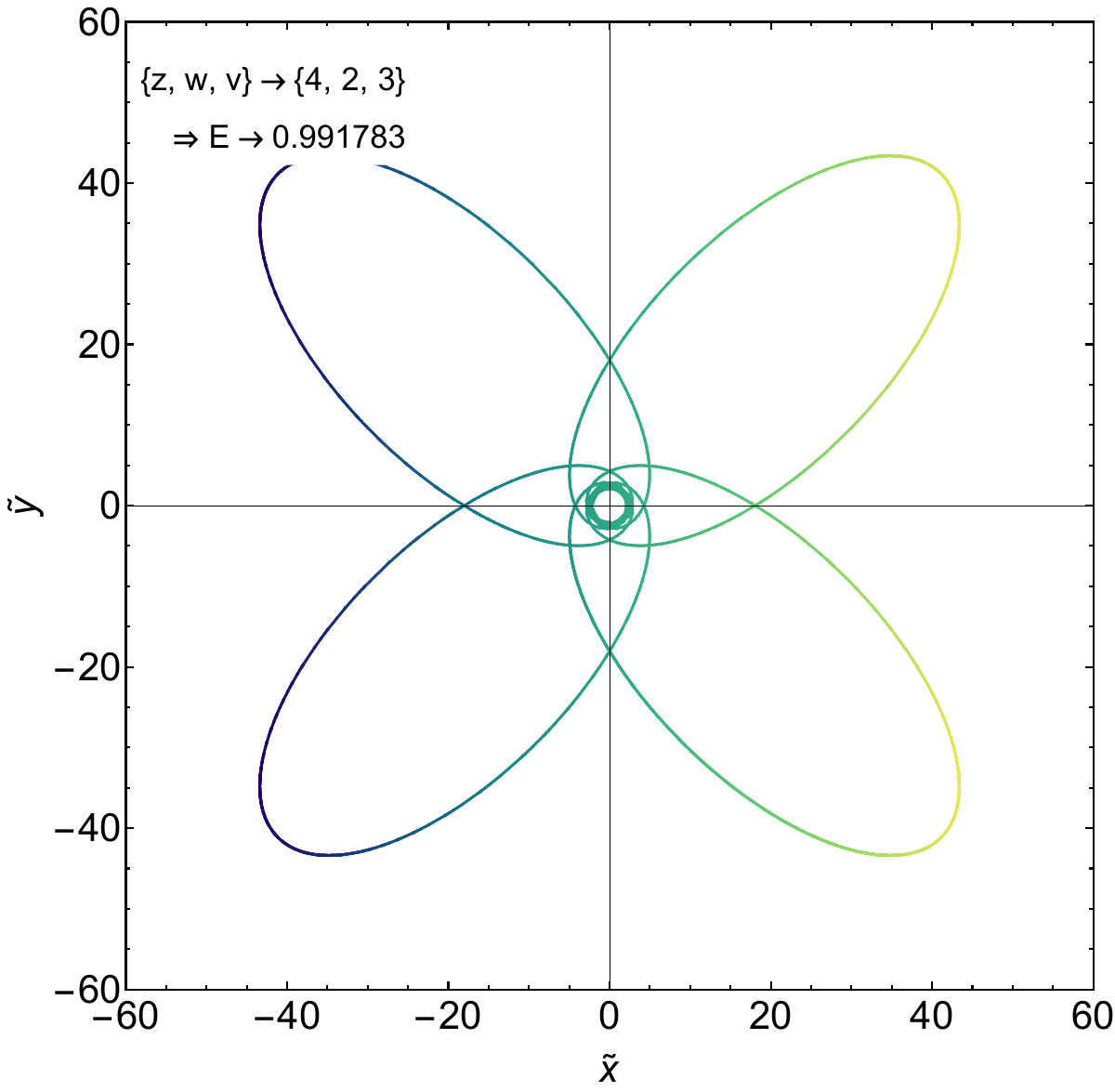} 
\captionsetup{justification=raggedright,singlelinecheck=false}
\caption{Periodic orbits for different values of $(z, w, v)$ (so that different $q$ as well as $E$) around a compact object characterized by the $\gamma$-metric \eqref{ds2}. For the above trajectories we are adopting the 2D coordinate system $({\tilde x}, {\tilde y})$, for which we have the dimensionless variables ${\tilde x} \equiv r/m \cos{\phi}$ and ${\tilde y} \equiv r/m \sin{\phi}$. Here we have set $\gamma=0.5$ and $\epsilon=0.9$.}
\label{periodic2}
\end{figure*}

Now we are in the position of obtaining the trajectories for the periodic motions of the particle around the central compact body characterized by the $\gamma$-metric \eqref{ds2}. To do so, one needs to first apply the Euler-Lagrange equation to the Lagrangian \eqref{Lagrangian} with respect to $r(\lambda)$. As a result, a 2nd-order ordinary differential equation (ODE) of $r(\lambda)$ can be obtained, with the substitution by using the first equation of \eqref{EOM}. On the other hand, notice that \eqref{EOM} also provides us with a 1st-order ODE for $\phi(\lambda)$. Combining these two, we actually have a set of coupled ODEs,
 \bqn
\lb{ODEs1}
0 &=&  \left(\tilde{r}-2\right) \tilde{r} \left(1-\frac{2}{\tilde{r}}\right)^{-\gamma } \left(m{\dot \phi} \right)-\frac{L}{m}, \nb\\
0 &=& \gamma  E^2 \left(\tilde{r}-2\right) \tilde{r}+\left(\tilde{r}-1\right)^4 \left[\frac{\left(\tilde{r}-2\right) \tilde{r}}{\left(\tilde{r}-1\right)^2}\right]^{\gamma ^2+1} \left(m^2 {\ddot {\tilde r}} \right) \nb\\
&& +\left(\tilde{r}-1\right) \left(\gamma ^2+\gamma -1\right) \left[\frac{\left(\tilde{r}-2\right) \tilde{r}}{\left(\tilde{r}-1 -\gamma  \tilde{r}\right)^2}\right]^{\gamma ^2} \left(m {\dot {\tilde r}} \right)^2 \nb\\
&& +\left(\tilde{r}-2\right)^2 \tilde{r}^2 \left(-\tilde{r}+\gamma +1\right) \left(m{\dot \phi} \right)^2,
\eqn
for which we have converted them into their dimensionless forms by using the dimensionless quantities $( m{\dot \phi} )$, $( m{\dot {\tilde r}} )$, $L/m$, and the dimensionless variable ${\tilde r}$ defined as ${\tilde r}\equiv r/m$. Such a treatment can technically reduce some of the confusions in selecting parameters like $m$ in practice.

To solve for \eqref{ODEs1}, as the first example, we choose to set $\gamma=0.5$ and $E=0.98$. At the same time, one can choose a specific $(z, w, v)$, viz., $q$, for the further calculations. As can be seen from, e.g., the Fig.\ref{qL1}, that will lead to a specific $L/m$\footnote{\textcolor{black}{For a numerical calculation, this can be done only up to a certain tolerance level of numerical error in practice. As will be seen later, our analysis to the GW signals of the EMRI systems under consideration needs several periods of the trajectory data. Therefore, a resultant $L/m$ needs to be accurate enough so that the periodic trajectory is stable within these periods. Overall speaking, the results are sensitive to the choices of $L/m$ and $E$. In practice, we learned \textit{a posteriori} that the above purpose can get satisfied once the $L/m$ is determined up to its, e.g., 10th digit, The similar principle also applies to the calculation of $E$. Interestingly, the trajectories are often found to own the tendency of being stable for different choices of parameters and it was realized that less digits for $L/m$ and $E$ will also work sometimes (although we prefer to increase a little the accuracy even for these cases so that the results can be more reliable). Similarly, for the other parts of our computational code (basically with {\it Mathematica}), the working precision is usually chosen to be bigger than, e.g., 30, to guarantee the accuracy of the produced results.}}. While for the initial conditions, without the loss of generality, set $\phi(\lambda=0)=0$, $r(\lambda=0)=r_1$ and ${\dot r}(\lambda=0)=0$. Thus, everything has been ready for solving \eqref{ODEs1} for $\{{\tilde r}, \phi\}$. These solutions form the trajectories for the cases under consideration and are plotted in Fig.\ref{periodic1}. In there all the trajectories are shown in the 2D coordinate system $({\tilde x}, {\tilde y})$, for which we have introduced the dimensionless variables ${\tilde x} \equiv r/m \cos{\phi}$ and ${\tilde y} \equiv r/m \sin{\phi}$. Notice that, during the whole procedure of solving \eqref{ODEs1}, one is not bothered by the arbitrariness of $m$ since the ODEs in \eqref{ODEs1} are in their dimensionless form (and in practice a dimensionless variable $\lambda/m$ can be borrowed to there expediently to eliminate those $m$'s). 

\textcolor{black}{
As noted, a critical step in the above calculations is to determine a sufficiently accurate value of $L/m$ for a given $q$. This is achieved using Fig.\ref{qL1}, which provides the numerical relation between $L/m$ and $q$. To minimize potential numerical errors, we have further verified the convergence of our results through two methods. First, we found that the trajectories (cf. Fig.\ref{periodic1}) remain stable, without exhibiting chaotic behavior—at least within the number of periods needed—when $L/m$ is specified up to a certain digit (typically less than 10, as mentioned earlier). We demonstrate that including additional digits in the numerically determined $L/m$ does not appreciably affect the results, confirming the adequacy of the chosen precision. Second, by supplying initial data with accuracy up to approximately the 10th digit (or less), we tested over selected intervals of $q$ (covering the cases in Figs.\ref{periodic1} and \ref{periodic2}) and verified that a fitting function of the form $L/m = \Sigma \alpha_n q^n$ (with $n = 0, 1, 2, \dots$ and constants $\alpha_n$) can reliably predict the required $L/m$ for trajectory calculations. A similar outcome was observed in deriving Fig.\ref{periodic2}. Together, these checks support the reliability of our results, which exhibit stable and convergent behavior (without significant numerical error or chaotic tendency) under finite input accuracy.
}

The second attempt can be choosing an $L/m$ together with a specific $(z, w, v)$, viz., $q$. For this purpose, we fix $\gamma=0.5$ and $\epsilon=0.9$.  As we have seen from, e.g., the Fig.\ref{qE1}, that will lead to a specific $E$  (up to a certain tolerance level of numerical error in practice, just like when we are fixing the $E$). Once again, everything is ready for solving \eqref{ODEs1} for $\{{\tilde r}, \phi\}$ after this operation. That brings us the desired trajectories for the cases under consideration. We plot out this part of the results in Fig.\ref{periodic2}. In there, all the trajectories are shown in the 2D coordinate system $({\tilde x}, {\tilde y})$ as in Fig.\ref{periodic1}. 
It is evident that $z$ describes the number of blade shapes for the orbit. As $z$ increases, the blade profile grows, and the trajectory becomes more complex (the similar phenomena are happening in Fig.\ref{periodic1}).
\begin{figure}[h]
\includegraphics[width=0.9
\linewidth]{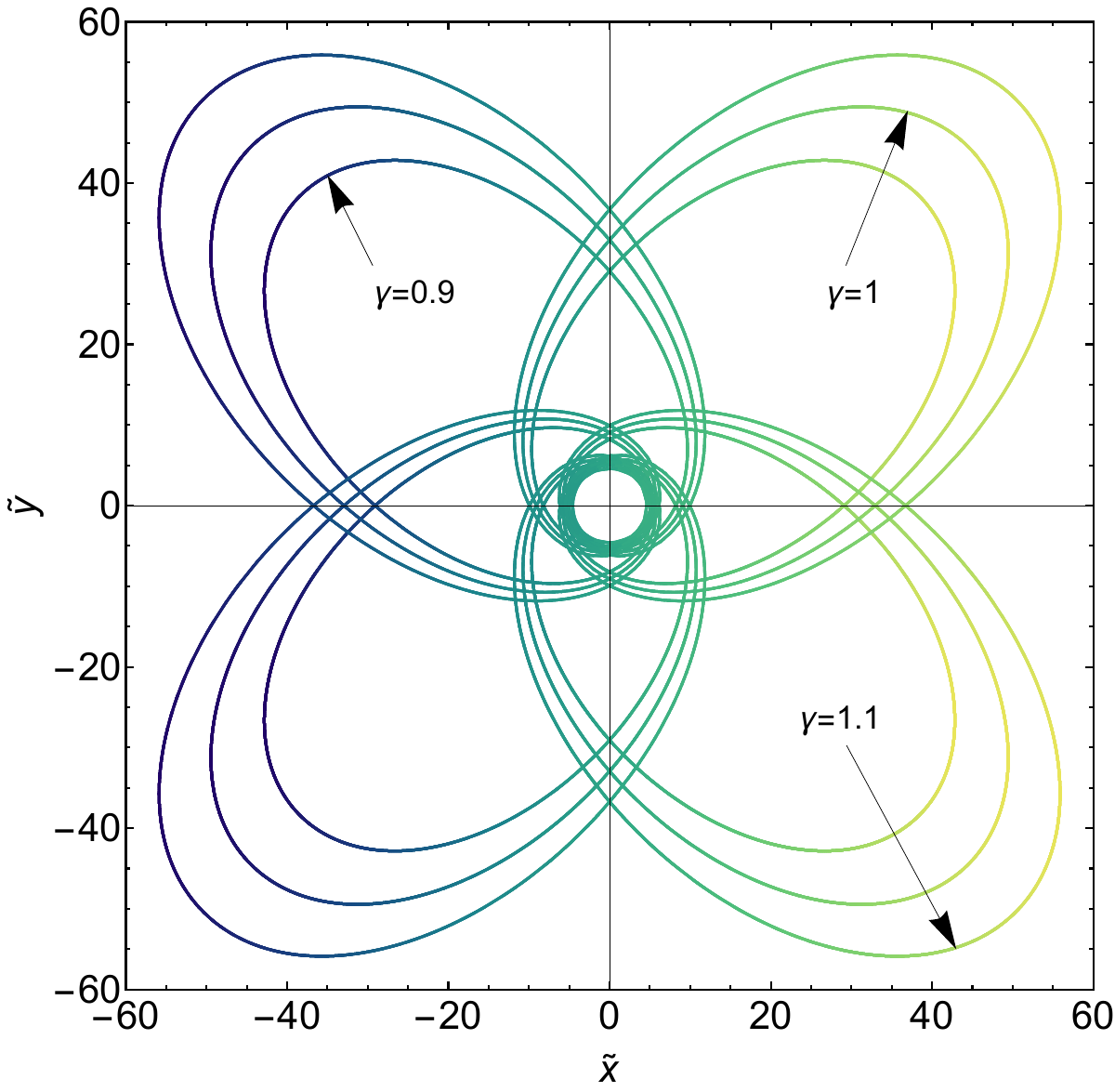} 
\caption{Periodic orbits for different values of $\gamma$  around a compact object characterized by the $\gamma$-metric \eqref{ds2}. Here we have set $(z, w, z)=(4, 0, 3)$ and $\epsilon=0.9$.
} 
\label{periodic3}
\end{figure} 

To better reflect the influence of $\gamma$ on the trajectories of the test part as well as for later convenience, we plot out the results of trajectories in Fig.\ref{periodic3} by setting $(z, w, z)=(4, 0, 3)$ and $\epsilon=0.9$. For the value of $\gamma$ we have chosen $\gamma=0.9$, $\gamma=1$ (the Schwarzschild case) and $\gamma=1.1$. This figure shows that, at the same angular phase position, the radius of the trajectory is getting increased explicitly with $\gamma$. Specially, that results in a significant discrepancy on the trajectory for the $\gamma \neq 1$ case in comparing to that of the Schwarzschild one. 


\section{Gravitational-wave radiation from Periodic orbits in $\gamma$-metric}
\renewcommand{\theequation}{5.\arabic{equation}} \setcounter{equation}{0}
\label{secV}


As mentioned in Sec.\ref{secI}, EMRIs are promising sources for future space-based GW detectors. These systems, typically comprising a stellar-mass compact object orbiting a supermassive compact object (e.g., a SMBH or its mimicker), emit GWs that encode rich information about the system's dynamics and the surrounding spacetime.  When the smaller object is in a periodic orbit around a supermassive object described by a certain metric, the emitted GW waveform offers a unique opportunity to study distinct features that can be exploited to probe the system's properties. This section outlines the theoretical framework for calculating gravitational waveforms from such periodic orbits.

\textcolor{black}{
In the EMRI system under consideration, the mass of the smaller object is much smaller than that of the central supermassive compact body. This large mass ratio justifies treating the small object as a perturbation to the background spacetime, which is described by the $\gamma$-metric. As a result, the orbital energy $E$ and angular momentum $L$ evolve slowly, with changes due to gravitational radiation becoming negligible over a few orbital periods. This condition validates the adiabatic approximation, which holds when the inspiral timescale is much longer than the orbital period (see, e.g., \cite{Tu:2023xab}). Within this approximation, the trajectory of the small object is well approximated by a geodesic of the fixed $\gamma$-metric background over multiple orbits, and the back-reaction of the emitted gravitational waves on the orbit can be safely ignored.
}


\textcolor{black}{We adopt the kludge waveform developed in \cite{Babak:2006uv} and the quadrupole formula to calculate the GW radiation from periodic orbits.}\footnote{\textcolor{black}{As shown in \cite{Babak:2006uv}, the kludge waveforms achieve overlaps exceeding 95\% with Teukolsky-based waveforms for orbits with periastra $r_p \gtrsim 5M$, and remain above 75\% even for inspirals terminating in the strong-field region. While this indicates reasonable accuracy for the exploratory purpose of the present work, i.e., to characterize how the deformation parameter $\gamma$ alters the taxonomy of periodic orbits and the qualitative morphology of gravitational waveforms in the $\gamma$-metric spacetime, we note that the quadrupole formula employed here represents a leading-order approximation. Since the orbits we consider sometimes approach regions as close as $r \approx 2.3m$ (cf. Fig.~\ref{periodic1}), a fully relativistic Teukolsky treatment would be necessary for obtaining the exact spectral power distribution and constructing accurate waveforms for gravitational radiation. This represents a very promising direction for future EMRI-related studies and we be taken into account in our future works.}} This approach involves a two-step process: first, the orbit of the small object is determined by numerically solving the geodesic equations of motion in the central massive body's spacetime (as we have done in the last section). Subsequently, the quadrupole formula for gravitational radiation is applied to this calculated orbit, generating the corresponding waveform. This approach allows for a preliminary investigation of the GW signals emitted by EMRIs and their potential to reveal insights into the properties of both the orbit and the central compact body, as well as the metric under consideration.


For a metric perturbation $h_{ij}$ representing the GW and the symmetric and trace-free (STF) mass quadrupole $I_{ij}$, the quadrupole formula for gravitational radiation is given by
\begin{equation}
  h_{ij}=\frac{1}{A} \frac{d^2I_{ij}}{d t^2},
\end{equation}
where $A=c^4 D_L/(2G_N)$, $G_N=1=c$, $D_L$ is the luminosity distance to the source. By numerically solving the geodesic equations of motion, the trajectory $Z_i(t)$ of the smaller object in the curved spacetime can be determined.  This trajectory is crucial for the subsequent calculation of the gravitational waveform. For a small object of mass $m_\star$ following a trajectory $Z^i(t)$, the $I_{ij}$ is given by \cite{Thorne:1980ru}
\begin{equation}\label{lvalue}
I^{ij}={m_\star} \int d^3 x^i x^j  \delta^3(x^i - Z^i(t)).
\end{equation}




The choice of coordinate system plays a crucial role in the calculation and interpretation of gravitational waveforms. While the geodesic equations are often solved in Boyer-Lindquist coordinates $(t, r, \theta, \phi)$, the waveform is typically expressed in a detector-adapted coordinate system $(X, Y, Z)$.  This transformation simplifies the analysis of the signal as measured by a GW detector.  The transformation from Boyer-Lindquist to Cartesian coordinates is given by \cite{Thorne:1980ru}
\begin{equation}\label{4.3}
x=r \sin\theta \cos\phi, \quad y=r\sin\theta\sin\phi,  \quad  z=r \cos\theta.
\end{equation}
This allows us to project the small object's trajectory onto a Cartesian grid.
The metric perturbations $h_{ij}$, representing the GWs, are then calculated using the second time derivative of the mass quadrupole moment $I_{ij}$ as
\begin{equation}\label{hij}
h_{ij}=
\frac{2 m_\star}{D_L}(a_i x_j+a_j x_i+2 v_i v_j),
\end{equation}
where $a_i$ and $v_i$ are its acceleration and velocity components, respectively. Here we denote $(x, y, z)$ as $x_i$ with $i=1, 2, 3$. Notice that the velocity and acceleration can be calculated through $v_i=d x_i/d t=(dx_i/d\lambda)/{\dot t}$ and $a_i=d v_i/d t=(dv_i/d\lambda)/{\dot t}$, respectively. 
Since $\dot t$ can be expressed as a function of $r(\lambda)$ as well as $\phi(\lambda)$ [cf. \eqref{EOM}], $x_i$, $v_i$ ,and $a_i$ are actually all functions of  $r(\lambda)$ as well as $\phi(\lambda)$ and their derivatives (As usual, we are adopting the choice of $\theta(\lambda)=\pi/2$ in this paper).  In other words,  $h_{ij}$ can be calculated by substituting the orbital information into \eqref{hij}.


\begin{figure*}
  \centering
  \begin{tabular}{ c }
    \includegraphics[width=0.4\textwidth]{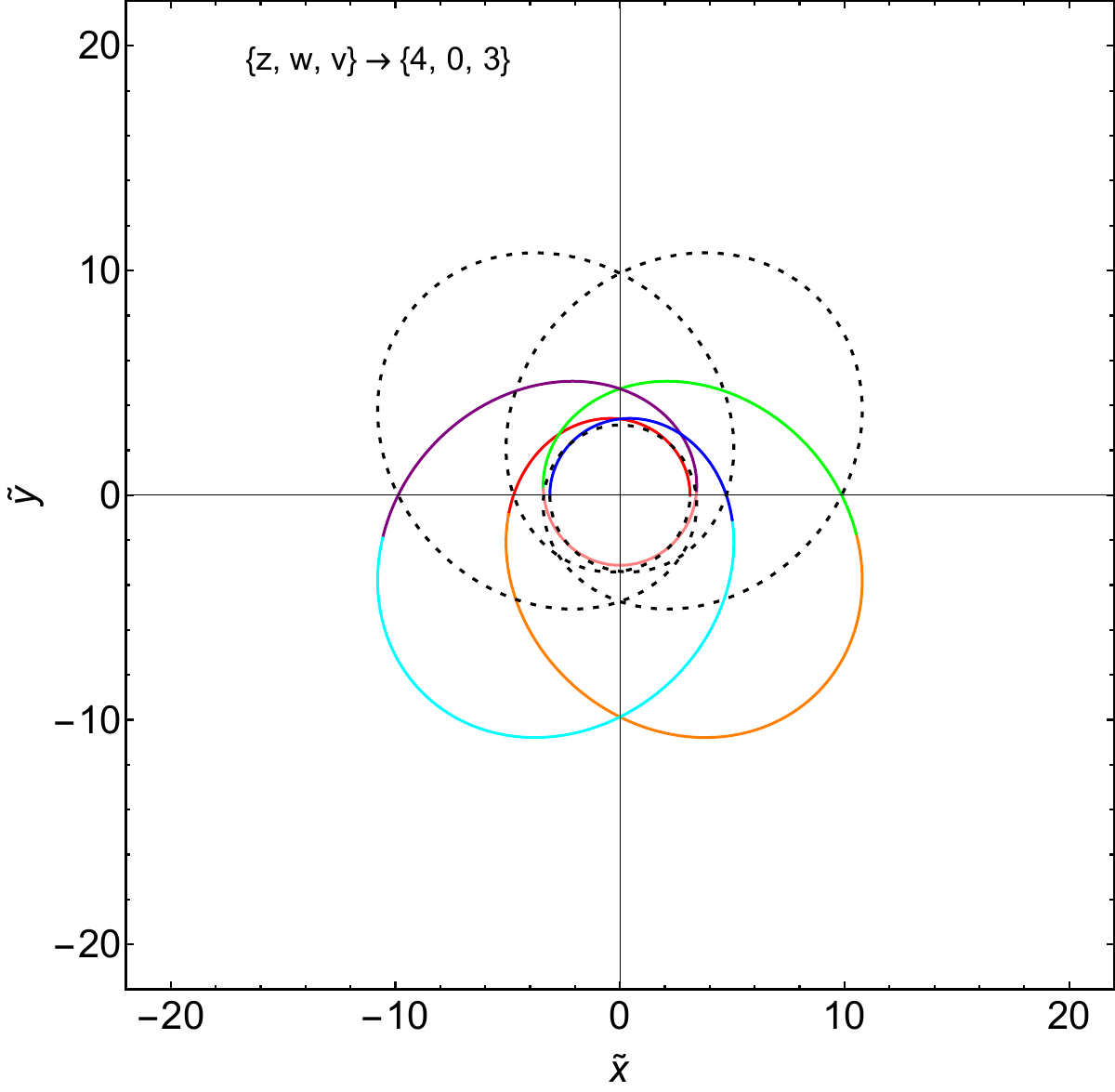}
  \end{tabular}%
  \begin{tabular}{ c c }
   \includegraphics[width=0.5\textwidth]{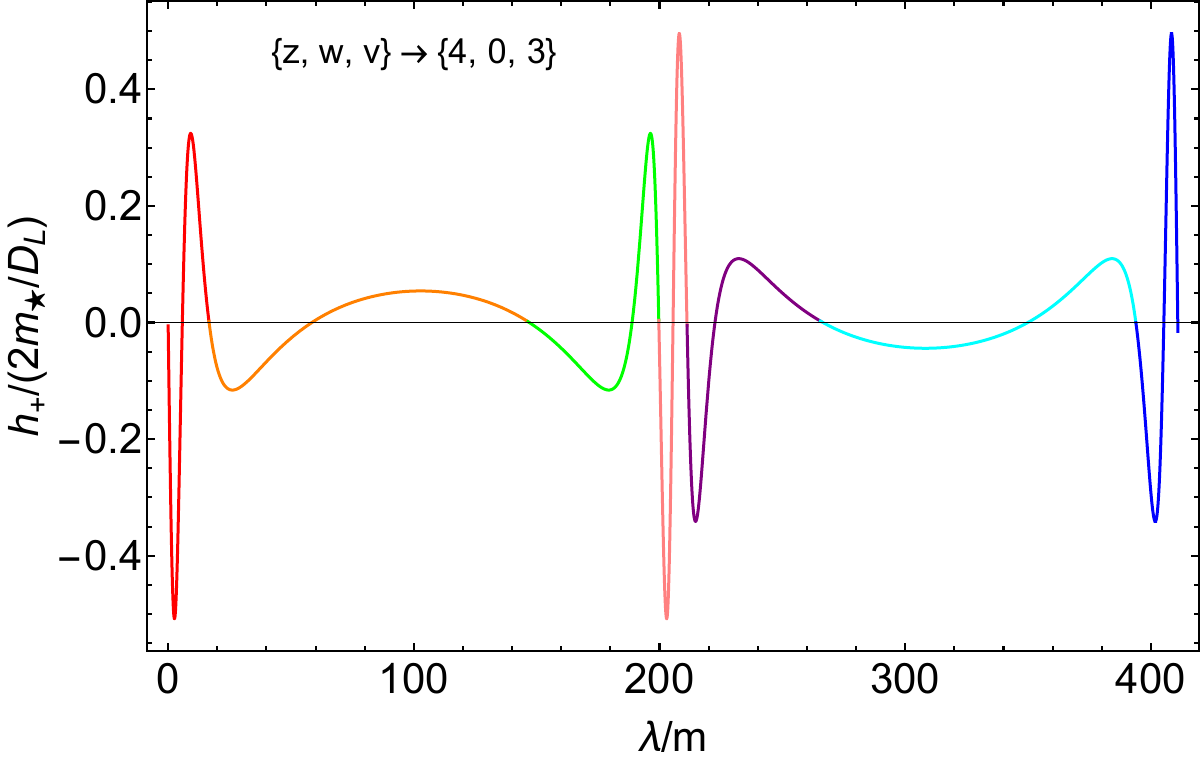}\\ 
    \includegraphics[width=0.5\textwidth]{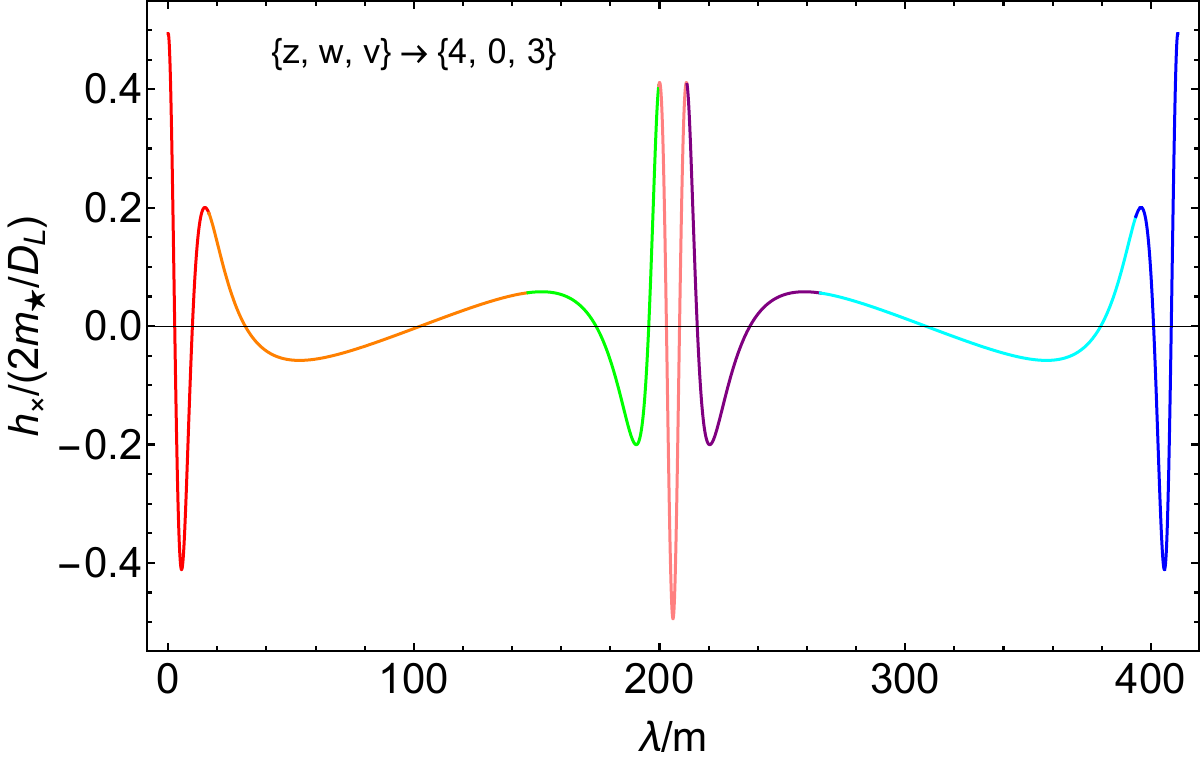} 
     \end{tabular}
\captionsetup{justification=raggedright,singlelinecheck=false}
\caption{The left-hand figure is a sketch figure which shows a typical periodic orbit of a test particle around a central compact object with $(z,w, v) = (4, 0, 3)$, $\gamma=0.5$, $\epsilon=0.9$ and $\zeta=\iota=\pi/4$. The top\textbackslash bottom panel of the right-hand figure shows the waveform $h_+$\textbackslash$h_\times$ for the corresponding orbit. Notice that, for the solid part of these curves, the same color means the three figures are undergoing the same period spanned by $\lambda$.}
\label{waveform1}
\end{figure*}

\begin{figure}[h]
\includegraphics[width=0.9
\linewidth]{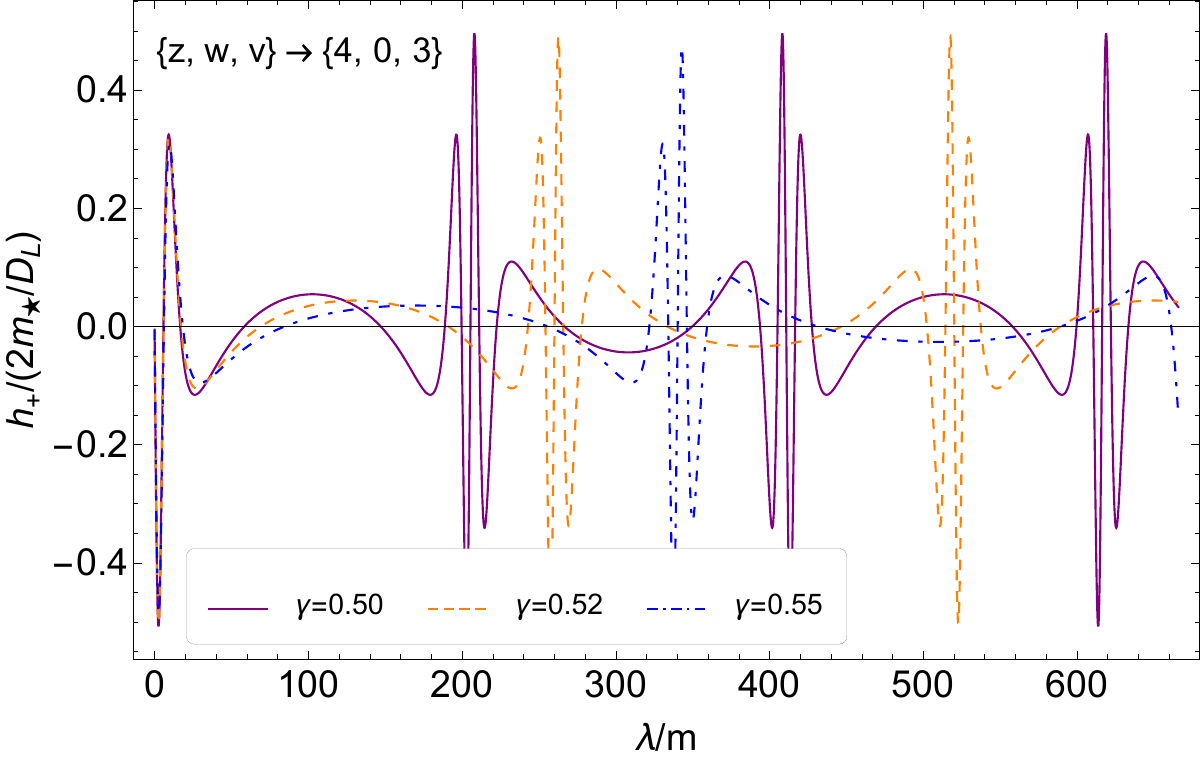} 
\includegraphics[width=0.9
\linewidth]{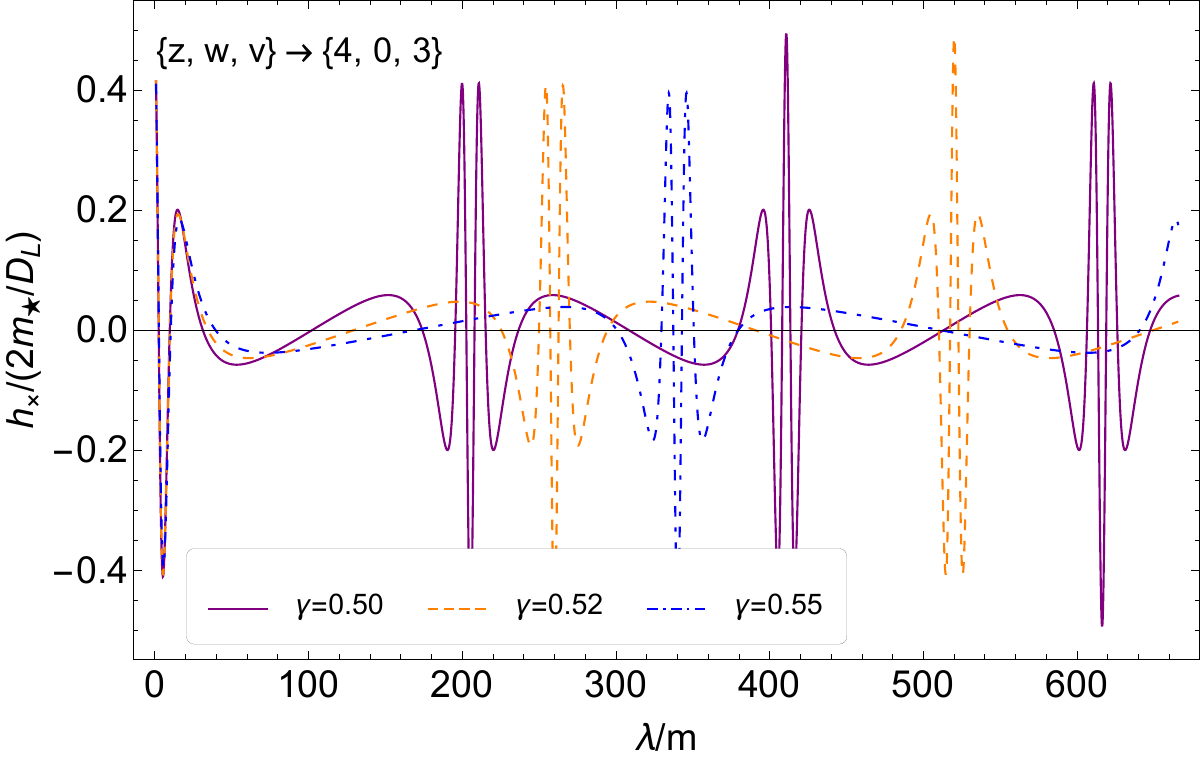} 
\caption{The  behaviors of the two polarizations $h_{+, \times}$ as functions of the dimensionless variable $\lambda/m$ for the typical case $(z,w, v) = (4, 0, 3)$ and $\gamma=0.50, 0.52, 0.55$. Here we have set $\epsilon=0.9$ and $\zeta=\iota=\pi/4$. 
} 
\label{waveform2}
\end{figure} 

\begin{figure}[h]
\includegraphics[width=0.9
\linewidth]{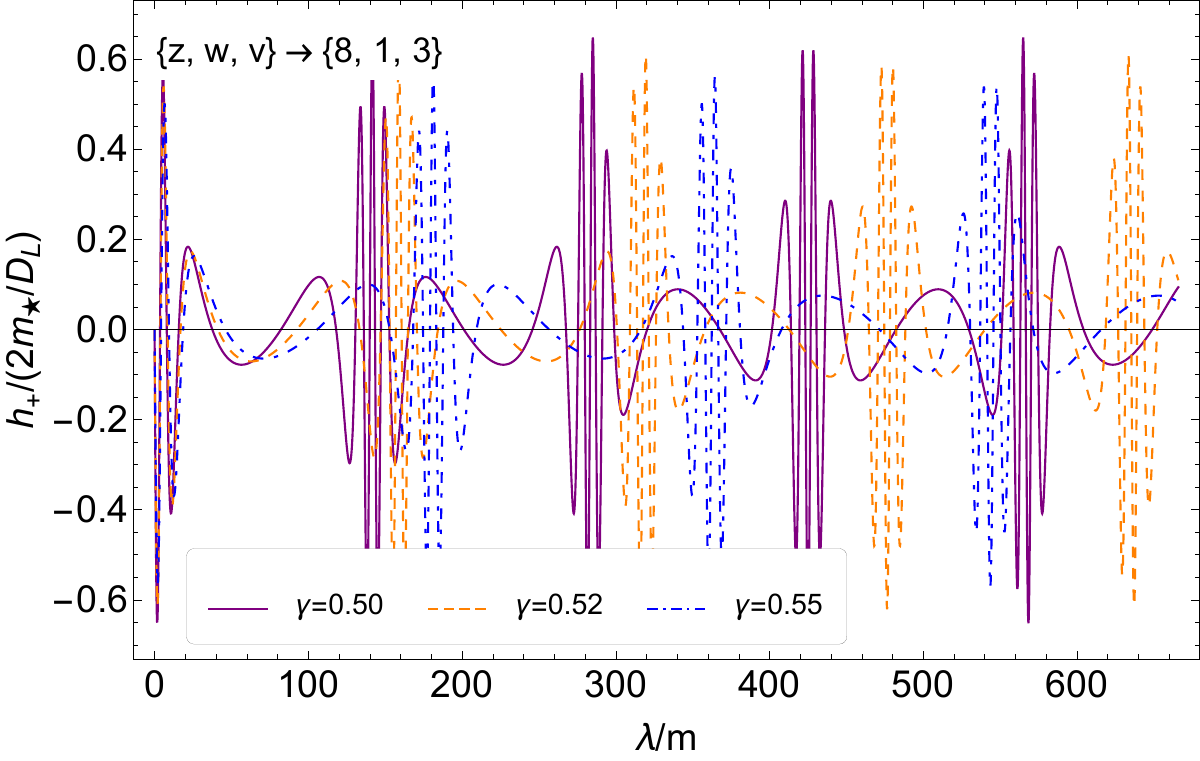} 
\includegraphics[width=0.9
\linewidth]{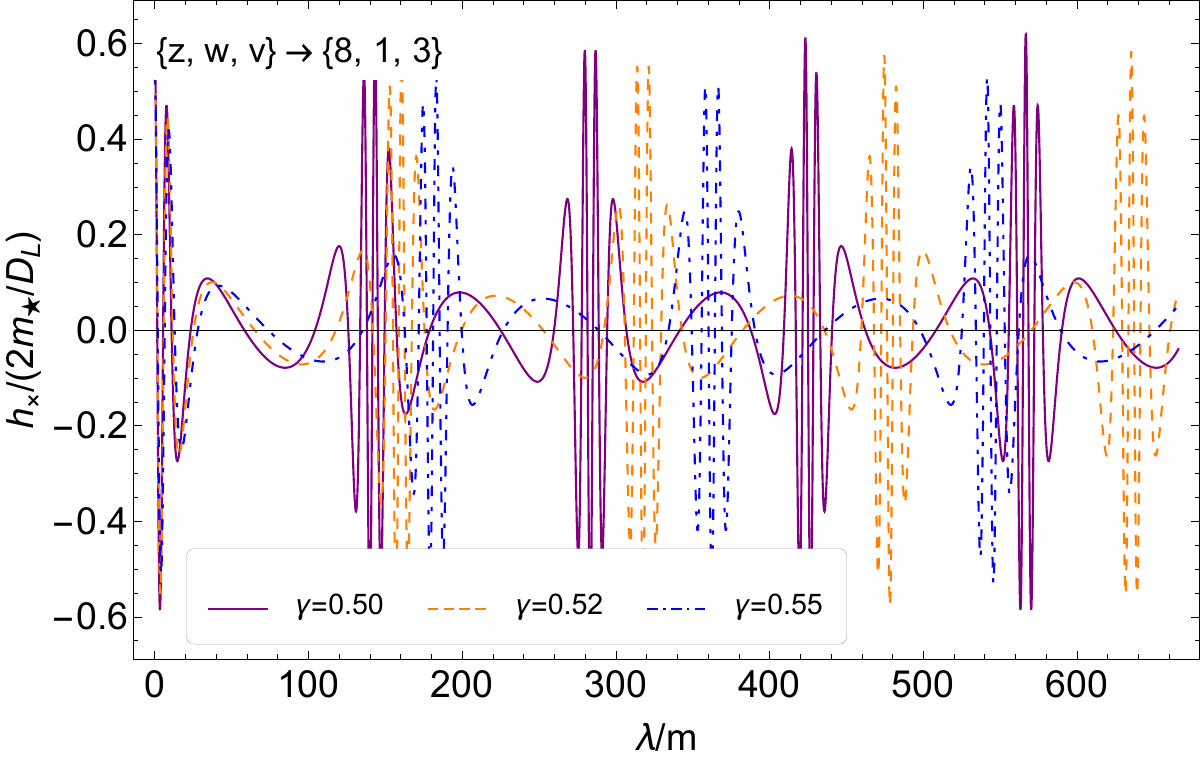} 
\caption{The  behaviors of the two polarizations $h_{+, \times}$ as functions of the dimensionless variable $\lambda/m$ for the typical case $(z,w, v) = (8, 1, 3)$ and $\gamma=0.50, 0.52, 0.55$. Here we have set $\epsilon=0.5$ and $\zeta=\iota=\pi/4$. 
} 
\label{waveform3}
\end{figure} 

By following \cite{Wang:2025hla, Haroon:2025rzx}, the polarizations of the waveforms, which are usually denoted as $h_+$ and $h_\times$ to distinguish these two different modes, can be written as the combination of $h_{ij}$ through the transformation from the coordinate $(x, y, z)$ to $(X, Y, Z)$. 
To be specific, these polarizations can be expressed in terms of components $h_{\zeta\zeta}$, $h_{\iota\iota}$, and $h_{\iota\zeta}$. Here we have introduced the inclination angle $\iota$ and the longitude of pericenter $\zeta$ \cite{Haroon:2025rzx}. 
As a result, the polarized waveforms $h_{+, \times}$ can be expressed as
\begin{eqnarray}\label{hplus}
h_+&=&\frac{1}{2}\big(h_{\zeta\zeta}-h_{\iota\iota}\big),\\\label{hcross}
h_{\times}&=&h_{\iota\zeta},
\end{eqnarray}
where \cite{Wang:2025hla}
\begin{widetext}
 \bqn
\lb{hzetaiota}
h_{\zeta \zeta} &=& \cos ^2(\zeta ) \left(\sin ^2(\iota ) h_{22}+\sin (2 \iota ) h_{12}+\cos ^2(\iota ) h_{11}\right)-\sin (2 \zeta ) \left(\sin (\iota ) h_{23}+\cos (\iota ) h_{13}\right)+\sin ^2(\zeta ) h_{33}, \nb\\
h_{\iota \zeta} &=& \cos (\zeta ) \left(-\frac{1}{2} \sin (2 \iota ) h_{11}+\frac{1}{2} \sin (2 \iota ) h_{22}+\cos (2 \iota ) h_{12}\right)+\sin (\zeta ) \left(\sin (\iota ) h_{13}-\cos (\iota ) h_{23}\right),\nb\\
h_{\iota \iota} &=& \sin ^2(\iota ) h_{11}-\sin (2 \iota ) h_{12}+\cos ^2(\iota ) h_{22}.
\eqn
\end{widetext}

The behaviors of the two modes $h_{+, \times}$ are reflected by the corresponding two dimensionless quantities $h_{+, \times}/(2 m_\star/D_L)$ and are plotted as functions of the dimensionless variable $\lambda/m$ in Fig.~\ref{waveform1} for one of the typical cases $(z, w, v)=(4, 0, 3)$. As an example, in there we set $\gamma=0.5$, $\epsilon=0.9$ and $\zeta=\iota=\pi/4$.  
In order to make that more explicit for the polarizations generated by the small object $m_\star$ when it is traveling at a certain sector of the orbit, in the three different figures of Fig.~\ref{waveform1}, we are using the same color on these curves to mark the corresponding parts spanned by $\lambda$. The zoom-whirl behaviors mentioned in Sec.\ref{secI} are actually quite clear in this figure. 

The behaviors of the two modes $h_{+, \times}$ can vary for different choices of parameters. Some of the examples are shown in Figs.\ref{waveform2} and \ref{waveform3} by choosing $\gamma=0.5, 0.52, 0.55$ and $\zeta=\iota=\pi/4$. For the former we consider the case $(z, w, v)=(4, 0, 3)$ and $\epsilon=0.9$ while for the latter we consider the case $(z, w, v)=(8, 1, 3)$ and $\epsilon=0.5$. Due to the increase of the zoom number $z$, the curves in Fig.\ref{waveform3} adopt a more complicated structure compared to that appearing in Fig.~\ref{waveform2}, as can be anticipated from, e.g., Fig.~\ref{periodic2}.
By looking at these two figures, we can determine how the waveforms are shifted with the changing $\gamma$. 
Our results show that the value of $\gamma$ different from one can lead to a noticeable shift in the phase of the waveform, along with a significant boost in amplitude. 

\begin{figure}[h]
\includegraphics[width=0.9
\linewidth]{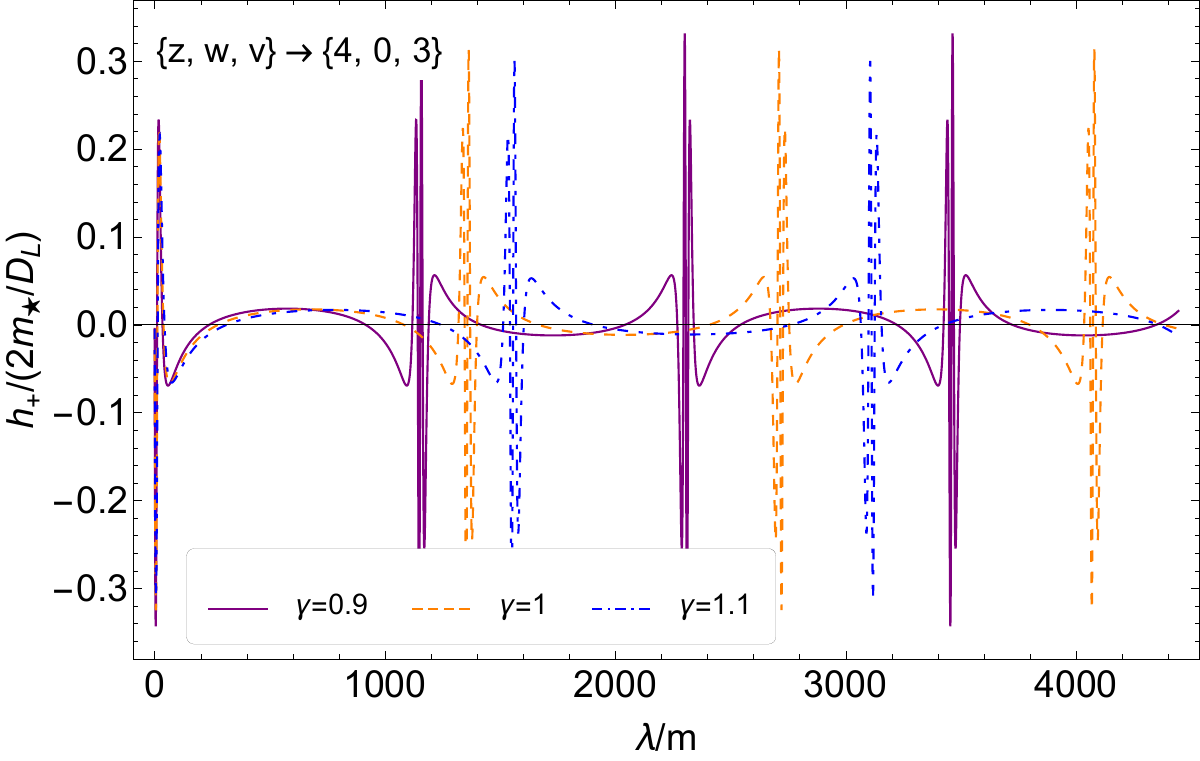} 
\includegraphics[width=0.9
\linewidth]{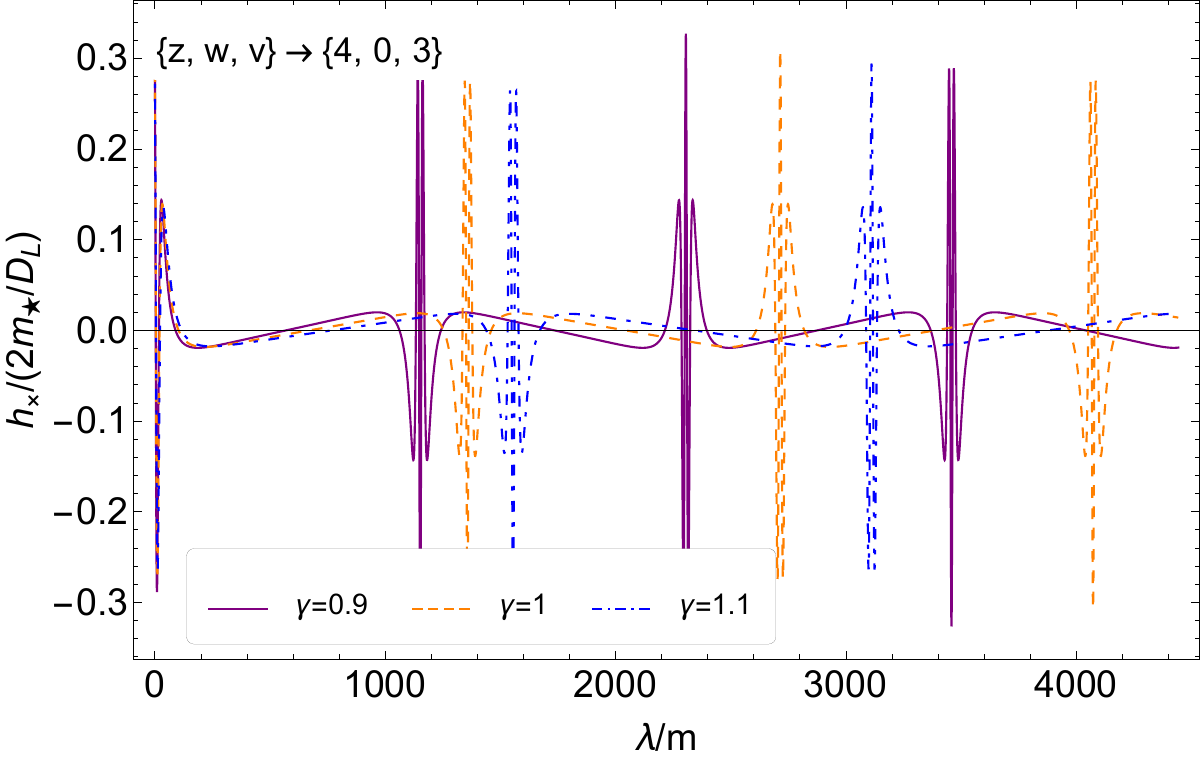} 
\caption{The  behaviors of the two polarizations $h_{+, \times}$ as functions of the dimensionless variable $\lambda/m$ for the typical case $(z,w, v) = (4, 0, 3)$ and $\gamma=0.9$, $1$ (the Schwarzschild case), $1.1$. Here we have set $\epsilon=0.9$ and $\zeta=\iota=\pi/4$. 
} 
\label{waveform4}
\end{figure} 

Knowing that Figs.\ref{waveform2} and \ref{waveform3} provide us with good examples for demonstrating the basic patterns of the modes $h_{+, \times}$, we want to further compare the results of the $\gamma \neq 1$ case with its Schwarzschild counterpart by, e.g., working on top of the results shown in Fig.\ref{periodic3}. Therefore, in Fig.\ref{waveform4} we calculate and plot out the behaviors of the physical quantity $h_{+, \times}/({2 m_\star}/{D_L})$ as a function of dimensionless variable $\lambda/m$ for $\gamma=0.9$, $1$ (the Schwarzschild case), $1.1$ in Fig.\ref{waveform4}. In there we also set  $(z,w, v) = (4, 0, 3)$,  $\epsilon=0.9$ and $\zeta=\iota=\pi/4$, just like in Fig.\ref{periodic3}. 
In comparing to Figs.\ref{waveform2} and \ref{waveform3}, Fig.\ref{waveform4} actually shows similar patterns on $h_{+, \times}$ and their amplitudes stay at the same level except that the overall period of the curves seems to be ``stretched'' in there, which is actually a predictable result since such a tendency has already been reflected by Fig.\ref{waveform2}.

In addition, by considering about one of the typical cases \cite{Babak:2017tow, Wang:2025hla, Yang:2024lmj}, e.g., choosing the mass of the central compact object as $M=3 \times 10^5 M_{\odot}$, the mass of the small object as $m_\star=10 M_{\odot}$ and the luminosity distance as $D_L = 100Mpc$ \cite{Baumann:2022mni}, where $M_{\odot} \approx 1.989 \times 10^{30} kg$ stands for the solar mass, one can calculate the corresponding dimensionless characteristic strain $h_c$ and comparing that with LISA \cite{Robson:2018ifk} as well as TianQin's \cite{Huang:2020rjf} characteristic sensitivity $\sqrt{f S_n(f)}$ to determine the detectability of the EMRI system under consideration. 

First of all, calculate the frequency-domain waveforms ${\tilde h}_{+, \times}(f)$ by working on top of the time-domain results \eqref{hplus} and \eqref{hcross}. Due to the complicity of our problem, here we apply the discrete Fourier transform (DFT) to obtain ${\tilde h}_{+, \times}(f)$. To achieve this, one needs to get the discrete values of $h_{+, \times}(t)$. As a representative example, in this paper we choose the time interval $t\in (0, 20,000)$ plus factors $(z,w, v) = (4, 0, 3)$, $\gamma=0.5, 0.9, 1$, $\epsilon=0.9$ and $\zeta=\iota=\pi/4$ for our calculations. The step size for the time is chosen as $\delta t=t_{\text{max}}/N=0.5$, where $t_{\text{max}}=20,000$ and $N=40,000$. Notice that, here we are adopting the unit system of $c=G_N=m=1$, so that the units for length, time and mass are completely fixed \footnote{In addition to $c=G_N=1$, the $m=1$ is also a natural choice for the unit system, as it brings the shapes of curves for, e.g., $L_{\text{mbo}}$ in Fig.\ref{plotmbo}, $r_{\text{isco}}$ in Fig.\ref{plotisco}, etc. to what they look like right now in their corresponding figures. That is, setting $m=1$ is almost equivalent to plotting out the dimensionless versions of these quantities from the technic point of view.}.  Therefore, a list of $\{h_0, h_1, ..., h_{N}\}$ can be obtained for $h_{+}$ (so does $h_{\times}$, for which we omit the detailed descriptions for simplicity), where $h_n \equiv h_+(\lambda(n \delta t))$ and the solutions of $t(\lambda)$ resulted from \eqref{EOM} is needed in here for finding the correspondence between variables $\lambda$ and $t$, with $n=0, 1, 2,... N$.

\begin{figure}[h]
\includegraphics[width=0.9
\linewidth]{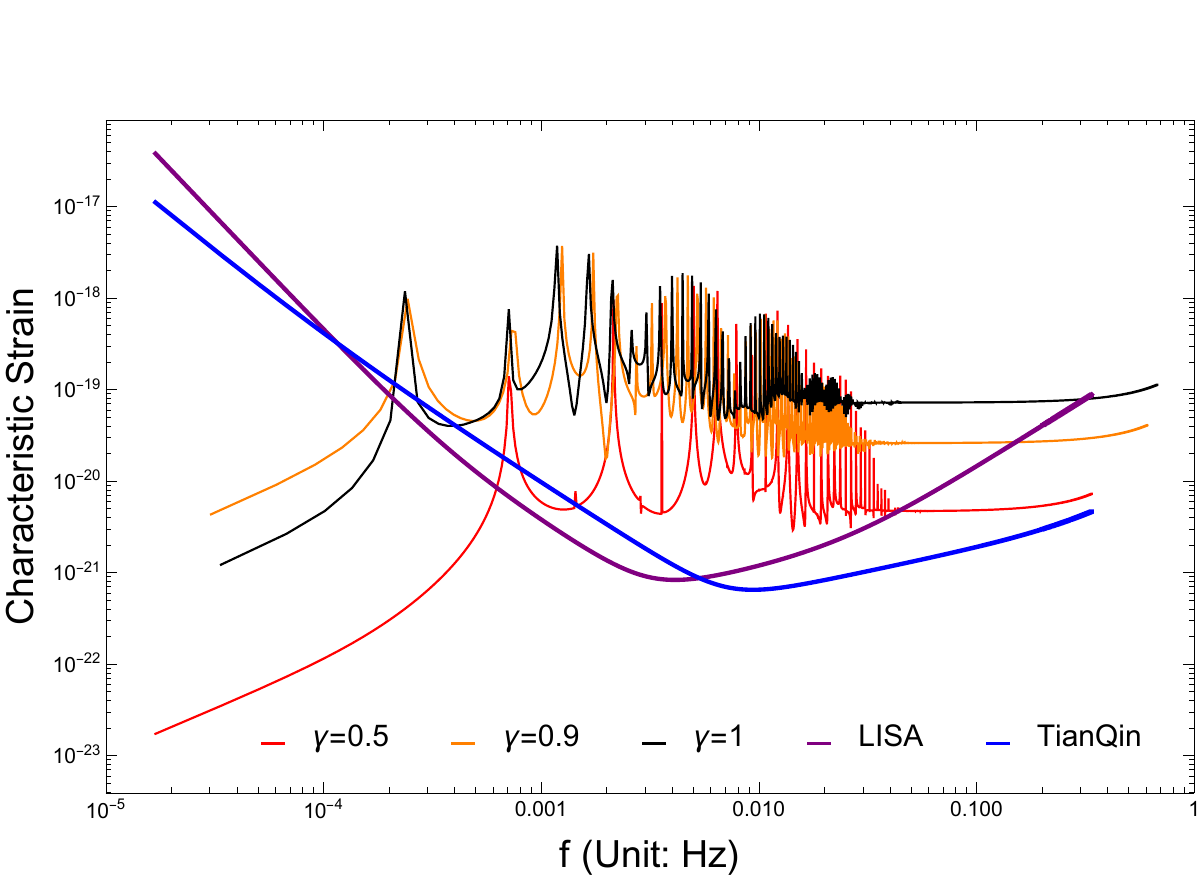} 
\caption{The behaviors of dimensionless quantities $h_c$ [for $\gamma=0.5$, $\gamma=0.9$ and $\gamma=1$ (the Schwarzschild case)] and  $\sqrt{f S_n(f)}$ (for LISA and TianQin detectors) as functions of frequency $f$. Here we are considering the case  $M=3 \times 10^5 M_{\odot}$,  $m_\star=10 M_{\odot}$, $D_L = 100Mpc$ $(z,w, v) = (4, 0, 3)$, $\epsilon=0.9$ and $\zeta=\iota=\pi/4$. The red line corresponds to $\gamma=0.5$, orange line corresponds to $\gamma=0.9$, black line corresponds to $\gamma=1$, purple line corresponds to LISA and blue line corresponds to TianQin. 
} 
\label{plothc}
\end{figure} 

After that, by using some build-in functions of {\it Mathematica} and mimicking what we did in \cite{Zhao:2019kif}, we obtain a list of discrete frequency-domain quantities for ${\tilde h}_{+, \times}(f)$ with DFT. As mentioned by \cite{Robson:2018ifk,  Yang:2024lmj}, we are interested in the dimensionless characteristic strain $h_c(f) \equiv 2 f \sqrt{|{\tilde h}_+|^2+|{\tilde h}_{\times}|^2}$. Therefore, we plot that out by mimicking \cite{Robson:2018ifk} in Fig. \ref{plothc}, together with the curve of dimensionless characteristic sensitivity $\sqrt{f S_n(f)}$ for the LISA and TianQin detectors provided in  \cite{Robson:2018ifk, Huang:2020rjf} (Notice that, in there the horizontal axis has been synchronized for these curves so that we are staying with the SI unit Hz). 

\textcolor{black}{
In Fig. \ref{plothc} the red, orange and black solid lines correspond to the characteristic strain function $h_c(f)$ for $\gamma=0.5$, $\gamma=0.9$ and $\gamma=1$ (the Schwarzschild case), respectively. The sensitivity curves of $\sqrt{f S_n(f)}$ are given by the purple and blue solid lines for the LISA and TianQin detectors.
Corresponding to Fig.\ref{waveform4}, one can observe the discrepancies between the $\gamma=1$ and $\gamma \neq 1$ cases from  Fig. \ref{plothc}. On the other hand, for the frequency regime $f \in (0.001, 0.01)Hz$, we see clearly that the resultant $h_c(f)$'s of the EMRI system under consideration are above the sensitivity curves of LISA and TianQin. This clearly reflects the detectability of these detectors on the GWs emitted by the EMRI system we studied under the $\gamma$-metric. 
}

\section{Discussions and Conclusions}
\label{secVI}


In this paper, we investigate the features of the $\gamma$-metric [cf. \eqref{ds2}] by considering the GWs emitted from a periodic orbit of a small object around a compact object (i.e., an EMRI system is considered) under the adiabatic approximation. This metric is characterized by three known functions, denoted by $F(r)$, $G(r)$, and $H(r)$ [cf., \eqref{FGH}]. Its deviation from that of Schwarzschild is mainly controlled by the dimensionless parameter $\gamma$, with which we further defined the factor $m$ through $m=\gamma M$, where $M$ is the total mass of the source. Setting  $\gamma=1$ brings the  $\gamma$-metric back to that of the Schwarzschild case. 

By scrutinizing \eqref{grr}, a predictable compact body can be obtained by concentrating on the case of $\gamma \in (1/2-\sqrt{5}/2, 1/2+\sqrt{5}/2)$.
To check the deviation of the $\gamma$-metric from that of Schwarzschild, we first plot out the functions of $F(r)$, $G(r)$, and $H(r)$ in Fig.\ref{plotFGH} (The Schwarzschild limitation is added in there as a comparison). According to the allowed choices of $\gamma$ mentioned earlier, in this figure, the cases of $\gamma=0.5$ and $\gamma=1.5$ are considered.  One can clearly discern deviations from the Schwarzschild metric in Fig.~\ref{plotFGH}. This implies the potential discrepancies from that of Schwarzschild for other physical quantities, including the gravitational waveforms (as to be shown later). 

To build up the problem for studying the GWs emitted by the EMRI system, we write down the Lagrangian [cf. \eqref{Lagrangian}], equation of motion [cf.\eqref{EOM}] as well as the geodesic equation [cf. \eqref{geodesic}] for a test particle around the center massive compact object. Throughout the paper, we restrict ourselves to equatorial motion, $\theta(\lambda)=\pi/2$, where $\lambda$ is the affine parameter of the world line of the test particle under consideration. With the help of the above equations, the effective potential can be determined, which is given by \eqref{Veff} and consists of factors of $L$ (the angular momentum, adopting the unit of $m$ under the unit system of $G_N=c=1$) as well as $\gamma$. We plot that out in Fig.\ref{plotveff} for different choices of $L$ and $\gamma$. Fig.\ref{plotveff} displays the effective potential for angular momenta lying between $L_{\text{mbo}}$ and $L_{\text{isco}}$, as we are considering the orbits bounded by the marginally bound orbit (MBO) as well as the innermost stable circular orbit (ISCO). The relative radii, energy (denoted as $E$), and the angular momentum themselves are plotted as functions of $\gamma$ in Figs.\ref{plotmbo} and \ref{plotisco}. Knowing the famous results of a Schwarzschild BH (viz., $r_{\text{isco}}=6M$ and $r_{\text{mbo}}=4M$), the resulting departure from the Schwarzschild limit is immediately apparent.
Fig.\ref{plotmbo} shows that, increasing the parameter $\gamma$ increases both the $r_{\text{mbo}}$ and the  $L_{\text{mbo}}$. Similar trends emerged for ISCOs, as shown in Fig.\ref{plotisco}.


Next, we studied the influence of the parameter $\gamma$ on the periodic orbits of a small particle around the central supermassive compact object.
Each orbit is labelled by a distinct combination of $q$ [viz., $(z, w, v)$, described by \eqref{qradial}], $L$ [viz., $\epsilon$, described by \eqref{Lexpression}] and $E$ (only two of them are independent). Their numerical relations are plotted in Figs.\ref{qE1} and \ref{qL1} for several typical choices of $\gamma$. We shall consult these figures when selecting parameters for calculating those periodic orbits.

With the help of the Euler-Lagrange equation [obtained from \eqref{Lagrangian}] and the equations of motion \eqref{EOM}, we arrive at the ODEs for determining the quantities $r(\lambda)$ and $\phi(\lambda)$ [cf., \eqref{ODEs1}].  Without the loss of generality, we set the initial conditions as $\phi(\lambda=0)=0$, $r(\lambda=0)=r_1$ [cf. \eqref{apsidal}] and ${\dot r}(\lambda=0)=0$ in solving \eqref{ODEs1}. Treating the $\gamma=0.5$ case as an example, we plot out the periodic orbits for different combinations of $q$, $E$, and $L$ in Figs.~\ref{periodic1} and \ref{periodic2}. Notice that: (i) To get rid of the arbitrariness of $m$ (as mentioned earlier), these orbits are plotted out in the 2D coordinate $({\tilde x}, {\tilde y})$ with the two dimensionless variables defined by ${\tilde x} \equiv r/m \cos{\phi}$ and ${\tilde y} \equiv r/m \sin{\phi}$. Similarly, in there we are considering the dimensionless forms of these parameters (e.g., $L/m$); (ii) As can be seen from \eqref{apsidal}, for a pair of given $q$ and $E$\textbackslash$L$, the corresponding $(m^{-1}L)$\textbackslash$E$ needs to be calculated numerically and then be inserted into \eqref{ODEs1} for further calculations.  Clearly, only values of $(m^{-1}L)$ or $E$ that satisfy the resonance condition within numerical precision produce a trajectory compatible with the prescribed $q$. That is why we need to calculate $(m^{-1}L)$\textbackslash$E$ to certain digits (as pointed out earlier in this paper). 
Figs.~\ref{periodic1} and \ref{periodic2} reveal that the orbital morphology depends sensitively on $q$ [as anticipated from the physical interpretation of $(z,w,v)$] and on the chosen values of other parameters. In addition, to better reflect the influence of $\gamma$ on trajectories, they are plotted out for various of $\gamma$ (including the Schwarzschild case) by setting $(z, w, z)=(4, 0, 3)$ and $\epsilon=0.9$ in Fig.\ref{periodic3}.
Overall, the presence of the parameter $\gamma$ significantly alters the characteristics of the periodic orbits compared to a standard Schwarzschild case.

With the orbital information in hand, we are ready to investigate the GWs for the corresponding EMRI system. The calculations can be fulfilled by inserting the orbital information into the equations for the two different modes of the waveforms [cf. \eqref{hplus} and \eqref{hcross}]. This part of results are exhibited in Figs.~\ref{waveform1}-\ref{waveform4}, for which we set $\zeta=\iota=\pi/4$ by mimicking, e.g., \cite{Wang:2025hla}. For the factors $\gamma$, $q$, $\epsilon$, etc., we adopted various values for different curves in these figures. Specially, in Fig.~\ref{waveform1} we use the same color to mark the same period spanned by the dimensionless variable $\lambda/m$ for the trajectory, $h_{+}$ and $h_{\times}$ so that one can determine the correspondence among them. In Figs.~\ref{waveform2} and \ref{waveform3}, we can find out the influence of the parameter $\gamma$ on the polarized waveforms. Clearly, the curves of $h_{+, \times}$ will shift on the axis of $\lambda$ with a small modification on $\gamma$. On the other hand, by comparing Figs.~\ref{waveform2} and \ref{waveform3}, we can see that an increase of zoom number $z$ can indeed bring us a more sophisticated structure of $h_{+, \times}$, as can be anticipated by looking at, e.g., Fig.~\ref{periodic2}. 
\textcolor{black}{To better reflect the influence of $\gamma \neq 1$ in comparing to the Schwarzschild case, the choices of $\gamma=0.9, 1, 1.1$ are considered in Fig.~\ref{waveform4}. By considering a representative case of $M=3 \times 10^5 M_{\odot}$, $m_\star=10 M_{\odot}$, $D_L = 100$~Mpc (see, e.g., \cite{Babak:2017tow}), $(z,w,v)=(4,0,3)$, $\epsilon=0.9$ and $\zeta=\iota=\pi/4$ for different choices of $\gamma$ (including the Schwarzschild limit), we show in Fig.~\ref{plothc} that the EMRI under consideration is well within the detectability of 3rd-generation detectors like LISA and TianQin. This fact provides us with a potential way for distinguishing a standard Schwarzschild BH from the compact object predicted by the $\gamma$-metric (with $\gamma \neq 1$), though we acknowledge that definitive discrimination would require comprehensive parameter estimation studies accounting for spin and other effects in future work.}


This work can be extended to several branches of deeper and further studies. For instance, it is worth investigating more about the behavior of the metric described by \eqref{ds2} across a wider range of $\gamma$, including regimes where potentially subtle behaviors may emerge. We also plan to explore off-equatorial orbits and three-dimensional dynamics, though this introduces considerable complexity including possible non-integrability and chaotic behaviors that require careful treatment. Secondly, one can exploit more combinations of $q$, $E$, and $L$ by starting from, e.g., \eqref{apsidal}, to gain a more comprehensive understanding of waveform morphology. At the same time, we will further study observational signatures for future 3rd-generation GW detectors, with particular attention to refining our waveform calculations. This includes incorporating self-force effects and comparing quadrupole results with Teukolsky-based approaches for strong-field regimes, as well as developing kludge waveforms that balance computational efficiency with physical accuracy. 
\textcolor{black}{In addition, plunging orbits represent a valuable extension, offering access to deeper strong-field regions near $r \approx 2m$. As $\gamma \neq 1$ leads to a curvature singularity at this radius, GWs propagating through such a highly curved spacetime may undergo relativistic effects such as redshift, lensing, and scattering that go beyond the Newtonian order. These propagation effects could imprint unique signatures in the observed waveforms and merit further investigation in future high-fidelity studies. }
Finally, breaking parameter degeneracies—specifically distinguishing $\gamma$ effects from mass, spin, inclination, and eccentricity variations—represents a major long-term endeavor. This requires multiple observational channels, multi-event population analyses, and comprehensive Bayesian model comparison against Kerr spacetime, constituting essential future work building upon the foundational predictions established here.

 
\section*{Acknowledgments}

C.Z. would like to thank Mr. LU Shuo for the valuable discussions. C.Z. is supported in part by the National Natural Science Foundation of China under Grant No. 12205254, and the Science Foundation of China University of Petroleum, Beijing, under Grant No.~2462024BJRC005. T.Z. is supported by the National Natural Science Foundation of China under Grant No.~12275238, No.~12542053, and No.~11675143, the National Key Research and Development Program under Grant No. 2020YFC2201503, and the Zhejiang Provincial Natural Science Foundation of China under Grants No. LR21A050001 and No. LY20A050002, and the Fundamental Research Funds for the Provincial Universities of Zhejiang in China under Grant No. RF-A2019015.





\end{document}